\documentclass[12pt,letterpaper]{article}

\usepackage{latexsym,multirow}
\usepackage{amssymb,amsmath, bm}
\usepackage{graphicx}
\usepackage{verbatim}
\usepackage{centernot}
\usepackage{multicol}
\usepackage{mathrsfs}
\usepackage[export]{adjustbox}
\usepackage{centernot}
\usepackage{ascmac}
\usepackage{caption}
\usepackage{enumitem}

\usepackage{tgtermes}

\usepackage{natbib}
\usepackage[pdftex, bookmarksopen=true, bookmarksnumbered=true,
pdfstartview=FitH, breaklinks=true, urlbordercolor={0 1 0},
citebordercolor={0 0 1}]{hyperref}
\hypersetup{hidelinks = true}
\usepackage{colortbl}
\usepackage{subfigure}
\usepackage{inputenc}
\usepackage{dcolumn}
\newcolumntype{.}{D{.}{.}{-1}}
\newcolumntype{d}[1]{D{.}{.}{#1}}

\usepackage{ntheorem}
\makeatletter
\renewtheoremstyle{plain} 
{\item{\theorem@headerfont ##1\ ##2\theorem@separator}~}
{\item{\theorem@headerfont ##1\ ##2\ (##3)\theorem@separator}~}
\makeatother

\theorembodyfont{\normalfont}
\theoremheaderfont{\normalfont\bfseries}
\theoremstyle{break}
\theoremstyle{plain}
\newtheorem{theorem}{Theorem}

\theoremstyle{break}
\newtheorem{assumption}{Assumption}
\theoremstyle{definition}
\theoremstyle{break}
\newtheorem{definition}{Definition}

\theoremstyle{break}
\theoremstyle{plain}
\newtheorem{lemma}{Lemma}

\theoremstyle{break}

\newcommand{\qed}{\hfill \ensuremath{\Box}}
\newcommand{\indep}{\mbox{$\perp\!\!\!\perp$}}

\usepackage{xr}
\usepackage{arydshln}

\usepackage[compact]{titlesec}

\usepackage{array}
\newcolumntype{L}[1]{>{\raggedright\let\newline\\\arraybackslash\hspace{0pt}}m{#1}}
\newcolumntype{C}[1]{>{\centering\let\newline\\\arraybackslash\hspace{0pt}}m{#1}}
\newcolumntype{R}[1]{>{\raggedleft\let\newline\\\arraybackslash\hspace{0pt}}m{#1}}

\usepackage{booktabs}

\usepackage{tcolorbox}

\usepackage{framed}

\allowdisplaybreaks 

\begin{document}

\newcommand\ud{\mathrm{d}}
\newcommand\dist{\buildrel\rm d\over\sim}
\newcommand\ind{\stackrel{\rm indep.}{\sim}}
\newcommand\iid{\stackrel{\rm i.i.d.}{\sim}}
\newcommand\logit{{\rm logit}}
\renewcommand\r{\right}
\renewcommand\l{\left}
  \newcommand\pre{{(t-1)}}
  \newcommand\cur{{(t)}}
  \newcommand\cA{\mathcal{A}}
  \newcommand\cB{\mathcal{B}}
  \newcommand\bone{\mathbf{1}}
  \newcommand\E{\mathbb{E}}
  \newcommand\Var{{\rm Var}}
  \newcommand\cD{\mathcal{D}}
  \newcommand\cK{\mathcal{K}}
  \newcommand\cP{\mathcal{P}}
  \newcommand\cT{\mathcal{T}}
  \newcommand\cX{\mathcal{X}}
  \newcommand\cXR{\mathcal{X,R}}
  \newcommand\wX{\widetilde{X}}
  \newcommand\wT{\widetilde{T}}
  \newcommand\wY{\widetilde{Y}}
  \newcommand\wZ{\widetilde{Z}}
  \newcommand\bX{\mathbf{X}}
  \newcommand\bx{\mathbf{x}}
  \newcommand\bT{\mathbf{T}}
  \newcommand\bt{\mathbf{t}}
  \newcommand\bwT{\widetilde{\mathbf{T}}}
  \newcommand\bwt{\tilde{\mathbf{t}}}
  \newcommand\bbT{\overline{\mathbf{T}}}
  \newcommand\bbt{\overline{\mathbf{t}}}
  \newcommand\ubT{\underline{\mathbf{T}}}
  \newcommand\ubt{\underline{\mathbf{t}}}
  \newcommand\bhT{\widehat{\mathbf{T}}}
  \newcommand\bht{\hat{\mathbf{t}}}

  \newcommand\cF{\mathcal{F}} 
  \newcommand\cC{\mathcal{C}} 
  \newcommand\cS{\mathcal{S}} 
  \newcommand\cN{\mathcal{N}} 
  \newcommand\bZ{\mathbf{Z}} 
  \newcommand\bz{\mathbf{z}} 

  \newcommand\bW{\mathbf{W}} 
  \newcommand\bY{\mathbf{Y}} 

  \newcommand\bC{\mathbf{C}} 
  \newcommand\bc{\mathbf{c}} 

  \newcommand\bV{\mathbf{V}} 
  \newcommand\bv{\mathbf{v}} 
  \newcommand\bbv{\mathbf{\bar{v}}} 
  \newcommand\bbx{\mathbf{\bar{x}}} 

  \newcommand\cw{\mathcal{w}} 
  \newcommand\cW{\mathcal{W}} 

  \newcommand\bg{\bar{g}} 
  \newcommand\pg{g^\prime} 

  \newcommand\bw{\mathbf{w}} 

  \newcommand\cG{\mathcal{G}} 
  \newcommand\cH{\mathcal{H}} 
  \newcommand\cU{\mathcal{U}} 

  \newcommand\bu{\mathbf{u}} 

  \newcommand\pS{\mathscr{S}}
  \newcommand\pP{\mathscr{P}}

  \newcommand\cM{\mathcal{M}} 
  \newcommand\cO{\mathcal{O}} 
  \newcommand\cJ{\mathcal{J}} 

  \newcommand\CDE{{\rm {\bf CDE}}}
  \newcommand\ADE{{\rm {\bf ADE}}}
  \newcommand\cADE{{\rm {\bf cADE}}}
  \newcommand\DE{{\rm {\bf DE}}}
  \newcommand\cANSE{{\rm {\bf cANSE}}}
  \newcommand\ANSE{{\rm {\bf ANSE}}}
  \newcommand\ASE{{\rm {\bf ASE}}}
  \newcommand\MSE{{\rm {\bf MSE}}}
  \newcommand\INT{{\rm {\bf INT}}}
  \newcommand\ATSE{{\rm {\bf ATSE}}}
  \newcommand\MDE{{\rm {\bf MDE}}}
  \newcommand\CSE{{\rm {\bf CSE}}}
  \newcommand\TSE{{\rm {\bf TSE}}}

  \newcommand\Prz{{\rm Pr}_{\mathbf{z}}}

  \newenvironment{changemargin}[2]{%
    \begin{list}{}{%
        \setlength{\topsep}{0pt}%
        \setlength{\leftmargin}{#1}%
        \setlength{\rightmargin}{#2}%
        \setlength{\listparindent}{\parindent}%
        \setlength{\itemindent}{\parindent}%
        \setlength{\parsep}{\parskip}%
      }%
    \item[]}{\end{list}}

  \newcommand\MR{{\rm MR}}
  \newcommand\RR{{\rm RR}}

  \newcommand\Nor{{\rm Normal}}

  \newcommand\cATIE{{\sc cATIE}} 
  \newcommand\ATIE{{\sc ATIE}}
  \newcommand\cg{\cellcolor[gray]{0.7}}

  \newcommand\mo{\mathbf{1}}
  \newcommand\PA{\mbox{PA}}
  \newcommand\PAo{\mbox{PA}^\ast}
  \newcommand\sign{\texttt{sign}}

  \newcommand{\argmax}{\operatornamewithlimits{argmax}}
  \newcommand{\argmin}{\operatornamewithlimits{argmin}}

  \newcommand{\minb}{\operatornamewithlimits{min}}
  \newcommand{\maxb}{\operatornamewithlimits{max}}

  \newcommand\spacingset[1]{\renewcommand{\baselinestretch}%
    {#1}\small\normalsize}

  \spacingset{1.25}

  \newcommand{\tit}{\vspace{-0.3in} {\fontsize{18}{15}\selectfont Identification of Causal
      Diffusion Effects  \\ Under Structural Stationarity}}



  \title{\bf \tit \thanks{I thank Peter Aronow, Eytan Bakshy, Matt
      Blackwell, Dean Eckles, Justin Grimmer, 
      Erin Hartman, Zhichao Jiang, Gary King, Dean Knox, 
      James Robins, Ilya Shpitser, Dustin Tingley, Tyler VanderWeele,
      Soichiro Yamauchi, and participants of
      the 2019 Atlantic Causal Inference Conference, for helpful
      comments and discussions. I am particularly grateful to Kosuke Imai, Rafaela
      Dancygier, and Brandon Stewart for their detailed feedback. 
      The earlier draft of this article was entitled, ``Identification of Causal Diffusion Effects
      Using Stationary Causal Directed Acyclic Graphs,'' (Egami, 2018),
      arXiv: \url{https://arxiv.org/abs/1810.07858v1} \vspace{0.1in}}}

  \author{Naoki Egami \thanks{Assistant Professor  (starting in 2020), Department of
      Political Science, Columbia University, New York NY 10027. Ph.D. Candidate, Department of Politics,
      Princeton University, Princeton NJ 08544. Email:
      \href{mailto:negami@princeton.edu}{negami@princeton.edu}, URL:
      \href{http://scholar.princeton.edu/negami}{http://scholar.princeton.edu/negami}}}

  \date{First Version: August 29, 2018 \\ 
    This Version: January 26, 2020}

  \maketitle




  \pdfbookmark[1]{Title Page}{Title Page}
  \vspace{-0.35in}
  \thispagestyle{empty} 
  \setcounter{page}{0} 
  \begin{abstract}
    Although social and biomedical scientists have long been interested in the process
    through which ideas and behaviors diffuse, the identification of causal
    diffusion effects, also known as peer and contagion effects, remains
    challenging. Many scholars consider the commonly used assumption of no
    omitted confounders to be untenable due to contextual confounding and
    homophily bias. To address this long-standing 
    problem, we examine the causal identification under a new assumption of {\it structural stationarity}, which
    formalizes the underlying diffusion process with a class of dynamic
    causal directed acyclic graphs. First, we develop a statistical test that can detect a wide range of
    biases, including the two types mentioned above. 
    We then propose a difference-in-differences style estimator that can directly
    correct biases under an additional parametric assumption. Leveraging
    the proposed methods, we study the spatial diffusion of hate crimes
    against refugees in Germany. After correcting large upward bias in existing studies,
    we find hate crimes diffuse only to areas that have a high proportion
    of school dropouts.  
  \end{abstract}

  \noindent%
  \vspace{0.2in}
  \vfill

  \newpage
  \spacingset{1.83} 


  \section{Introduction}
  Scientists have long been interested in how ideas and behaviors diffuse across space, networks,
  and time. For example, social scientists have studied the diffusion of 
  policies and voting behaviors in political science \citep{sinclair2012social,
    graham2013diffusion, jones2017social}, educational outcomes and crimes in
  economics \citep{glaeser1996crime, sacerdote2001peer,
    duflo2011peer}, and innovations and job attainment in sociology
  \citep{rogers1962diffusion, granovetter1973strength}. Epidemiologists
  and researchers in public health have
  focused on the spread of infectious disease \citep{halloran1995causal,
    crawford2018risk, cai2019identification} and health behavior \citep{fowler2013review}. In
  each of these research areas, a growing number of scholars aim to estimate the
  causal impact of diffusion dynamics, that is, how much an outcome of one unit causes, not just
  correlates with, an outcome of another unit. 

  Despite its importance, the identification of causal diffusion
  effects, also known as peer effects, contagion effects, or social influence, is challenging
  \citep{manski1993identification, vanderweele2013social}. Although commonly-used statistical methods, including spatial econometric 
  models \citep[e.g.,][]{anselin2013spatial}, require the assumption of no
  omitted confounders, this assumption is often untenable due to two 
  well-known types of confounding; contextual confounding and
  homophily bias \citep{ogburn2017challenges}. When there
  exist some unobserved contextual factors that affect multiple units, we suffer from {\it
    contextual confounding} --- we cannot distinguish whether units affect one another through diffusion
  processes or units are jointly affected by the shared unobserved contextual variables.
  {\it Homophily bias} arises when the spatial or network proximity is
  affected by some unobserved characteristics. We cannot discern whether
  units close to one another exhibit similar outcomes because of diffusion or because they selectively become
  closer in space or networks with others who have similar unobserved characteristics. Emphasizing concerns
  over these biases, influential papers across disciplines criticize existing diffusion
  studies \citep[e.g.,][]{cohen2008obesity, lyons2011spread, angrist2014perils}. In fact, 
  causal diffusion effects are often found to be overestimated by a
  large amount, for example, by 300 -- 700\% \citep{aral2009distinguishing, eckles2017bias}. 
  \cite{shalizi2011homophily} argue that it is nearly impossible to
  credibly estimate causal diffusion effects from observational studies
  by relying on the conventional assumption of no omitted confounders.

  To address this long-standing challenge, we examine the
  identification of causal diffusion effects under a new assumption of
  {\it structural stationarity}, which formalizes diffusion processes 
  with a causal directed acyclic graph (DAG) approach
  \citep{pearl2000causality, ogburn2014DAG}. In particular, we assume that the
  underlying causal DAG belongs
  to a class of dynamic causal DAGs \citep{dean1989model,
    pearl2001bayesian}, which repeat nonparametric causal
  substructure over time (Section~\ref{subsec:dagIden}). Thus, the structural stationarity assumption
  requires the existence of causal
  relationships among variables --- not the effect or sign of such
  relationships --- to be stable over time. This is in contrast
  to a usual DAG-based approach that assumes a specific causal DAG and
  the full knowledge of its structure, which may be difficult to
  justify in applied contexts. Instead, we propose methodologies that
  have the same statistical guarantees for any causal DAG within the
  general class of dynamic causal DAGs. 


  Under the structural stationarity, we first develop a placebo test
  that uses a lagged dependent variable to detect a wide class of
  biases, including contextual confounding and 
  homophily bias (Section~\ref{subsec:placebo}). It assesses whether a lagged dependent variable is 
  conditionally independent of the treatment variable. We prove statistical
  properties of the test based on a new theorem, which states
  that under the structural stationarity, the no omitted confounders assumption is equivalent to the 
  conditional independence of a lagged dependent variable and the
  treatment variable. This proof exploits the structure of back-door
  paths \citep{pearl1995causal} and the graphical
  representation of the no omitted confounders assumption
  \citep*{shpitser2012validity} under the structural stationarity.

  In addition, we propose a bias-corrected estimator that can
  directly remove biases under an additional parametric assumption (Section~\ref{subsec:calibrate}). In
  its basic form, it subtracts the bias detected by the placebo
  test from a biased estimator. We prove unbiasedness of this estimator
  under a parametric assumption that the effect and imbalance of
  unobserved confounders are constant over time. We describe its
  connection to the widely-used difference-in-differences
  estimator \citep{angrist2008mostly, tchetgen2016negative}. 

  Applying the proposed methods, we study the spatial diffusion of hate crimes against
  refugees in Germany. Facing the biggest refugee crisis since the Second World War, 
  Germany has recently registered more than 1 million asylum applications,
  making them the largest refugee-hosting country in Europe
  \citep{unhcr2017}. During this time period, the number of
  hate crimes against refugees has substantially increased, a close to
  200\% increase from 2015 to 2016. A clear, {\it descriptive} pattern is that the incidence of
  hate crimes was spatially clustered and the number grew over time as
  waves (see Section~\ref{sec:mot}). However, what is the {\it causal} process behind this dynamic spatial pattern?
  Understanding the causal impact of hate crime diffusion is of policy and scientific interest to prevent further spread
  of hate crimes. We leverage the proposed placebo test and
  bias-corrected estimator to tackle concerns about unmeasured contextual
  confounding. See Section~\ref{sec:mot} for the details of the data and Section~\ref{sec:app} for empirical analysis.


  This article builds on a growing literature of causal diffusion
  effects \citep{shalizi2011homophily, imbens2013social,
    ogburn2017challenges}.\footnote{\spacingset{1}{\footnotesize Related
      but different literature is on causal inference with interference. The
      difference is that while interference
      focuses primarily on the causal effect of others' {\it treatments},
      diffusion (a.k.a, peer and contagion effects)
      considers the causal effect of others' {\it outcomes}
      \citep{ogburn2014DAG}. See \citet{halloran2016review} for a review of the interference literature.}} 
  In addition to research on the use of experimental 
  or quasi-experimental design \citep{bramoulle2009identification,
    omalley2014diff, an2015instrumental, eckles2017Online,
    basse2019peer, jagadeesan2019designs, li2019randomization}, 
  a series of papers address problems of omitted
  confounders by deriving tests or bounds \citep[e.g.,][]{anagnostopoulos2008influence}. \cite{vanderweele2012why} show that after controlling for
  homophily bias and contextual confounding, the spatial autoregressive model can
  be used to test the existence of diffusion effects. To compute bounds for
  diffusion effects, \cite{versteeg2010difftest,
    versteeg2013statistical} examine a specific causal DAG only with homophily and
  diffusion, and \cite{vanderweele2011sen} proposes sensitivity analysis
  methods. This paper shares concerns about the no omitted confounders
  assumption. However, instead of testing the existence of diffusion
  effects or deriving bounds, this paper focuses on the point
  identification and estimating the magnitude of causal diffusion effects. 

  This paper also draws upon emerging literature of negative controls \citep{tchetgen2010negative,
    tchetgen2013control}. In particular, this paper
  extends recent studies using negative controls in panel data
  settings \citep{tchetgen2016negative, flanders2017new,
    tchetgen2017invited} to the identification of causal diffusion effects. The 
  proposed methods differ from the previous literature in that we use
  the structural stationarity, which assumes a class of
  dynamic causal DAGs rather than one specific causal DAG. This class of dynamic causal
  DAGs \citep{pearl2001bayesian} is a causal extension of the dynamic bayesian networks (DBN) popular in
  the probabilistic graphical modeling literature
  \citep[e.g.,][]{murphy2002dynamic}. The key difference is that while the DBN often 
  assumes the parameters of conditional probability distributions are
  time-invariant, the dynamic causal DAG only assumes the stability of
  the nonparametric causal structure and allows for any higher-order
  Markov model. Finally, causal DAGs \citep{pearl2000causality} are useful not only for causal
  identification but also for asymptotic statistical
  inference. \cite{van2014causal} and \cite{ogburn2017causal}  
  offer one of the first foundations to use causal directed acyclic 
  graphs for network data. \cite{tchetgen2017auto} provide an alternative approach using
  chain graphs. Because we focus on the
  identification of causal diffusion effects, our proposed methods are complementary to these
  recent papers that develop theories of statistical inference in a network asymptotic regime.


  \section{A Motivating Empirical Application:  \vspace{-0.2in}\\ Spatial Diffusion of Hate Crimes against Refugees}
  \label{sec:mot}
  Research across the social sciences has shown 
  that many types of violence are contagious \citep{wilson1982broken,myers2000diffusion}. One small act of violence can trigger another act of violence, which again induces another, and can lead to waves of violence \citep{hill1986contagion,
    gleditsch2008conflict}. Without taking into account how violent behaviors spread across space, it is difficult to
  explain when, where, and why some areas experience violence and to prevent further spread of violence. 

  In this paper, we investigate the spatial diffusion of hate crimes
  against refugees in Germany, one of the most pressing problems in the country.
  Over the last few years, Germany has experienced a record influx of
  refugees \citep{bamf2019report}, and during the same time period, the number of hate crimes
  against refugees has increased substantially. Our primary data source of
  hate crimes is a project, Mut gegen rechte Gewalt
  (courage against right-wing violence), by the Amadeu Antonio
  Foundation and the weekly magazine {\it Stern}, which has been
  documenting anti-refugee violence in Germany since the beginning of
  2014. 
  This data source has
  been recently analyzed by several papers  \citep[e.g.,][]{benvcek2016refugees, jackle2016dark}.  
  The dataset we analyze in this paper is compiled by 
  \cite{dancygier2018hate}, who extended this hate crime data by
  merging in other variables, such as the number of refugees, the population
  size, a proportion of school dropouts and unemployment rates, collected from the Federal Statistical Office
  in Germany. 
  
\begin{figure}[!t]
    \begin{center}
      \begin{tabular}{c}
        \hspace{-0.3in}
        \begin{minipage}[b]{0.3\textwidth}   
          \begin{center}
            \hspace{-0.2in}
            \includegraphics[height=6.3cm]{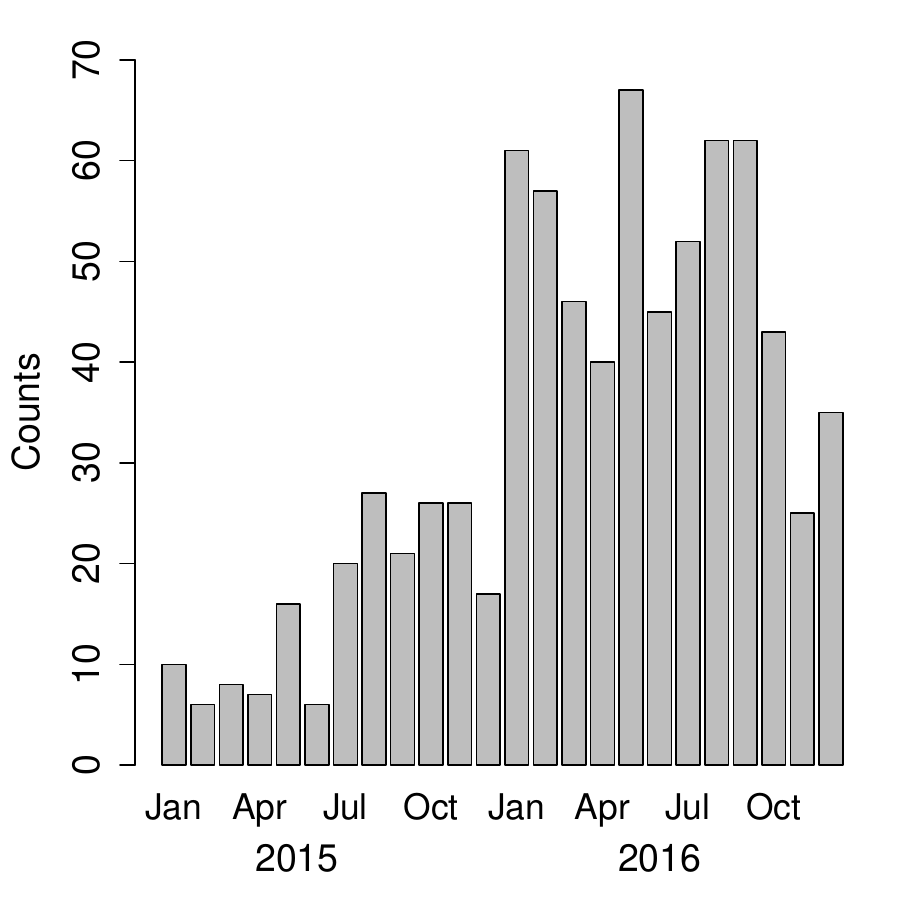}\\
          \end{center}
        \end{minipage}      
        \hspace{0.5in}   
        \begin{minipage}[b]{0.3\textwidth}   
          \begin{center}
            \includegraphics[height=6cm]{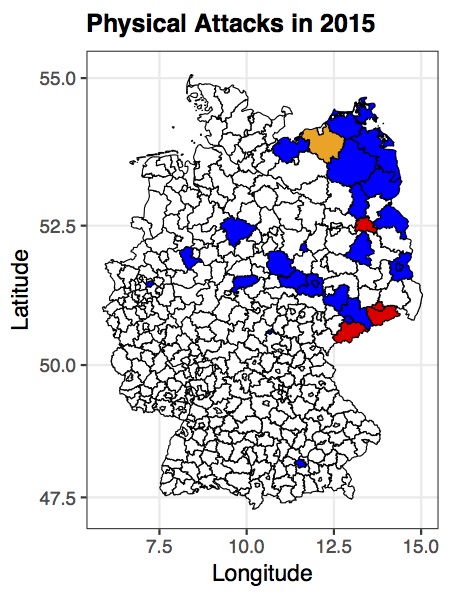}\\
          \end{center}
        \end{minipage}   
        \hspace{-0.05in}  
        \begin{minipage}[b]{0.33\textwidth}   
          \begin{center}
            \includegraphics[height=6cm]{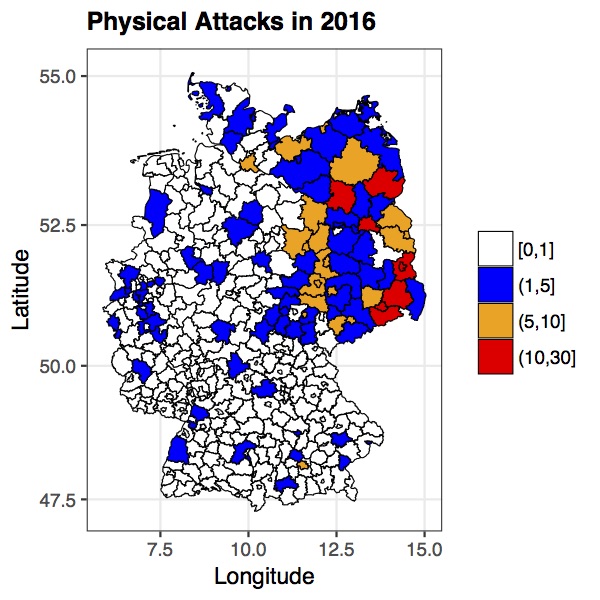}
          \end{center}
        \end{minipage}   
      \end{tabular}
    \end{center}
    \vspace{-0.2in} \hspace{0.7in} (a) Time Trend  \hspace{1.85in} (b) Spatial Patterns
    \spacingset{1}{\caption{Temporal and Spatial Patterns of Hate Crimes
        in Germany. {\small Note: The left figure shows the number of
          physical attacks each month. In the middle and right figures, we show the
          number of physical attacks in each county in 2015 and 2016,
          respectively. Each of 402 counties is 
          colored in white, blue, orange, or red if the number of hate crimes in a given year is less
          than or equal to 1, 5, 10, or greater than 10, respectively.}\vspace{-0.1in}} \label{fig:hatemap}}
  \end{figure}

  Figure~\ref{fig:hatemap} (a) reports 
  the number of physical attacks against refugees each month, from
  the beginning of 2015 to the end of 2016. While there were about 15
  hate crimes on average in each month of 2015, this rose to more
  than 40 in 2016, a close to 200\% increase. Figure~\ref{fig:hatemap} (b) presents the spatial patterns over the two years. 
  Two empirical patterns are worth noting. First, hate crimes were spatially clustered in East 
  Germany. Second, the number of counties that experience hate
  crimes grew over time as waves. This dynamic spatial pattern is consistent with the spatial diffusion
  theory which argues that hate crimes diffuse from one county to
  another spatially proximate county over time
  \citep{myers2000diffusion, braun2011diffusion}. Indeed,
  \cite{jackle2016dark} found that the incidence of hate crimes in one county predicts that of 
  hate crimes in its spatially proximate counties using the data from Germany in 2015. 

  However, it is challenging to estimate the causal impact of this
  spatial diffusion process because there exist well-known concerns of contextual
  confounding: many unobserved confounders can be spatially correlated. For example, the number of refugees 
  increased substantially during this period and is also spatially
  correlated. Even if we collect a long list of covariates, it
  is difficult to assess whether a selected set of
  control variables is sufficient for removing contextual confounding. 
  To address this type of pervasive concerns over bias, we develop a
  placebo test to detect bias and a bias-corrected estimator to remove
  bias. The main empirical analysis appears in
  Section~\ref{sec:app}. Although our empirical application 
  focuses on the spatial diffusion problem, the proposed approach
  is also applicable to network diffusion settings where homophily bias is a common concern.



  \section{The Setup for Causal Diffusion Analysis}
  \label{sec:setup}
  Causal diffusion, also known as peer and contagion effects, 
  refers to a process in which an outcome of one unit influences an outcome of another unit over time
  \citep{shalizi2011homophily, vanderweele2012why}. This section
  introduces the setup for analyzing such causal diffusion. We define the
  average causal diffusion effect and then describe challenges for its identification. 

  \subsection{Average Causal Diffusion Effect}
  \label{subsec:potential}
  Consider $n$ units over $T$ time periods. Let $Y_{it}$ be the outcome for unit $i$ at time $t$ 
  for $i \in \{1, \ldots, n\}$ and $t \in \{0, 1, \ldots, T\}.$ Use $\bY_t$ 
  to denote a vector $(Y_{1t}, \ldots, Y_{nt})$, which contains the
  outcomes at time $t$ for $n$ units. To encode spatial or network
  connections between these $n$ units, we follow the standard spatial statistics literature
  \citep{cressie2015statistics} and use a distance matrix $\bW$ where
  $\bW$ can be an asymmetric, weighted matrix. 
  In the motivating application, it is of interest to estimate how much hate crimes in
  one county diffuse to other spatially proximate counties. Here, the
  distance matrix $\bW$ could encode physical distance between counties
  where $W_{ij}$ might be an inverse of the distance between 
  district $i$ and $j$. In network diffusion settings, $W_{ij}$ could
  represent a directed tie, e.g., whether unit $i$ follows unit $j$ in a
  Twitter network. 
  Define {\it neighbors} $\cN_i$ to be other units who are connected
  with a given unit $i$, i.e., $\cN_i \equiv \{j : W_{ij} \neq 0\}.$ In spatial diffusion analysis, researchers often assign $0$ to $W_{ij}$
  when the distance between two units is greater than a certain
  threshold, e.g., 100 km. 

  We rely on potential outcomes \citep{neyman1923,
    rubin1974causal} to formally define causal diffusion effects. 
  Based on the tradition of spatial econometrics \citep{anselin2013spatial,
    franzese2007spatial}, this paper focuses on the
  weighted average of the neighbors' outcomes $\bW_{i}^\top
  \bY_t$ as the treatment variable. Although we keep this setup throughout the
  paper, the methods in this paper can be easily
  applied to other definitions of the treatment variable. We use $D_{it} \equiv \bW_{i}^\top \bY_t$ to denote the treatment variable
  and let $Y_{i,t+1} (d)$ represent the potential outcome variable of 
  unit $i$ at time $t+1$ if the unit receives the treatment $D_{it} = d$. 

  We are interested in the {\it average causal diffusion effect} (ACDE) at
  time $t+1$, which is defined as the average causal effect of the treatment variable 
  $D_{it}$ on the outcome at time $t+1$ \citep{ogburn2014DAG, ogburn2017challenges}. It is the comparison between 
  the potential outcome under a higher value of the treatment
  $D_{it}= d^H$ and the potential outcome under a lower value of the
  treatment $D_{it}= d^L.$
  \vspace{0.1in}
  \begin{definition}[Average Causal Diffusion Effect]
    \label{acde} \spacingset{1.2}{
      The average causal diffusion effect (ACDE) at time $t+1$ is defined as, 
      \begin{equation} 
        \tau_{t+1}(d^H, d^L)  \equiv \E[Y_{i,t+1} (d^H) - Y_{i,t+1} (d^L)],  \label{eq:ACDE}
      \end{equation}
      where $d^H$ and $d^L$ are two constants specified by researchers. 
    } \vspace{-0.1in}
  \end{definition}
  For example, the ACDE could quantify how much the 
  risk of having hate crimes in the next month changes if we see more
  hate crimes in neighboring counties this month. This captures how
  much hate crimes diffuse across space over time. 


  Finally, we introduce an assumption about the measurement of 
  outcomes. We assume that we observe one of the potential outcomes at
  every time period $t = 1, \ldots, T$. 
  \vspace{0.1in}
  \begin{assumption}[Sequential Consistency] 
    \label{seq}  \spacingset{1.2}{
      For every unit at every time period $t=1, \ldots, T$, one of the potential outcome
      variables is observed, and the realized outcome variable for unit $i$ at time $t+1$ is denoted by  
      \begin{equation} 
        Y_{i, t+1} = Y_{i,t+1} (D_{it}).
      \end{equation}}
  \end{assumption}
  This is a simple extension of the consistency assumption widely used
  in the cross-sectional settings \citep{vanderweele2009concerning}
  to the diffusion setup. The assumption means that we avoid the temporal
  aggregation problem \citep{granger1988some} that can mask the dynamics of the underlying diffusion process. Its violation implies simultaneity bias, that
  is, the treatment variable and the outcome variable simultaneously
  cause each other \citep{danks2013learning, hyttinen2016causal}. 
  In the literature of causal diffusion analysis, this assumption is
  essential because, without it, the causal order 
  of the treatment and outcome becomes ambiguous, and causal diffusion 
  effects are no longer well-defined \citep{lyons2011spread,
    ogburn2014DAG, ogburn2017challenges}. See \cite{joffe2011control} for a similar problem in the structural
  nested model and g-estimation.  In practice, researchers can make
  this assumption more plausible by measuring outcomes 
  frequently. For example, the assumption could be more tenable
  when we can measure the incidence of hate crimes monthly rather
  than annually. We maintain this assumption throughout the paper given its essential role
  in defining the ACDE, but in
  Appendix~\ref{subsec:joint}, we also 
  discuss the connection between its violation and the proposed placebo test. 

  
  \subsection{{\fontsize{13.2}{10}\selectfont Identification under No Omitted Confounders Assumption}}
  \label{subsec:iden}
  We now describe the widely used identification assumption of no omitted
  confounders and explain pervasive concerns about its violation. This
  assumption states that all relevant confounders are in a selected set of control
  variables. Formally, the potential outcomes at time $t+1$ are independent of a joint distribution of
  neighbors' outcomes at time $t$ given control variables. 
  \vspace{0.1in}
  \begin{assumption}[No Omitted Confounders]
    \label{ig} \spacingset{1.2}{
      For $i=1, 2, \ldots, n$, \vspace{-0.05in}
      \begin{eqnarray}
        Y_{i,t+1} (d) \ \indep \ \{Y_{jt}\}_{j \in \cN_i} \ \mid \bC_{i,t+1},
      \end{eqnarray}
      for $d \in \cD$ where $\cD$ is the support of $D_{it}$, and $\bC_{i,t+1}$ is a set of
      pretreatment variables, which we call a {\it control set}.} Note
    that control set $\bC_{i,t+1}$ can include time-independent variables and
    time-dependent variables measured at time $t+1$ or before $t+1$.  \vspace{-0.1in}
  \end{assumption}
  Under the assumption of no omitted confounders, the ACDE is identified
  as follows. 
  \begin{eqnarray}
    \hspace{-0.8in} && \tau_{t+1} (d^H, d^L)  =  \hspace{-0.1in}\int_{\mathcal{C}}  \biggl\{
                       \E[Y_{i,t+1} | D_{it} = d^H, \bC_{i,t+1} =\bc ] -
                       \E[Y_{i,t+1} | D_{it} = d^L,
                       \bC_{i,t+1}  = \bc ] \biggr\} d
                       F_{\bC_{i,t+1}} (\bc), \label{eq:iden}
  \end{eqnarray}
  where $F_{\bC_{i,t+1}} (\bc)$ is the cumulative distribution function of
  $\bC_{i,t+1}$ and the standard overlap assumption is made:
  $\Pr (D_{it}=d^H|\bC_{i,t+1}=\bc) > 0 $ and $\Pr
  (D_{it}=d^L|\bC_{i,t+1}=\bc) > 0$  for $i=1, \ldots, n$ and all $\bc \in
  \mathcal{C}$ where $\cC$ is the support of $\bC_{i,t+1}$. 
  We can estimate the ACDE by
  estimating the conditional expectation $\E[Y_{i,t+1} | D_{it}, \bC_{i,t+1}]$ and
  then averaging it over the empirical distribution of control variables
  $\bC_{i,t+1}$. 

  Although many empirical studies of diffusion make the assumption of no
  omitted confounders, it is widely known that the assumption is often questionable in
  practice \citep{manski1993identification, shalizi2011homophily,
    vanderweele2013social}. This concern is pervasive mainly because it implies the absence
  of two well-known types of biases: contextual confounding and homophily
  bias. {\it Contextual confounding} -- the primary focus of the spatial
  diffusion literature -- can exist when units share some 
  unobserved contextual factors. For example, in the motivating
  application of hate crime diffusion,  the
  risk of having hate crimes is likely to be affected by some economic policies, which often affect
  multiple counties at the same time. In this case, researchers might
  observe spatial clusters of hate crimes even without diffusion. Another well-known type of bias is {\it homophily bias} -- the main
  concern in the network diffusion literature. This bias arises when units become connected due to their unobserved
  characteristics. For example, voters who are connected to each other
  can have similar political opinions without any diffusion or
  social influence because people who have similar political views might
  become friends in the first place \citep{fowler2011causality}. We
  discuss the causal DAG representation of these biases when we
  introduce our proposed methods in Section~\ref{sec:method}.




  \section{The Proposed Methodology}
  \label{sec:method}
  In this section, we examine the identification of causal diffusion
  effects under a new assumption of 
  structural stationarity. After introducing the assumption (Section~\ref{subsec:dagIden}), we
  first develop a statistical placebo test to detect a wide range of
  biases (Section~\ref{subsec:placebo}) and then propose a bias-corrected
  estimator (Section~\ref{subsec:calibrate}).

  \subsection{Structural Stationarity}
  \label{subsec:dagIden}
  We formalize the underlying diffusion process with a causal directed
  acyclic graph (DAG) framework \citep{pearl2000causality}. In
  particular, we assume the {\it structural stationarity}, which states the
  underlying causal DAG belongs to a general class of dynamic causal
  DAGs \citep{dean1989model, pearl2001bayesian}. It requires that the existence of causal relationships
  between variables, not the effect or sign of such relationships, to be
  stable over time. A class of dynamic
  causal DAGs and the structural stationarity are formally defined as
  follows. We review basic causal DAG terminologies in Appendix~\ref{sec:dag_review}. 
  \vspace{0.01in} 
  \begin{definition}[Dynamic Causal DAGs \citep{dean1989model, pearl2000causality}]
    \label{d-dag} \vspace{0.1in}
    \spacingset{1.2}{
      Consider variables in a causal DAG $\cG$ that
      have more than one child or have at least one parent. Among these variables, distinguish two types; the
      time-independent variable $Z_i$ and the time-dependent variable
      $X_{it}$. A class of dynamic causal DAGs is
      any causal DAG $\cG$ that satisfies the following conditions. {\small
        \begin{align*}
          (2.1) \ \ & X_{it} \in \PA(X_{i,t+1}) \ \ \mbox{ for }  i \in
                      \{1, \ldots, n\} \mbox{ and } \ t= 0, \ldots, T-1.  \notag \\
          (2.2) \ \ & \mbox{For} \ i, i^\prime \in \{1, \ldots, n\}, \ \exists \ t, k \ \mbox{s.t. }
                      X_{it} \in \PA(\widetilde{X}_{i^\prime, t+k}) \Rightarrow \
                      X_{it^\prime} \in \PA(\widetilde{X}_{i^\prime, t^\prime+k}) \
                      \mbox{ for all } t^\prime = 0, \ldots, T - k. \notag \\
          (2.3) \ \ & \mbox{For} \ i, i^\prime \in \{1, \ldots, n\}, \ \exists \ t \ \mbox{s.t. }
                      Z_i \in \PA(X_{i^\prime t})
                      \ \Rightarrow \  Z_i \in \PA(X_{i^\prime t^\prime})
                      \ \mbox{ for all } t^\prime = 0, \ldots, T,
        \end{align*}} 
      where $ A \in \PA(B)$ indicates that variable $A$ is a parent of
      variable $B$.} 
  \end{definition}
  \vspace{0.01in} 
  \begin{assumption}[Structural Stationarity]
    \label{ss} 
    \spacingset{1.2}{
      The distribution over outcome $Y$, treatment $D$, and control
      variables $\bC$ is faithful to one of the dynamic causal DAGs.    
    }
  \end{assumption}

  \begin{figure}[!t]
    \begin{center}
      \begin{tabular}{c}
        \hspace{-0.3in}
        \begin{minipage}{0.45\textwidth}   
          \begin{center}
            \hspace{-0.2in}
            \includegraphics[width=7.25cm]{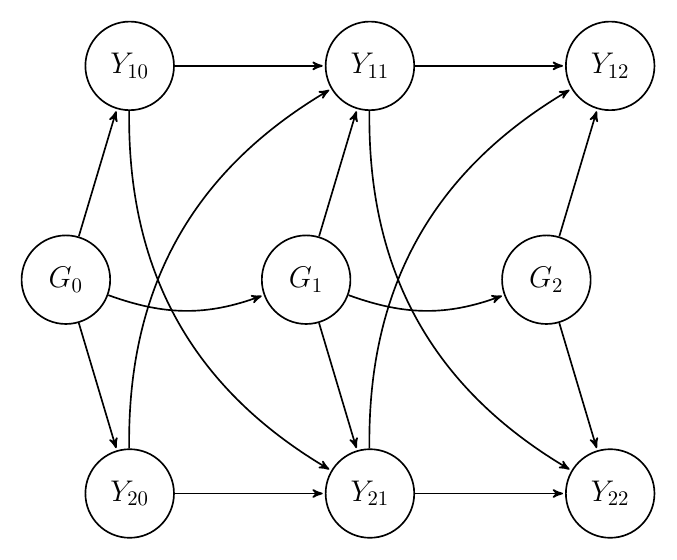} \\
            (a) Structural Stationarity
          \end{center}
        \end{minipage}
        \hspace{0.3in}
        \begin{minipage}{0.45\textwidth}   
          \begin{center}
            \hspace{-0.2in}
            \includegraphics[width=7.25cm]{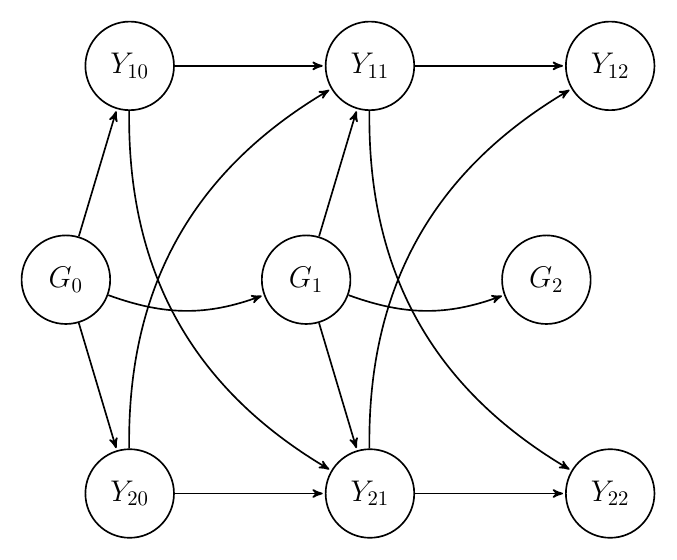} \\
            (b) Violation of Structural Stationarity
          \end{center}
        \end{minipage}      
      \end{tabular}
    \end{center}
    \vspace{-0.2in} 
    \spacingset{1}{\caption{Illustration of Structural Stationarity. {\small
          Note: Six nodes $Y_{it}$ represent outcome
          variables for two individuals $ i \in \{1, 2\}$ over three time periods
          $ t \in \{0, 1, 2\}.$ Three nodes $G_{t}$ are contextual variables
          for $t \in \{0, 1, 2\}.$ In the first panel, the causal structure between
          variables $Y$ and $G$ are stable over time. In the second
          panel, variable $G$ has no effect on $Y$ at $t=2$ and thus the
          structural stationarity is violated.}} \label{fig:ss}}
  \end{figure}

  The faithfulness is defined as follows. If a distribution is
  faithful to causal directed acyclic graph  $\cG$, variables $A$ and $B$ are independent if and
  only if the variables are d-separated in $\cG$
  \citep{spirtes2000causation}. 
  Condition 2.1 of Definition~\ref{d-dag} requires that all time-dependent variables that have at
  least one parent be affected by their own lagged variables. This
  condition is more plausible when the time intervals are 
  shorter. Condition 2.2 means that if two time-dependent variables have a
  child-parent relationship at one time period, the same causal
  relationship should exist for all other time periods. Similarly,
  Condition 2.3 requires that if a time-independent variable is a parent of a time-dependent variable
  at one time period, the same child-parent relationship should exist at all other time periods. The last
  two requirements are the core -- the
  existence of causal relationships should be stable over
  time. Importantly, the effect of each variable can be changing over
  time; the only requirement is the time-invariant existence of the
  causal relationships. Figure~\ref{fig:ss} visualizes examples of the
  structural stationarity and its violation. 

  In our motivating application, suppose that the unemployment rate is a
  confounder in one month. Then, the structural stationarity requires that the unemployment rate 
  should remain a confounder during the time periods we analyze. The assumption is
  violated when a set of confounders changes over time. The
  effect of the unemployment rate can be changing over time.  

  Several points are worth noting. First, the structural stationarity
  only assumes a {\it class} of dynamic causal DAGs rather than a
  specific dynamic causal DAG. This is in contrast to conventional DAG
  approaches that assume one particular DAG and require full knowledge
  of its DAG structure. Thus, researchers can rely on the structural
  stationarity assumption even when they cannot justify their full
  knowledge of the underlying DAG structure, as far as the existence of
  causal relationships is time-invariant. 

  Second, the structural stationarity is often a natural
  requirement in applied contexts.  In fact, causal DAGs in several 
  important papers about causal diffusion effects
  \citep{shalizi2011homophily, omalley2014diff, ogburn2014DAG} are
  examples of dynamic causal DAGs. Causal DAGs in the causal discovery literature
  often impose a similar but stronger condition \citep{danks2013learning, hyttinen2016causal}. They often assume that
  variables are affected only by one-time lag (also known as the
  first-order Markov assumption) and this structure is time-invariant. In
  contrast, the structural stationarity allows for any higher-order
  temporal dependence (see Condition 2.2 of
  Definition~\ref{d-dag}). Finally, when the underlying causal structure changes at some time,
  the structural stationarity is violated. However, if researchers know the time when the
  underlying structure changes, we can still make use of the structural
  stationarity assumption separately, before and after this time point.

  \subsection{Placebo Test to Detect Bias}
  \label{subsec:placebo}
  Under the structural stationarity, we propose a placebo test -- using a lagged dependent
  variable as a general placebo outcome -- that can detect a wide class of 
  biases, including contextual confounding and homophily bias. 
  This placebo test helps the credible identification of causal diffusion effects by statistically 
  assessing the validity of the confounder adjustment. We focus on 
  theories and methodologies of the placebo test in this section, and we
  provide a simulation study calibrated to the hate crime data in Appendix~\ref{subsec:simpl}.


  \subsubsection{Equivalence Theorem}
  The proposed placebo test exploits a lagged dependent variable as a
  placebo outcome. It tests the assumption of no omitted confounders by assessing
  whether a lagged dependent variable is conditionally independent of
  the treatment variable. This placebo test is formally justified based on
  the equivalence theorem, which states that, under the structural stationarity, the assumption of no omitted
  confounders is equivalent to the conditional independence of the
  simultaneous outcomes given {\it a placebo set} defined below. This theorem and the placebo test are formally written as follows.
  \begin{theorem}[\small Equivalence between No Omitted Confounders
    Assumption and Conditional Independence of Simultaneous Outcomes]
    \label{placebo} \spacingset{1.3}{
      Under Assumption~\ref{seq} and Assumption~\ref{ss},
      \begin{equation}
        Y_{i, t+1}(d)   \ \indep \ \{Y_{jt}\}_{j \in \cN_i} \mid \bC_{i,t+1} \ \Longleftrightarrow 
        Y_{it} \ \indep \  \{Y_{jt}\}_{j \in \cN_i} \mid \bC_{i,t+1}^P, \label{eq:placebotest}   
      \end{equation}
      where a placebo set $\bC^P$ is defined as 
      \begin{equation}
        \bC_{i,t+1}^P \equiv   \{ \bC_{i,t+1}, \bC_{i,t+1}^{(-1)}, \{Y_{j,t-1}\}_{j \in \cN_i} \}\setminus \mbox{Des}(Y_{it}), \label{eq:placeboset}   
      \end{equation}
      where  $\bC_{i,t+1}^{(-1)}$ is a lag of the time-dependent variables in $\bC_{i,t+1}$, $\{Y_{j,
        t-1}\}_{j \in \cN_i}$ is a lag of the treatment variable, 
      and $\mbox{Des}(Y_{it})$ is a descendant of $Y_{it}$, i.e.,
      variables affected by $Y_{it}$. As a regularity condition, we assume that the violation
      of the no omitted confounders assumption, if any, is due to
      unobserved confounders, i.e., the change in the lag-structure of
      the selected control set cannot remove the bias (see Appendix~\ref{subsec:placebo-proof} for details).
    }
  \end{theorem}
  \vspace{-0.2in}
  \begin{framed}
    \spacingset{1.2}{
      \noindent  {\bf Placebo Test:} For a given control set $\bC$, the following test statistically
      assesses whether the control set contains all confounders, i.e., Assumption~\ref{ig}. 
      {\setlength{\leftmargini}{10pt}
        \begin{enumerate}   
        \item[] {\bf Step 1}:  Derive placebo set $\bC_{i,t+1}^P$ from control set $\bC_{i,t+1}$ based on equation~\eqref{eq:placeboset}.
        \item[] {\bf Step 2}:  Test the conditional independence, $Y_{it} \ \indep \  \{Y_{jt}\}_{j \in \cN_i} \mid \bC_{i,t+1}^P.$     \vspace{-0.1in}
        \end{enumerate}    
      }
      \vspace{0.05in}
      \noindent Note: the first step follows a deterministic rule to derive
      placebo set $\bC_{i,t+1}^P$. \\ (1) add lags of existing control variables and a
      lag of the treatment variable to the original control set $\bC$, and (2) remove all the variables affected by outcomes at time $t$.
    }
  \end{framed}
  The proof of Theorem~\ref{placebo} (in
  Appendix~\ref{subsec:placebo-proof}) exploits the structure of back-door
  paths \citep{pearl1995causal} and the graphical
  representation of the no omitted confounders assumption
  \citep*{shpitser2012validity} under the structural stationarity. 
  In equation~\eqref{eq:placebotest}, the assumption of no 
  omitted confounders (the left-hand side) is proven to be
  equivalent to the conditional independence of the observed outcome of
  individual $i$ and her neighbors' outcomes at the same time period
  given a placebo set (the right-hand side). Because this right-hand
  side is observable and testable, this theorem directly implies that we can
  statistically assess the assumption of no omitted confounders by the
  placebo test of the conditional independence of the simultaneous
  outcomes $Y_{it} \ \indep \  \{Y_{jt}\}_{j \in \cN_i} \mid \bC_{i,t+1}^P$. 

  The basic idea behind the theorem is as follows: under the structural
  stationarity, back-door paths between the main outcome and the
  treatment are similar to those between the lagged dependent variable
  and the treatment. 
  The difference between control set $\bC$ and placebo set $\bC^P$ is to
  formally guarantee that unblocked back-door paths between the main outcome and the treatment 
  are the same (from a causal graph perspective) 
  to those between the placebo outcome and the treatment. To derive this placebo set, we only
  need to know which variables in the control set are time-dependent and
  which variables are affected by outcomes at time $t$. The former 
  information is often readily available, and the latter one is the same
  as the information used to avoid post-treatment bias in the standard causal inference settings. 


  \begin{figure}[!t]
    \begin{center}
      \begin{tabular}{c}
        \hspace{-0.3in}
        \begin{minipage}{0.45\textwidth}   
          \begin{center}
            \hspace{-0.2in}
            \includegraphics[width=7.25cm]{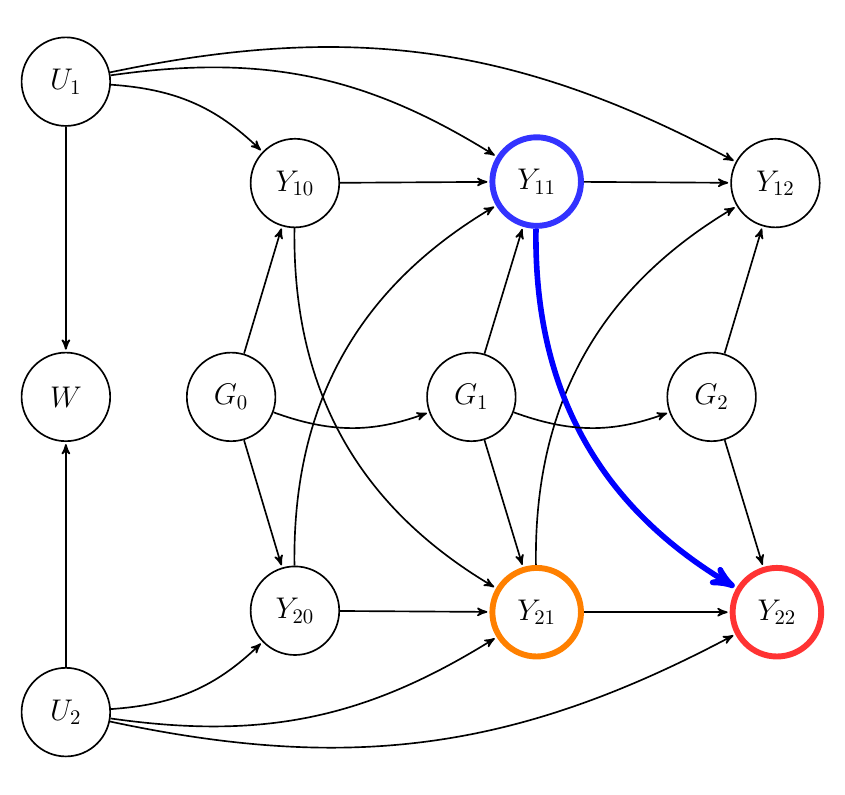} \\
            \hspace{-0.1in} (a) Example of Placebo Test
          \end{center}
        \end{minipage}      
        \begin{minipage}{0.5\textwidth}   
          \vspace{0.6in}
          \begin{center}
            {\footnotesize
              \begin{tabular}{|l|c|c|c|}
                \hline
                & $C$ & $C^P$ &  Placebo Test \\    [0.15ex]
                \hline
                No Bias & $Y_{21}, U_2, G_2$ & $Y_{20}, Y_{10}, U_2, G_2, G_1$
                              & Accept\\ [0.4ex] 
                \cdashline{1-4}
                Contextual & \multirow{2}{*}{$Y_{21}, U_2$} 
                      & \multirow{2}{*}{$Y_{20}, Y_{10}, U_2$} 
                              & \multirow{2}{*}{Reject} \\ [0.2pt]
                Confounding &  & & \\ [0.2pt] \cdashline{1-4}
                Homophily Bias& $Y_{21}, G_2, G_1$  &$Y_{20}, Y_{10},G_2, G_1,G_0$  & Reject\\
                [0.4ex]  \cdashline{1-4} 
                Both & $Y_{21}, Y_{20} $  &$Y_{20}, Y_{10}$  & Reject\\ [0.3ex]  \cdashline{1-4}
                \hline
              \end{tabular}}        
          \end{center}
          \centering         \vspace{0.2in}
          \hspace{0.6in} (b) Control and Placebo Sets
        \end{minipage}   
      \end{tabular}
    \end{center}
    \vspace{-0.2in} 
    \spacingset{1}{\caption{Illustration of Placebo Test. {\small
          Note: We focus on the ACDE of 
          $Y_{11}$ on $Y_{22}$ where $Y_{11}$ is the treatment variable (blue),
          $Y_{22}$ is the outcome variable (red), and the causal arrow of
          interest $Y_{11} \rightarrow Y_{22}$ is colored blue. 
          The placebo outcome $Y_{21}$ is colored orange. }} \label{fig:placeboDAG}}
  \end{figure}

  \subsubsection{Illustrations with Causal DAGs}
  Although the proposed placebo test is applicable to any causal DAGs that satisfy the
  structural stationarity, we consider a causal DAG in Figure~\ref{fig:placeboDAG} (a) as one concrete example. 
  The causal DAG has twelve nodes in total; six nodes $Y_{it}$ representing outcome
  variables for two individuals $ i \in \{1, 2\}$ over three time periods
  $ t \in \{0, 1, 2\},$ three nodes $G_{t}$ representing contextual variables
  for $t \in \{0, 1, 2\},$ two nodes $U_i$ representing individual-level
  characteristics for $i \in \{1, 2\},$ and finally variable $W$
  indicating the connection of two individuals, taking $1$ if they are
  connected and $0$ otherwise. Suppose we are interested in the ACDE of 
  $Y_{11}$ on $Y_{22}$ where $Y_{11}$ is the treatment variable (blue),
  $Y_{22}$ is the outcome variable (red), and the causal arrow of
  interest $Y_{11} \rightarrow Y_{22}$ is colored blue.  The placebo outcome $Y_{21}$ is colored orange.  

  Based on this causal DAG in Figure~\ref{fig:placeboDAG} (a), 
  Table in Figure~\ref{fig:placeboDAG} (b)
  shows four different scenarios: no bias, contextual confounding,
  homophily bias, and both types of biases. For each set of control variables, the
  placebo test checks conditional independence, $Y_{11} \ \indep \ 
  Y_{21} \mid \bC^P$ where we derive a placebo set $\bC^P$ from a
  chosen control set $\bC$ using equation~\eqref{eq:placeboset}. These scenarios show 
  how the placebo test detects biases by exploiting the structural stationarity. 

  First, when we control for three variables $\{Y_{21}, U_2, G_2\},$
  the ACDE of interest is identified (``No Bias''). Without 
  knowledge of the entire causal DAG, we can assess the absence of bias 
  by implementing the placebo test. Following
  equation~\eqref{eq:placeboset}, we derive a placebo set $\bC^P =
  \{Y_{20}, Y_{10}, U_2, G_2, G_1\}$ and then the placebo test checks
  $Y_{11} \indep Y_{21} | \bC^P.$ In Figure~\ref{fig:placeboDAG} (a),  
  there is no unblocked back-door path between $Y_{11}$ and $Y_{21}$, and
  the conditional independence holds as Theorem~\ref{placebo} implies. 

  Second, we consider a typical form of contextual
  confounding. When we control for two variables $\{Y_{21}, U_2\},$
  the ACDE is not identified due to a back-door path
  ($Y_{11} \leftarrow G_1 \rightarrow G_2 \rightarrow Y_{22}$). We now verify
  that the placebo test correctly detects this bias. We first derive
  a placebo set as $\bC^P = \{Y_{20}, Y_{10}, U_2\}$ and then assess
  whether there is any unblocked back-door path between $Y_{11}$ and $Y_{21}.$ In
  fact, we correctly reject the placebo test; $Y_{11} \centernot{\indep}
  Y_{21} | \bC^P$ due to a back-door path $(Y_{11} \leftarrow G_1
  \rightarrow Y_{21})$.

  Finally, we investigate homophily bias. When we control for three
  variables $\{Y_{21}, G_2, G_1\},$ the ACDE is not identified due to a back-door path
  ($Y_{11} \leftarrow U_1 \rightarrow \fbox{$W$} \leftarrow U_2
  \rightarrow Y_{22}$) where the square box means that connection
  variable $W$ is adjusted for. As shown in \cite{shalizi2011homophily}, $W$ is always,
  often implicitly, adjusted for in any causal diffusion analysis because researchers need to compare observations with
  similar spatial/network pre-treatment characteristics. In this case, a placebo set is $\bC^P = \{Y_{20}, Y_{10}, G_2, G_1,
  G_0\}$ and we can verify that $Y_{11} \centernot{\indep}
  Y_{21} | \bC^P$ due to a back-door path ($Y_{11} \leftarrow U_1
  \rightarrow \fbox{$W$} \leftarrow U_2 \rightarrow Y_{21}$). The placebo test
  correctly detects homophily bias. If we follow the same logic, it is straightforward to verify
  that the placebo test can also detect biases even when contextual confounding and homophily bias coexist. 



  \subsubsection{Connection to Spatial Autoregressive Model}
  \label{subsubsec:sar}


  Although there are many ways to implement the second step of the
  placebo test, one approach is a parametric test based on the spatial autoregressive (SAR) model
  \citep[e.g.,][]{anselin2013spatial, cressie2015statistics}. 
  For example,  when outcomes are continuous, we can implement the placebo test by the following linear spatial autoregressive model. 
  \begin{eqnarray}
    Y_{it}  & = & \alpha_0 + \delta \bW_i^\top \bY_t + \gamma_0^\top
                  \bC_{i,t+1}^P + \epsilon_{it}, \label{eq:sar}
  \end{eqnarray}
  where $\bW_i^\top \bY_t \equiv D_{it}$ is the treatment variable,
  $\bC_{i,t+1}^P$ is a placebo set, and $\epsilon_{it}$ is an error
  term. In the motivating application (Section~\ref{sec:app}), we employ
  logistic spatial autoregressive model in a similar way. It is
  important to note that the equivalence theorem (Theorem~\ref{placebo})
  is nonparametric, so researchers can combine the theorem with any
  nonparametric or parametric models in applied settings. 

  Theorem~\ref{placebo} implies that the placebo outcome $Y_{it}$ is conditionally independent of the
  treatment variable when the assumption of no omitted confounders
  (Assumption~\ref{ig}) holds. 
  Therefore, the spatial autoregressive coefficient $\delta$ serves as a test 
  statistic of the placebo test. By testing whether this spatial
  autoregressive coefficient is zero, researchers can assess the no
  omitted confounders assumption and thus detect biases, including
  contextual confounding and homophily bias. In Appendix~\ref{subsec:simpl}, 
  we investigate the statistical power of the proposed placebo
  test through simulation studies and show that its power is comparable
  to a theoretical upper bound. 

  This use of the SAR model as a placebo test differs from 
  existing approaches in the spatial econometrics literature that are
  designed to capture spatial correlations \citep[e.g.,][]{anselin2013spatial}. While researchers
  conventionally interpret the spatial autoregressive
  coefficient as the strength of the spatial correlation, the proposed placebo test uses
  the spatial autoregressive coefficient to detect biases rather than to 
  estimate diffusion effects. For the estimation of the
  ACDE, we estimate the conditional expectation $\widehat{\E}[Y_{i,t+1} \mid D_{it}, \bC_{i,t+1}]$ 
  and then uses the identification formula in equation~\eqref{eq:iden}. 


  It is important to note that if the parametric assumptions of
  the model are violated, the spatial autoregressive coefficient in
  equation~\eqref{eq:sar} can be zero even when unmeasured confounding
  remains. Like any other statistical tests, a specific parametric placebo
  test can fail if its underlying parametric assumptions do not
  hold. A key advantage of the proposed approach is that the
  equivalence theorem (Theorem~\ref{placebo}) is nonparametric. The
  theorem implies that when there exist no omitted confounders, the placebo outcome and
  the treatment are conditionally independent in any parametric
  and nonparametric tests. Therefore, in practice, researchers can 
  verify the conditional independence of the placebo outcome and the
  treatment variable using additional non- or semiparametric
  conditional independence tests
  \citep[e.g.,][]{su2008nonpara, zhang2012kernel}. 

  \subsection{Bias-Corrected Estimator}
  \label{subsec:calibrate}
  If the placebo test detects bias, one may want to 
  collect more data and improve the selection of control variables. This strategy
  might, however, be infeasible in many applied settings. To help researchers in
  such common situations, this section considers how to correct biases by introducing an
  additional parametric assumption. We start with a simple example of
  linear models (Section~\ref{subsubsec:example}) and then provide
  general results in Sections~\ref{subsubsec:BCtheory}
  and~\ref{subsubsec:BCtheory2}. We provide simulation evidence in Appendix~\ref{subsec:simbc}.

  \subsubsection{An Example with Linear Models}
  \label{subsubsec:example}
  To develop an intuition for a bias-corrected  estimator, we first
  consider a simple example with linear models. We assume here that a
  selected set of control variables is time-independent and the same as
  its corresponding placebo set. A general result is provided in the
  following subsections. 

  Suppose we fit a linear model in which we regress the outcome at time
  $t+1$ on the treatment variable and the selected control set. \vspace{-0.05in}
  \begin{eqnarray}
    Y_{i,t+1}  & = & \alpha + \beta D_{it} + \gamma^\top \bC_{i, t+1} +
                     \tilde{\epsilon}_{i,t+1}, \label{eq:main} \vspace{-0.1in}
  \end{eqnarray}
  where $D_{it}$ is the treatment variable, $\bC_{i, t+1}$ is the selected control set, and $\tilde{\epsilon}_{i,t+1}$ is an error
  term. If the assumption of no omitted confounders (Assumption~\ref{ig}) holds, $\hat{\beta} \times
  (d^H - d^L)$ is an unbiased estimator of the ACDE given
  that the linear model specification is correct. In contrast, when the no omitted
  confounders assumption is violated, this estimator is biased. We would like to assess whether the
  assumption of no omitted confounders holds and also correct biases, if any. 

  To assess the assumption of no omitted confounders, suppose we 
  run a parametric placebo test using the following linear spatial  
  autoregressive model as in equation~\eqref{eq:sar}. 
  \begin{eqnarray*}
    Y_{it}  & = & \alpha_0 + \delta D_{it} + \gamma_0^\top
                  \bC_{i, t+1}^P + \epsilon_{it}, \vspace{-0.1in}
  \end{eqnarray*}
  where $\bC_{i, t+1}^P$ is a placebo set and $\epsilon_{it}$ is an error term. If the assumption of no omitted
  confounders holds, the spatial autoregressive coefficient $\delta$
  should be zero (Theorem~\ref{placebo}). In contrast, if the assumption
  of no omitted confounders does not hold, an estimated coefficient 
  $\hat{\delta}$ then serves as a bias-correction term.

  In this simple example, a proposed bias-corrected estimator is given
  by subtracting the bias-correction term $\hat{\delta}$ from an original
  biased estimator $\hat{\beta}$. \vspace{-0.05in}
  \begin{equation}
    \hat{\tau}_{BC}(d^H, d^L) \ \equiv \ (\hat{\beta}  - \hat{\delta})
    \times (d^H - d^L). \vspace{-0.1in}
  \end{equation}
  This bias-corrected estimator is unbiased for the ACDE for the
  treated under an additional parametric assumption we discuss in
  detail in the next subsection (Assumption~\ref{timeInv}). Note that when the assumption of no omitted confounders holds, the expected value
  of $\hat{\delta}$ is zero, meaning no bias correction.

  \subsubsection{Assumption}
  \label{subsubsec:BCtheory}
  To describe a general bias-corrected estimator, we begin by defining the average causal diffusion 
  effect for the treated (ACDT). We will show in
  Theorem~\ref{biascorrect} that the proposed bias-corrected estimator 
  is unbiased for the ACDT. The formal definition is as follows. 
  \begin{equation} 
    \tau^{d^H}_{t+1} (d^H, d^L)  \equiv \E[Y_{i,t+1} (d^H) - Y_{i,t+1} (d^L) \mid D_{it} = d^H].  \label{eq:ACDET}
  \end{equation}
  This is the average causal diffusion effect for units who received the
  higher level of the treatment. This quantity could represent the
  causal diffusion effect of hate crimes for counties in a higher risk
  neighborhood, i.e., $d^H\%$ of neighboring counties had hate
  crimes in month $t$. 

  To introduce necessary assumptions, we divide a control
  set into three types of variables $\bC_{i,t+1} \ \equiv \ \{ \bX_{i,t+1}, \bV_{i,t+1}, \bZ_{i} \}$ 
  where (1) $\bX_{i,t+1}$, the time-dependent variables
  that are descendants of $Y_{it}$, (2) $\bV_{i,t+1}$, the time-dependent variables
  that are not descendants of $Y_{it}$, and (3) $\bZ_i$, the
  time-independent variables.  Then, we can write a corresponding
  placebo set as $\bC_{i,t+1}^P \ \equiv \ \{ \bX_{it}, \bV_{i, t+1},
  \bV_{it}, \bZ_{i}, \{Y_{j, t-1}\}_{j \in \cN_i} \}.$ 

  Without loss of generality, first define an unobserved confounder
  $U$ such that the no omitted confounder assumption holds conditional on $U_{i,t+1}$ and the
  original control set $\bC_{i,t+1}$, i.e., $Y_{i,t+1} (d^L) \  \indep   \  \{Y_{jt}\}_{j \in \cN_i}
  \mid U_{i,t+1}, \bC_{i,t+1}.$ For simpler illustrations, we assume here that this $U_{i,t+1}$ is a descendant of
  $Y_{it}$ (general results are in Appendix~\ref{subsec:proof-correct}). 
  Theorem~\ref{placebo} then implies that observed
  simultaneous outcomes are independent conditional on $U_{it}$ and
  $\bC_{i,t+1}^P,$ i.e., $Y_{it}  \ \indep \   \  \{Y_{jt}\}_{j \in \cN_i} \mid U_{it}, \bC_{i,t+1}^P$.  


  With this setup, we introduce an assumption necessary for the bias
  correction; the effect and imbalance of unobserved confounders are constant over time. This is an extension of
  the structural stationarity (Assumption~\ref{ss}): while
  the structural stationarity only requires that the existence of causal
  relationships among outcomes and confounders be time-invariant, this
  additional parametric assumption requires that some of such causal
  relationships should have the same effect size over time. 
  \vspace{0.1in}
  \begin{assumption}[Time-Invariant Effect and Imbalance of Unobserved Confounder]
    \label{timeInv} \spacingset{1.2}{
      \begin{enumerate}
      \item Time-invariant effect of unobserved confounder $U$: For all $u_1, u_0, \bx$ and $\bc,$
        \begin{eqnarray*}
          \hspace{-0.6in} && \E[Y_{i, t+1} (d^L) | U_{i,t+1} = u_1, \bX_{i,t+1}=\bx, \bC_{i,t+1}^B=\bc] - \E[Y_{i, t+1} (d^L) | U_{i,t+1} = u_0, \bX_{i,t+1}=\bx, \bC_{i,t+1}^B=\bc] \\
          \hspace{-0.6in} & = & \E[Y_{it} (d^L) | U_{it} = u_1, \bX_{it}=\bx, \bC_{i,t+1}^B=\bc] - \E[Y_{it} (d^L) | U_{it} = u_0, \bX_{it}=\bx, \bC_{i,t+1}^B=\bc].
        \end{eqnarray*}
      \item Time-invariant imbalance of unobserved confounder $U$: For all $u, \bx$ and $\bc,$
        \begin{eqnarray*}
          \hspace{-0.6in} && \Pr (U_{i,t+1} \leq u \mid D_{it} = d^H, \bX_{i,t+1}=\bx,
                             \bC_{i,t+1}^B=\bc) - \Pr (U_{i,t+1} \leq u \mid D_{it} = d^L, \bX_{i,t+1}=\bx, 
                             \bC_{i,t+1}^B=\bc)  \\ 
          \hspace{-0.6in} & = &   \Pr (U_{it} \leq u \mid D_{it} = d^H, \bX_{it}=\bx,
                                \bC_{i,t+1}^B=\bc) -\Pr (U_{it} \leq u \mid D_{it} = d^L, \bX_{it}=\bx,
                                \bC_{i,t+1}^B=\bc).
        \end{eqnarray*}
      \end{enumerate}
      where $\bC_{i,t+1}^B \ \equiv \ \{ \bV_{i,t+1}, \bV_{it}, \bZ_{i}, \{Y_{j, t-1}\}_{j \in \cN_i} \}.$} \vspace{-0.1in}
  \end{assumption}
  Assumption~\ref{timeInv}.1 requires that the effect of unobserved confounders on 
  the potential outcomes be stable over time. This assumption is more
  plausible when we can control for a variety of observed time-varying
  confounders  $\bX_{i, t+1}$ and $\bX_{it}$. However, this assumption
  might be violated when the change in the effect of $U$ is quick and cannot be
  explained by observed covariates $\bX$. Suppose that the unemployment rate is the unobserved confounder in our
  motivating application. This assumption then implies that
  the effect of the unemployment rate on the incidence of hate
  crimes is the same over time. In the causal DAG in
  Figure~\ref{fig:placeboDAG}, this means that the effect of $G_2$ on
  $Y_{22}$ is the same
  as the effect of $G_1$ on $Y_{21}.$ 

  Assumption~\ref{timeInv}.2 requires that the imbalance of unobserved confounders be stable over time. In
  other words, the strength of association between the treatment variable and unobserved
  confounders is the same at time $t$ and $t+1$. Importantly, it does not require that the
  distribution of confounders is the same across different treatment
  groups. Instead, it requires that the difference between treatment
  groups be stable over time. For example, this means 
  that an association between the incidence of hate crimes in neighborhoods (treatment) and the
  unemployment rate is stable over. In the causal DAG in
  Figure~\ref{fig:placeboDAG}, this assumption implies that the association between $G_2$ and 
  $Y_{11}$ is the same as the one between $G_1$ and $Y_{11}$. This assumption substantively means the stability of omitted
  confounder $G$. 

  In practice, both conditions are more likely to hold when the interval
  between time $t$ and $t+1$ is shorter because $U_{i,t+1} \approx 
  U_{it}$ and $\bX_{i, t+1} \approx \bX_{it}.$ In particular, when all 
  confounders are time-invariant between time $t$ and $t+1$, Assumption~\ref{timeInv}.2 holds exactly. Even
  when confounders are time-varying, we can make these
  assumptions more plausible by adjusting for observed time-varying confounders $\bX_{i,t+1}$ and $\bX_{it}.$ 

  In a special case where there is no descendant of $Y_{it}$ in the control set,
  i.e., $\bX_{i,t+1} = \bX_{it} = \emptyset,$ Assumption~\ref{timeInv} is 
  equivalent to the parallel trend assumption required for the standard difference-in-differences estimator \citep{angrist2008mostly}.  By allowing for time-varying
  confounders, Assumption~\ref{timeInv} extends the parallel trend 
  assumption. It is also closely connected to the change-in-change
  method \citep{athey2006identification, tchetgen2016negative}. 
  Specifically, Assumption~\ref{timeInv}.2 (time-invariant imbalance) is
  a direct extension of Assumption 3.3 in \cite{athey2006identification} to
  the diffusion setting.

  \subsubsection{Estimator and Identification}
  \label{subsubsec:BCtheory2}
  We introduce a general bias-corrected estimator under
  Assumption~\ref{timeInv}. Intuitively, it subtracts bias detected by
  the proposed placebo test from an estimator that we would use under the no omitted confounders assumption. 
  \vspace{0.1in}
  \begin{definition}[Bias-Corrected Estimator]
    \spacingset{1.2}{A bias-corrected estimator $\hat{\tau}_{\text{BC}}$ is 
      the difference between two estimators $\hat{\tau}_{\text{Main}}$ and $\hat{\delta}_{\text{Placebo}}$.
      \begin{eqnarray}
        \hat{\tau}_{\text{BC}}  \equiv  \hat{\tau}_{\text{Main}}  -
        \hat{\delta}_{\text{Placebo}}  \label{eq:bc}
      \end{eqnarray}
      where {\small
        \begin{eqnarray*}
          \hspace{-0.6in} && \hat{\tau}_{\text{Main}} \equiv  \int \bigl\{ \widehat{\E} [Y_{i, t+1} \mid D_{it}=d^H,
                             \bX_{i,t+1}, \bC_{i,t+1}^B]  - \widehat{\E}[ Y_{i, t+1}
                             \mid  D_{it} = d^L, \bX_{i,t+1}, \bC_{i,t+1}^B] \bigr\} d F_{\bX_{i,t+1}, \bC^B_{i,t+1} \mid D_{it}=d^H}(\bx, \bc), \\ 
          \hspace{-0.6in} && \hat{\delta}_{\text{Placebo}} \equiv  \int \bigl\{ \widehat{\E}[Y_{it} \mid D_{it}=d^H, \bX_{it}, \bC_{i,t+1}^B]  -  \widehat{\E}[
                             Y_{it} \mid D_{it}=d^L, \bX_{it},
                             \bC_{i,t+1}^B] \bigr\} d F_{\bX_{i,t+1}, \bC_{i,t+1}^B \mid D_{it}=d^H}(\bx, \bc),
        \end{eqnarray*}}
      \noindent where $\widehat{\E}[\cdot]$ is any unbiased estimator of $\E[\cdot]$, and 
      researchers can use regression, weighting, matching or other
      techniques to obtain such an unbiased estimator. Note
      that both estimators are marginalized over the same conditional
      distribution $F_{\bX_{i,t+1}, \bC_{i,t+1}^B \mid D_{it}=d^H}(\bx, \bc).$} 
  \end{definition}
  This bias-corrected estimator consists of two parts,
  $\hat{\tau}_{\text{Main}}$ and $\hat{\delta}_{\text{Placebo}}$. The
  first part is an estimator unbiased for the ACDT under
  the no omitted confounders assumption. However, $\hat{\tau}_{\text{Main}}$
  suffers from bias when this identification assumption is violated. 
  The purpose of the second part $\hat{\delta}_{\text{Placebo}}$ is
  to correct this bias. It is closely connected to the proposed
  placebo test; when the assumption of no omitted confounders holds,
  $\E[\hat{\delta}_{\text{Placebo}}] = 0$ and there is no bias correction. When the assumption is instead violated,
  $\hat{\delta}_{\text{Placebo}}$ serves as an estimator of the bias. We
  rely on $\widehat\Var(\hat{\tau}_{\text{Main}}) +
  \widehat\Var(\hat{\delta}_{\text{Placebo}})$ as a conservative variance
  estimator of the bias-corrected estimator given that
  $\hat{\tau}_{\text{Main}}$ and $\hat{\delta}_{\text{Placebo}}$ are
  often positively correlated. 

  The theorem below shows that under Assumption~\ref{timeInv}, 
  the bias-corrected estimator is unbiased for the ACDT. 

  \vspace{0.1in}
  \begin{theorem}[Identification with A Bias-Corrected Estimator]
    \label{biascorrect} \spacingset{1.2}{
      Under Assumptions~\ref{seq} and~\ref{timeInv},  
      the proposed  bias-corrected estimator is unbiased for the ACDT. 
      \begin{eqnarray*}
        \E[\hat{\tau}_{\text{BC}}]  & = & \tau^{d^H}_{t+1} (d^H, d^L).
      \end{eqnarray*}}
  \end{theorem}
  The proof is in Appendix~\ref{subsec:proof-correct}. It is also true that this estimator is unbiased for the  
  ACDT when the no omitted confounders assumption holds. Through
  a simulation study calibrated to the hate crime data, we show that the proposed
  bias-corrected estimator can reduce the bias and root mean squared error
  even when the required time-invariance assumption (Assumption~\ref{timeInv}) is slightly
  violated (Appendix~\ref{subsec:simbc}).  

  In Appendix~\ref{subsec:bc-ex}, we consider two extensions of the
  bias-corrected estimator. First, we introduce a sensitivity analysis 
  to investigate the robustness of the bias-corrected estimates to the
  potential violation of the time-invariance assumption
  (Assumption~\ref{timeInv}). Second, while this section considers the ACDT as the
  causal estimand following the standard
  difference-in-differences literature \citep{angrist2008mostly}, we
  discuss modification of Assumption~\ref{timeInv} sufficient for the
  identification of the ACDE.

  \section{Empirical Analysis}
  \label{sec:app}
  Applying the proposed methods, we estimate the ACDE of
  hate crimes against refugees in Germany. We begin with the setup of data analysis
  (Section~\ref{subsec:setup}) and then turn to the
  estimation of the ACDE (Section~\ref{subsec:acde}) and heterogeneous
  effects (Section~\ref{subsec:h-acde}).


  \subsection{Setup}
  \label{subsec:setup}
  As one of the most well-studied outcomes, we focus on physical attacks
  against refugees as the main dependent variable. Formally, we define the outcome variable $Y_{it}$ to be
  binary, taking the value 1 if there exists any physical attack against refugees at county $i$ in month $t$, and
  taking the value $0$ otherwise. The outcomes are defined for 402 counties in Germany every
  month from the beginning of 2015 to the end of 2016. 
  Averaging over all counties in Germany during this period, the sample mean of the outcome
  variable is $6.4\%$. This means that $6.4\%$ of counties experienced
  at least one physical attack in a typical month. In Saxony, a state with the largest number of 
  hate crimes, the sample mean of the outcome variable is $34\%$. 



  We use a distance matrix to encode the physical proximity between
  counties. In particular, we construct an initial distance matrix $\widetilde{\bW}$ using 
  an inverse of the straight distance between counties
  $i$ and $j$ as $\widetilde{W}_{ij}$. We then row-standardize the
  initial matrix $\widetilde{\bW}$ and obtain a final distance matrix
  $\bW$. For the outcome variable in month $t+1$, the treatment variable is
  defined to be $D_{it} \equiv \bW_{i}^\top \bY_{t}$, the weighted proportion of neighboring counties that experience the 
  incidence of physical attacks in month $t$. The first causal quantity
  of interest is the ACDE, which quantifies how much the probability of
  having hate crimes changes due to the increase in the 
  proportion of neighboring counties that have experienced hate crimes
  last month. 

  To investigate how the proposed methods detect and correct biases, we consider five different sets of control variables in
  order (summarized in Table~\ref{tab:control-s}). As the first set of control
  variables, we include one-month lagged dependent and treatment
  variables. We also adjust for basic summary statistics of $\bW_{i}$,
  i.e., the number of neighbors and variance of $\bW_i$, in order to
  compare observations with similar spatial characteristics. These
  lagged variables and basic summary statistics of the spatial distance are sufficient
  for the identification if the spatial diffusion is the only mechanism through which neighboring counties 
  exhibit similar outcomes. Then, as the second set of control
  variables, we add two-month lagged dependent variables to see whether
  adjusting for a longer history of past outcomes can reduce bias
  \citep[e.g.,][]{fowler2013review, eckles2017bias}. 
  The third set of control variables add state fixed effects. 
  Although the state fixed effects are often excluded from 
  existing studies \citep[e.g.,][]{jackle2016dark}, we show how much these fixed effects help remove
  biases. Then, the fourth set adds a list of contextual variables
  related to the number of refugees, demographics, education, general crimes, economic
  indicators, and politics. Finally, the fifth set controls for the time
  trend using third-order polynomials. We provide details of the five control sets
  and the corresponding placebo sets in Appendix~\ref{sec:app_em}. 

  For the proposed placebo test, we rely on the structural stationarity
  assumption (Assumption~\ref{ss}). For example, if  
  discussions of the refugee crisis in newspapers, which we do not
  measure, are confounders, the structural stationarity requires that
  such discussions in newspapers remain confounders throughout 2015 and
  2016. Importantly, the placebo test is valid even when the tone of
  discussions is changing over time (unmeasured time-varying confounders) and the effect of
  discussions changes over time. For the bias-corrected estimator, the
  time-invariance assumption (Assumption~\ref{timeInv}) requires a stronger parametric assumption,
  similar to the difference-in-differences literature
  \citep{athey2006identification, angrist2008mostly,
    tchetgen2016negative}, that the effect of newspapers is stable over
  time and the imbalance of unobserved discussions in newspapers 
  is stable over time after controlling for observed time-varying confounders. 

  \begin{table}[!t]
    \centering \small
    \begin{tabular}{|l|l|}
      \hline
      C1   &  $Y_{it}, D_{i,t-1}$, summary statistics of $\bW_i
             (|\mathcal{N}_i|, \Var(\bW_i))$    \\
      \hdashline
      C2  & C1 + $Y_{i,t-1}$\\
      \hdashline
      C3  & C2 + state fixed-effects \\
      \hdashline
      C4  & C3 + contextual variables studied in the literature \\
      \hdashline
      C5  & C4 + time trend (third-order polynomials) \\
      \hline
    \end{tabular}
    \spacingset{1}{\caption{Five Different Control Sets.}\label{tab:control-s}}
  \end{table}

  \subsection{Estimation of Average Causal Diffusion Effect}
  \label{subsec:acde}
  To estimate the ACDE, we use the following
  logistic regression to model the main outcome variable $Y_{i,t+1}$
  with the treatment variable and each of the five control sets. 
  \begin{equation}
    \mbox{logit} (\Pr(Y_{i,t+1} = 1\mid D_{it}, \bC_{i,t+1}))  \ = \alpha + \beta
    D_{it} + \gamma^\top \bC_{i,t+1}, \label{eq:main1}
  \end{equation}
  where $D_{it}$ is the treatment variable and $\bC_{i,t+1}$ is a specified set
  of control variables. Under the assumption of no omitted confounders,
  the difference in the estimated probabilities of $Y_{i,t+1}$ under
  $D_{it}=d^H$ and $D_{it}=d^L$ serves as an
  estimator for the ACDE. In particular, we estimate the ACDE that compares the following two treatment values;
  $d^H = 27\%$, the treatment received by the average counties in Saxony (a
  state with the largest number of hate crimes) 
  and $d^L = 0\%$, none of the neighbors experiencing hate crimes (common for
  safe areas in West Germany). Formally, $\hat{\tau} \equiv \int
  \{\widehat{\Pr}(Y_{i,t+1} = 1\mid D_{it} = 0.27, \bC_{i,t+1}) -
  \widehat{\Pr}(Y_{i,t+1} = 1 \mid D_{it} = 0, \bC_{i,t+1})\} d F_{\bC_{i,t+1}} (\bc).$

  To assess the no omitted confounders assumption, we also estimate the
  following placebo logistic regression. 
  \begin{equation}
    \mbox{logit} (\Pr(Y_{it} = 1 \mid D_{it}, \bC_{i,t+1}^P))  \ = \alpha_0
    + \rho D_{it} + \gamma_0^\top \bC_{i,t+1}^P, \label{eq:placebo1}
  \end{equation} 
  where $Y_{it}$ is the placebo outcome and $\bC_{i,t+1}^P$ is a placebo set corresponding to the control set
  $\bC_{i,t+1}$. When the no omitted confounders assumption holds,
  Theorem~\ref{placebo} implies that $\rho = 0$. We use 
  the difference in the estimated probabilities of $Y_{it}$ under 
  $D_{it}=d^H$ and $D_{it}=d^L$ as a test statistic of the placebo
  test.  Formally, $\hat{\delta} \equiv \int
  \{\widehat{\Pr}(Y_{it} = 1 \mid D_{it} = 0.27, \bC_{i,t+1}^P) -
  \widehat{\Pr}(Y_{it} = 1 \mid D_{it} = 0, \bC_{i,t+1}^P)\} d F_{\bC^P_{i,t+1}} (\bc^P).$

\begin{figure}[!t]
    \begin{center}
      \includegraphics[width=\textwidth]{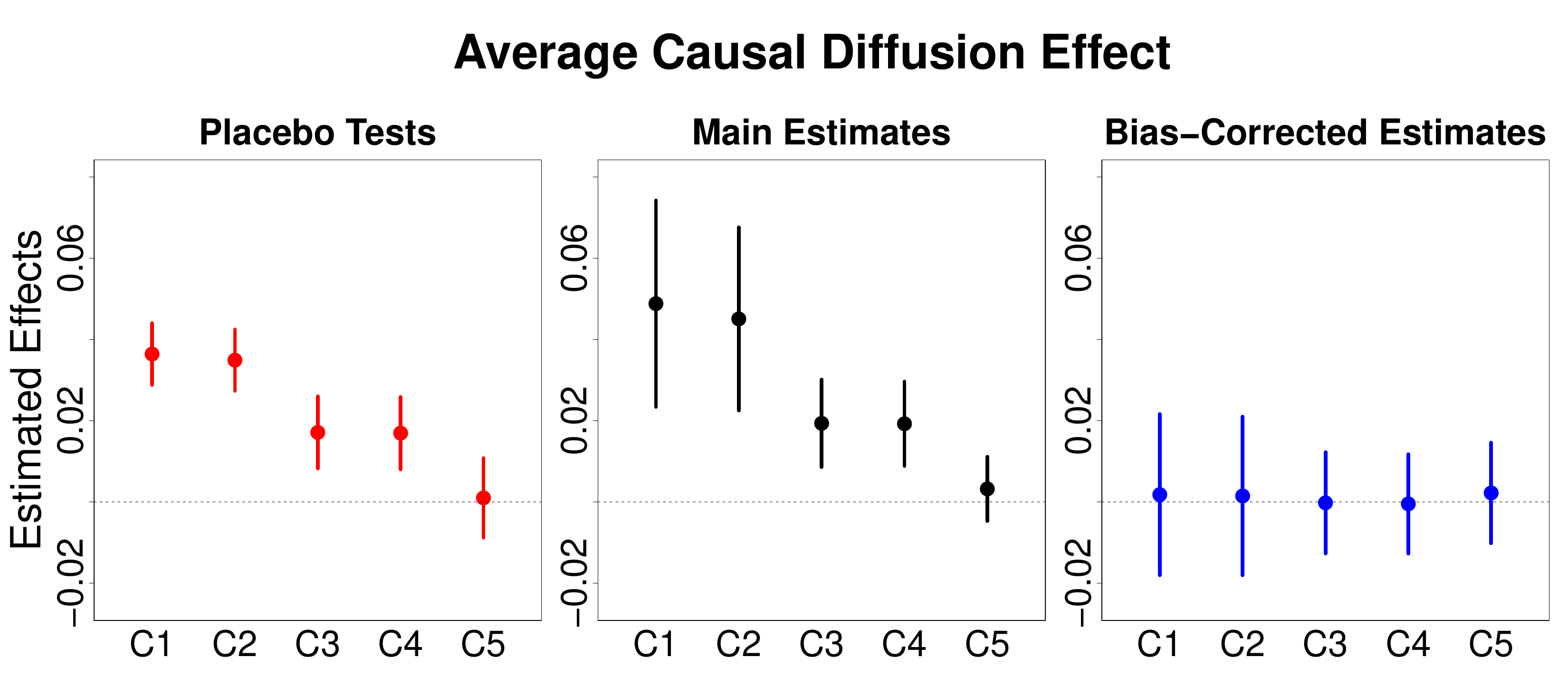} \\
      \vspace{-0.1in}
      \begin{tabular}{c} \begin{minipage}{0.33\textwidth} \centering
          \hspace{0.25in} (a) \end{minipage} 
        \begin{minipage}{0.33\textwidth} \centering \hspace{0.05in} (b) \end{minipage} 
        \begin{minipage}{0.33\textwidth} \centering \hspace{-0.2in} (c) \end{minipage}
      \end{tabular}
    \end{center}
    \vspace{-0.2in}
    \spacingset{1}{\caption{Placebo Tests, Main Estimates, and Bias-Corrected
        Estimates of the ACDE. \\ {\small Note: Figures (a), (b) and (c) present results from the placebo
          tests, estimates of the ACDE under the no omitted confounders
          assumption, and estimates from bias-corrected estimators with 95\% confidence intervals,
          respectively. 
        }}\label{fig:marginal}}
  \end{figure}

  Figures~\ref{fig:marginal} (a) and (b) present results from the placebo
  tests (equation~\eqref{eq:placebo1}) and estimates from the main model
  (equation~\eqref{eq:main1}) with 95\% confidence intervals,
  respectively. All standard errors are clustered at the state level. 
  C1, C2, C3, C4, and C5 refer to the five different control
  sets we introduced before. When a given set of control variables
  satisfies the no omitted confounders assumption, estimates from the
  placebo tests should be close to
  zero. Figure~\ref{fig:marginal} (a) shows that while the first four sets of control variables are not
  sufficient, the fifth set (C5) successfully adjusts for
  confounders; a placebo estimate is close to zero and its 95\% confidence
  interval covers zero. It is not enough to control for lagged dependent 
  variables and contextual variables and it is critical to
  control for the time trend flexibly. 

  On the basis of these results from the placebo tests, we can now
  investigate estimates of the ACDE from the main model 
  (equation~\eqref{eq:main1}) in Figure~\ref{fig:marginal}
  (b). For the first two cases (C1 and C2), estimates are as large as $5$ percentage points, but the placebo tests suggest that these estimates are heavily biased. Similarly, while the next
  two cases show point estimates of around $2$ percentage points, they
  are also likely to be biased. When we focus on the fifth control set, which produces a
  placebo estimate close to zero, a point estimate of the ACDE is
  smaller than $1$ percentage point, and its 95\% confidence interval
  covers zero. The comparison between
  this more credible estimate and the one from the fourth set shows that
  an estimate of the ACDE can suffer from $100$\% bias by missing just one variable. This demonstrates the importance of bias
  detection in causal diffusion analysis. 

  Although the proposed placebo tests suggest that the fifth control
  successfully adjusts for relevant confounders in this analysis, it is often infeasible to find such control 
  sets in many other applications. To address these common scenarios, 
  we now examine whether researchers could obtain similar results using a
  bias-corrected estimator even with control sets that reject the null
  hypothesis of the placebo test. 

  Figure~\ref{fig:marginal} (c) shows that 
  bias-corrected estimates are similar regardless of the selection of
  control variables and they all cover the most credible point estimate from the fifth
  control set. Even though the proposed placebo test detected a large
  amount of bias, researchers can obtain credible estimates by
  correcting the biases in this example. 

  These results suggest that, in contrast to existing studies
  \citep{braun2011diffusion, jackle2016dark}, the ACDE on the incidence of hate crimes is small when 
  averaging over all counties in Germany. In the next subsection, we show
  that the spatial diffusion of hate crimes is concentrated among a
  small subset of counties that have a higher proportion of school dropouts. 

  \subsection{Heterogeneous Diffusion Effects by Education}
  \label{subsec:h-acde}
  Now, we extend the previous analysis by
  considering the types of counties that are more susceptible to the
  diffusion of hate crimes. In particular, we examine the role of
  education. Given rich qualitative and quantitative evidence that 
  hate crime is often a problem of young people, it is critical to take into 
  account one of the most important institutional contexts around them,
  i.e., schooling. The literature has discussed at least three mechanisms
  through which education can reduce the risk of hate crimes. First,
  education increases economic returns to current and future
  legitimate work, thereby raising the opportunity cost of committing
  hate crimes \citep[e.g.,][]{lochner2004effect}. Second, education
  may change the psychological costs
  associated with hate crimes. More educated people tend to have lower
  levels of ethnocentrism and place more emphasis on cultural
  diversity \citep{hainmueller2007educated}. Finally, schooling has incapacitation effects -- keeping adolescents busy and off
  the street, thereby directly reducing the chances of committing crimes
  \citep{jacob2003idle}. 

  Building on the literature above, we investigate whether local educational
  contexts condition the spatial diffusion dynamics of hate crimes. We use a proportion of school dropouts without a
  secondary school diploma as a measure of local educational performance. 
  To better disentangle the education explanation, we analyze East Germany
  and West Germany separately because they have substantially different
  distributions of proportions of school dropouts (counties in East
  Germany have much higher proportions of school dropouts). Here we report results from East Germany and provide those
  for West Germany in Appendix~\ref{sec:app_em}. In particular, we estimate the conditional average causal diffusion
  effects (conditional ACDEs) for counties that have high and low
  proportions of school dropouts without a secondary school diploma. We use
  $9\%$ as a cutoff for high and low proportions of school dropouts, which is approximately
  the median value in East Germany. We add an interaction term between the treatment 
  variable and this indicator variable to the original model in 
  equation~\eqref{eq:main1} and to the original placebo model in
  equation~\eqref{eq:placebo1}. 

  \begin{figure}[!t]
    \begin{center}
      \includegraphics[width=\textwidth]{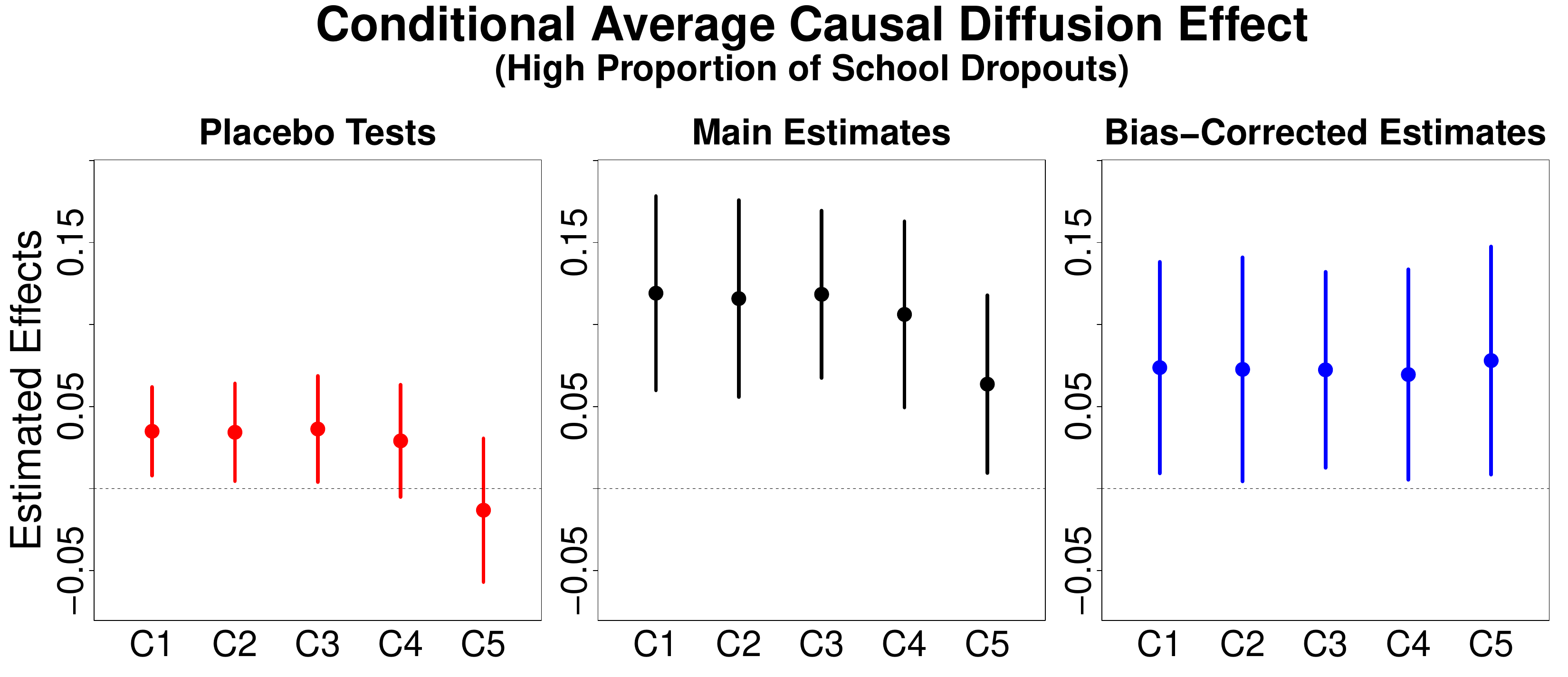}\\
      \vspace{-0.1in}
      \begin{tabular}{c} \begin{minipage}{0.33\textwidth} \centering
          \hspace{0.25in} (a) \end{minipage} 
        \begin{minipage}{0.33\textwidth} \centering \hspace{0.05in} (b) \end{minipage} 
        \begin{minipage}{0.33\textwidth} \centering \hspace{-0.2in} (c) \end{minipage}
      \end{tabular}
    \end{center}
    \vspace{-0.2in}
    \spacingset{1}{\caption{Placebo Tests, Main Estimates, and Bias-Corrected
        Estimates of the conditional ACDE for counties with a high
        proportion of school dropouts. {\small Note: Figures (a), (b) and (c) present results from the placebo
          tests, estimates of the conditional ACDE under the no omitted confounders
          assumption, and estimates from bias-corrected estimators with 95\% confidence intervals,
          respectively. 
        }}\label{fig:marCE}}
  \end{figure}

  Figure~\ref{fig:marCE} presents results for the conditional
  ACDE for counties that have a higher proportion of school dropouts. Similar to the case of the ACDE
  estimation, Figure~\ref{fig:marCE} (a) shows strong concerns of biases in the first four sets
  of control variables. Even though a 95\% confidence interval of the
  fourth estimate covers zero, its point estimate is far from zero
  (around $4$ percentage points). In contrast, the placebo test suggests that the fifth control
  set adjusts for relevant confounders where a placebo estimate is close
  to zero. 

  Based on results from the placebo tests, we examine estimates from
  the main model in Figure~\ref{fig:marCE} (b). The first four sets,
  likely to be biased, exhibit large point estimates, larger than $10$ percentage points. More 
  interestingly, even with the most credible fifth control set, a point estimate is as large as $6$
  percentage points and is statistically significant. This effect size
  is substantively important given that it is about
  one-fourth of the sample average outcome in this subset ($26\%$). 
  Bias-corrected estimates in Figure~\ref{fig:marCE} (c) confirm
  that the conditional ACDE for counties with a higher proportion of school
  dropouts is large and similar regardless of the selection of
  control sets. 

  When we estimate the conditional ACDE for counties that have a lower
  proportion of school dropouts, effects are close to zero and their 95\%
  confidence intervals cover zero, as the education hypothesis expects
  (see Appendix~\ref{sec:app_em}). Causal diffusion effects are also precisely
  estimated to be zero in West Germany, where the proportions of school dropouts are much lower than East Germany. 
  This additional analysis suggests that the spatial diffusion dynamics
  of hate crimes operate only in places with low educational
  performance  and thus, prevention policies can have positive multiplier
  effects only when targeting areas with low educational performance.

  \section{Concluding Remarks}
  \label{sec:con} 
  Causal diffusion dynamics have been an integral part of
  many social and biomedical science theories. Given that spatial and network panel
  data have become increasingly common, it is essential to develop
  methodologies to draw causal inference for diffusion
  effects. However, causal diffusion analysis has been challenging due
  to two well-known types of biases, i.e., contextual confounding and homophily bias.  
  Recognizing that causal inference for diffusion effects is generally impossible
  without further assumptions \citep{shalizi2011homophily,
    vanderweele2013social, ogburn2017challenges}, this paper examines the identification
  of causal diffusion effects under a new assumption of structural
  stationarity. This structural stationarity requires the existence of causal relationships among variables --- not
  the effect or sign of such relationships --- to be stable over
  time. Importantly, our approach based on the structural stationarity
  differs from a traditional DAG-based approach in that we only assume
  a class of dynamic causal DAGs, instead of a specific causal DAG. In
  particular, we
  develop methodologies valid for any causal DAGs within this general,
  large class of dynamic causal DAGs. Thus, the
  structural stationarity allows us to clearly encode assumptions about the
  underlying diffusion process without sacrificing its practical applicability. 

  Under the structural stationarity, we first propose a statistical
  placebo test that can detect a wide class of biases, including  
  contextual confounding and homophily bias. Then, we develop a
  difference-in-differences style estimator that can directly correct
  biases under an additional parametric assumption.
  Applying the proposed methods to geo-coded hate crime data, we examined the spatial diffusion of hate
  crimes in Germany. After removing upward bias in previous studies, we 
  found that the average effect of spatial diffusion is small, in contrast to recent quantitative
  analyses \citep{braun2011diffusion, jackle2016dark}. The investigation
  of heterogeneous effects, however, revealed that the spatial
  diffusion effect of hate crimes is large in areas that have a high 
  proportion of school dropouts. This empirical analysis demonstrates the large differences in
  substantive conclusions that can result from contextual
  confounding. By directly accounting for these biases, the proposed
  placebo test and bias-corrected estimator help researchers make more
  credible causal inference for diffusion studies.

  There are a number of possible future extensions. First, whereas we propose an extension of the
  difference-in-differences estimator to causal diffusion analysis,
  future research should also investigate how to incorporate into causal
  diffusion analysis other popular tools developed for estimating the average
  treatment effect in panel data settings, such as synthetic control
  methods \citep{abadie2010synthetic}. In addition, to further disentangle different
  channels of diffusion effects, it is of interest to study the
  intersection of the causal mediation analysis \citep{robins1992identifiability,
    pearl2001direct, imai2010identification, vander2014explanation} and
  the causal diffusion analysis \citep[e.g.,][]{ogburn2014DAG}. With
  this extension, researchers can analyze, 
  for example, micromechanisms of hate crime diffusion.

  \vspace{0.1in}
  \spacingset{1.2}
  {\small
    \pdfbookmark[1]{References}{References}
    \bibliography{egami,polsci}

\begin{thebibliography}{}

\bibitem[Abadie {\em et~al.}(2010)Abadie, Diamond, and
  Hainmueller]{abadie2010synthetic}
Abadie, A., Diamond, A., and Hainmueller, J. (2010).
\newblock {Synthetic Control Methods for Comparative Case Studies: Estimating
  the Effect of California's Tobacco Control Program}.
\newblock {\em Journal of the American Statistical Association\/}, {\bf
  105}(490), 493--505.

\bibitem[An(2015)An]{an2015instrumental}
An, W. (2015).
\newblock {Instrumental Variables Estimates of Peer Effects in Social
  Networks}.
\newblock {\em Social Science Research\/}, {\bf 50}, 382--394.

\bibitem[Anagnostopoulos {\em et~al.}(2008)Anagnostopoulos, Kumar, and
  Mahdian]{anagnostopoulos2008influence}
Anagnostopoulos, A., Kumar, R., and Mahdian, M. (2008).
\newblock {Influence and Correlation in Social Networks}.
\newblock In {\em {Proceedings of the 14th ACM SIGKDD International Conference
  on Knowledge Discovery and Data Mining}\/}, pages 7--15. ACM.

\bibitem[Angrist(2014)Angrist]{angrist2014perils}
Angrist, J.~D. (2014).
\newblock {The Perils of Peer Effects}.
\newblock {\em Labour Economics\/}, {\bf 30}, 98--108.

\bibitem[Angrist and Pischke(2008)Angrist and Pischke]{angrist2008mostly}
Angrist, J.~D. and Pischke, J.-S. (2008).
\newblock {\em Mostly Harmless Econometrics: An Empiricist's Companion\/}.
\newblock Princeton University Press, Princeton, NJ.

\bibitem[Anselin(2013)Anselin]{anselin2013spatial}
Anselin, L. (2013).
\newblock {\em Spatial Econometrics: Methods and Models\/}.
\newblock Springer.

\bibitem[Aral {\em et~al.}(2009)Aral, Muchnik, and
  Sundararajan]{aral2009distinguishing}
Aral, S., Muchnik, L., and Sundararajan, A. (2009).
\newblock {Distinguishing Influence-based Contagion from Homophily-driven
  Diffusion in Dynamic Networks}.
\newblock {\em Proceedings of the National Academy of Sciences\/}, {\bf
  106}(51), 21544--21549.

\bibitem[Athey and Imbens(2006)Athey and Imbens]{athey2006identification}
Athey, S. and Imbens, G.~W. (2006).
\newblock {Identification and Inference in Nonlinear Difference-in-Differences
  Models}.
\newblock {\em Econometrica\/}, {\bf 74}(2), 431--497.

\bibitem[Basse {\em et~al.}(2019)Basse, Ding, Feller, and
  Toulis]{basse2019peer}
Basse, G., Ding, P., Feller, A., and Toulis, P. (2019).
\newblock {Randomization Tests for Peer Effects in Group Formation
  Experiments}.
\newblock {\em arXiv preprint arXiv:1904.02308\/}.

\bibitem[Ben{\v{c}}ek and Strasheim(2016)Ben{\v{c}}ek and
  Strasheim]{benvcek2016refugees}
Ben{\v{c}}ek, D. and Strasheim, J. (2016).
\newblock {Refugees Welcome? A Dataset on Anti-Refugee Violence in Germany}.
\newblock {\em Research \& Politics\/}, {\bf 3}(4).

\bibitem[Bramoull{\'e} {\em et~al.}(2009)Bramoull{\'e}, Djebbari, and
  Fortin]{bramoulle2009identification}
Bramoull{\'e}, Y., Djebbari, H., and Fortin, B. (2009).
\newblock {Identification of Peer Effects through Social Networks}.
\newblock {\em Journal of Econometrics\/}, {\bf 150}(1), 41--55.

\bibitem[Braun(2011)Braun]{braun2011diffusion}
Braun, R. (2011).
\newblock {The Diffusion of Racist Violence in the Netherlands: Discourse and
  Distance}.
\newblock {\em Journal of Peace Research\/}, {\bf 48}(6), 753--766.

\bibitem[Buhaug and Gleditsch(2008)Buhaug and Gleditsch]{gleditsch2008conflict}
Buhaug, H. and Gleditsch, K.~S. (2008).
\newblock {Contagion or Confusion? Why Conflicts Cluster in Space}.
\newblock {\em International Studies Quarterly\/}, {\bf 52}(2), 215--233.

\bibitem[{Bundesamt f{\"u}r Migration und Fl{\"u}chtlinge}(2019){Bundesamt
  f{\"u}r Migration und Fl{\"u}chtlinge}]{bamf2019report}
{Bundesamt f{\"u}r Migration und Fl{\"u}chtlinge} (2019).
\newblock {Migrationsbericht 2016/2017}.
\newblock
  \href{http://www.bamf.de/SharedDocs/Anlagen/DE/Publikationen/Migrationsberichte/migrationsbericht-2016-2017.pdf?__blob=publicationFile}{http://www.bamf.de}.

\bibitem[Cai {\em et~al.}(2019)Cai, Loh, and Crawford]{cai2019identification}
Cai, X., Loh, W.~W., and Crawford, F.~W. (2019).
\newblock {Identification of Causal Intervention Effects Under Contagion}.
\newblock {\em arXiv preprint arXiv:1912.04151\/}.

\bibitem[Christakis and Fowler(2013)Christakis and Fowler]{fowler2013review}
Christakis, N.~A. and Fowler, J.~H. (2013).
\newblock {Social Contagion Theory: Examining Dynamic Social Networks and Human
  Behavior}.
\newblock {\em Statistics in Medicine\/}, {\bf 32}(4), 556--577.

\bibitem[Cohen-Cole and Fletcher(2008)Cohen-Cole and
  Fletcher]{cohen2008obesity}
Cohen-Cole, E. and Fletcher, J.~M. (2008).
\newblock {Is Obesity Contagious? Social Networks vs. Environmental Factors in
  the Obesity Epidemic}.
\newblock {\em Journal of health economics\/}, {\bf 27}(5), 1382--1387.

\bibitem[Cressie(2015)Cressie]{cressie2015statistics}
Cressie, N. (2015).
\newblock {\em {Statistics for Spatial Data}\/}.
\newblock John Wiley \& Sons.

\bibitem[Dancygier {\em et~al.}(2019)Dancygier, Egami, Jamal, and
  Rischke]{dancygier2018hate}
Dancygier, R.~M., Egami, N., Jamal, A.~A., and Rischke, R. (2019).
\newblock {Hating and Mating: Fears over Mate Competition and Violent Hate
  Crime against Refugees}.
\newblock Available at SSRN: https://ssrn.com/abstract=3358780.

\bibitem[Danks and Plis(2013)Danks and Plis]{danks2013learning}
Danks, D. and Plis, S. (2013).
\newblock {Learning Causal Structure from Undersampled Time Series}.
\newblock In {\em NIPS 2013 Workshop on Causality\/}.

\bibitem[Dean and Kanazawa(1989)Dean and Kanazawa]{dean1989model}
Dean, T. and Kanazawa, K. (1989).
\newblock {A Model for Reasoning About Persistence and Causation}.
\newblock {\em Computational intelligence\/}, {\bf 5}(2), 142--150.

\bibitem[Duflo {\em et~al.}(2011)Duflo, Dupas, and Kremer]{duflo2011peer}
Duflo, E., Dupas, P., and Kremer, M. (2011).
\newblock {Peer Effects, Teacher Incentives, and The Impact of Tracking:
  Evidence from a Randomized Evaluation in Kenya}.
\newblock {\em American Economic Review\/}, {\bf 101}(5), 1739--74.

\bibitem[Eckles and Bakshy(2017)Eckles and Bakshy]{eckles2017bias}
Eckles, D. and Bakshy, E. (2017).
\newblock {Bias and High-Dimensional Adjustment in Observational Studies of
  Peer Effects}.
\newblock {\em arXiv preprint arXiv:1706.04692\/}.

\bibitem[Flanders {\em et~al.}(2017)Flanders, Strickland, and
  Klein]{flanders2017new}
Flanders, W.~D., Strickland, M.~J., and Klein, M. (2017).
\newblock {A New Method for Partial Correction of Residual Confounding in
  Time-Series and Other Observational Studies}.
\newblock {\em American Journal of Epidemiology\/}, {\bf 185}(10), 941--949.

\bibitem[Fowler {\em et~al.}(2011)Fowler, Heaney, Nickerson, Padgett, and
  Sinclair]{fowler2011causality}
Fowler, J.~H., Heaney, M.~T., Nickerson, D.~W., Padgett, J.~F., and Sinclair,
  B. (2011).
\newblock {Causality in Political Networks}.
\newblock {\em American Politics Research\/}, {\bf 39}(2), 437--480.

\bibitem[Franzese and Hays(2007)Franzese and Hays]{franzese2007spatial}
Franzese, R.~J. and Hays, J.~C. (2007).
\newblock {Spatial Econometric Models of Cross-Sectional Interdependence in
  Political Science Panel and Time-Series-Cross-Section Data}.
\newblock {\em Political Analysis\/}, {\bf 15}(2), 140--164.

\bibitem[Glaeser {\em et~al.}(1996)Glaeser, Sacerdote, and
  Scheinkman]{glaeser1996crime}
Glaeser, E.~L., Sacerdote, B., and Scheinkman, J.~A. (1996).
\newblock {Crime and Social interactions}.
\newblock {\em The Quarterly Journal of Economics\/}, {\bf 111}(2), 507--548.

\bibitem[Goldsmith-Pinkham and Imbens(2013)Goldsmith-Pinkham and
  Imbens]{imbens2013social}
Goldsmith-Pinkham, P. and Imbens, G.~W. (2013).
\newblock {Social Networks and the Identification of Peer Effects}.
\newblock {\em Journal of Business \& Economic Statistics\/}, {\bf 31}(3),
  253--264.

\bibitem[Graham {\em et~al.}(2013)Graham, Shipan, and
  Volden]{graham2013diffusion}
Graham, E.~R., Shipan, C.~R., and Volden, C. (2013).
\newblock {The Diffusion of Policy Diffusion Research in Political Science}.
\newblock {\em British Journal of Political Science\/}, {\bf 43}(03), 673--701.

\bibitem[Granger(1988)Granger]{granger1988some}
Granger, C.~W. (1988).
\newblock {Some Recent Development in A Concept of Causality}.
\newblock {\em Journal of Econometrics\/}, {\bf 39}(1-2), 199--211.

\bibitem[Granovetter(1973)Granovetter]{granovetter1973strength}
Granovetter, M.~S. (1973).
\newblock {The Strength of Weak Ties}.
\newblock {\em American Journal of Sociology\/}, {\bf 78}(6), 1360--1380.

\bibitem[Hainmueller and Hiscox(2007)Hainmueller and
  Hiscox]{hainmueller2007educated}
Hainmueller, J. and Hiscox, M.~J. (2007).
\newblock {Educated Preferences: Explaining Attitudes toward Immigration in
  Europe}.
\newblock {\em International Organization\/}, {\bf 61}(2), 399--442.

\bibitem[Halloran and Hudgens(2016)Halloran and Hudgens]{halloran2016review}
Halloran, M.~E. and Hudgens, M.~G. (2016).
\newblock {Dependent Happenings: a Recent Methodological Review}.
\newblock {\em Current Epidemiology Reports\/}, {\bf 3}(4), 297--305.

\bibitem[Halloran and Struchiner(1995)Halloran and
  Struchiner]{halloran1995causal}
Halloran, M.~E. and Struchiner, C.~J. (1995).
\newblock {Causal Inference in Infectious Diseases}.
\newblock {\em Epidemiology\/}, {\bf 6}(2), 142--151.

\bibitem[Hill and Rothchild(1986)Hill and Rothchild]{hill1986contagion}
Hill, S. and Rothchild, D. (1986).
\newblock {The Contagion of Political Conflict in Africa and the World}.
\newblock {\em Journal of Conflict Resolution\/}, {\bf 30}(4), 716--735.

\bibitem[Holland {\em et~al.}(1983)Holland, Laskey, and
  Leinhardt]{holland1983stochastic}
Holland, P.~W., Laskey, K.~B., and Leinhardt, S. (1983).
\newblock {Stochastic Blockmodels: First Steps}.
\newblock {\em Social Networks\/}, {\bf 5}(2), 109--137.

\bibitem[Hyttinen {\em et~al.}(2016)Hyttinen, Plis, J{\"a}rvisalo, Eberhardt,
  and Danks]{hyttinen2016causal}
Hyttinen, A., Plis, S., J{\"a}rvisalo, M., Eberhardt, F., and Danks, D. (2016).
\newblock {Causal Discovery from Subsampled Time Series Data by Constraint
  Optimization}.
\newblock In {\em Proceedings of the 8th International Conference on
  Probabilistic Graphical Models (PGM)\/}, pages 216--227.

\bibitem[Imai {\em et~al.}(2010)Imai, Keele, and
  Yamamoto]{imai2010identification}
Imai, K., Keele, L., and Yamamoto, T. (2010).
\newblock {Identification, Inference and Sensitivity Analysis for Causal
  Mediation Effects}.
\newblock {\em Statistical Science\/}, {\bf 25}(1), 51--71.

\bibitem[J{\"a}ckle and K{\"o}nig(2016)J{\"a}ckle and
  K{\"o}nig]{jackle2016dark}
J{\"a}ckle, S. and K{\"o}nig, P.~D. (2016).
\newblock {The Dark Side of the German `Welcome Culture': Investigating the
  Causes behind Attacks on Refugees in 2015}.
\newblock {\em West European Politics\/}, {\bf 40}(2), 223--251.

\bibitem[Jacob and Lefgren(2003)Jacob and Lefgren]{jacob2003idle}
Jacob, B.~A. and Lefgren, L. (2003).
\newblock {Are Idle Hands the Devil's Workshop? Incapacitation, Concentration,
  and Juvenile Crime}.
\newblock {\em American Economic Review\/}, {\bf 93}(5), 1560--1577.

\bibitem[Jagadeesan {\em et~al.}(2019)Jagadeesan, Pillai, and
  Volfovsky]{jagadeesan2019designs}
Jagadeesan, R., Pillai, N., and Volfovsky, A. (2019).
\newblock {Designs for Estimating the Treatment Effect in Networks with
  Interference}.
\newblock {\em Annals of Statistics\/}.

\bibitem[Jones {\em et~al.}(2017)Jones, Bond, Bakshy, Eckles, and
  Fowler]{jones2017social}
Jones, J.~J., Bond, R.~M., Bakshy, E., Eckles, D., and Fowler, J.~H. (2017).
\newblock {Social Influence and Political Mobilization: Further Evidence from a
  Randomized Experiment in the 2012 US Presidential Election}.
\newblock {\em PloS one\/}, {\bf 12}(4), e0173851.

\bibitem[Li {\em et~al.}(2019)Li, Ding, Lin, Yang, and
  Liu]{li2019randomization}
Li, X., Ding, P., Lin, Q., Yang, D., and Liu, J.~S. (2019).
\newblock {Randomization Inference for Peer Effects}.
\newblock {\em Journal of the American Statistical Association\/}, pages 1--31.

\bibitem[Lipsitch {\em et~al.}(2010)Lipsitch, Tchetgen~Tchetgen, and
  Cohen]{tchetgen2010negative}
Lipsitch, M., Tchetgen~Tchetgen, E.~J., and Cohen, T. (2010).
\newblock {Negative Controls: A Tool for Detecting Confounding and Bias in
  Observational Studies}.
\newblock {\em Epidemiology\/}, {\bf 21}(3), 383.

\bibitem[Lochner and Moretti(2004)Lochner and Moretti]{lochner2004effect}
Lochner, L. and Moretti, E. (2004).
\newblock {The Effect of Education on Crime: Evidence from Prison Inmates,
  Arrests, and Self-Reports}.
\newblock {\em American Economic Review\/}, {\bf 94}(1), 155--189.

\bibitem[Lyons(2011)Lyons]{lyons2011spread}
Lyons, R. (2011).
\newblock {The Spread of Evidence-Poor Medicine via Flawed Social-Network
  Analysis}.
\newblock {\em Statistics, Politics, and Policy\/}, {\bf 2}(1).

\bibitem[Manski(1993)Manski]{manski1993identification}
Manski, C.~F. (1993).
\newblock {Identification of Endogenous Social Effects: The Reflection
  Problem}.
\newblock {\em The Review of Economic Studies\/}, {\bf 60}(3), 531--542.

\bibitem[Miao and Tchetgen~Tchetgen(2017)Miao and
  Tchetgen~Tchetgen]{tchetgen2017invited}
Miao, W. and Tchetgen~Tchetgen, E.~J. (2017).
\newblock {Invited Commentary: Bias Attenuation and Identification of Causal
  Effects with Multiple Negative Controls}.
\newblock {\em American Journal of Epidemiology\/}, {\bf 185}(10), 950--953.

\bibitem[Morozova {\em et~al.}(2018)Morozova, Cohen, and
  Crawford]{crawford2018risk}
Morozova, O., Cohen, T., and Crawford, F.~W. (2018).
\newblock {Risk Ratios for Contagious Outcomes}.
\newblock {\em Journal of The Royal Society Interface\/}, {\bf 15}(138),
  20170696.

\bibitem[Murphy(2002)Murphy]{murphy2002dynamic}
Murphy, K.~P. (2002).
\newblock {\em {Dynamic Bayesian Networks: Representation, Inference and
  Learning}\/}.
\newblock Ph.D. thesis, University of California, Berkeley.

\bibitem[Myers(2000)Myers]{myers2000diffusion}
Myers, D.~J. (2000).
\newblock {The Diffusion of Collective Violence: Infectiousness,
  Susceptibility, and Mass Media Networks}.
\newblock {\em American Journal of Sociology\/}, {\bf 106}(1), 173--208.

\bibitem[Neyman(1923)Neyman]{neyman1923}
Neyman, J. (1923).
\newblock {On the Application of Probability Theory to Agricultural
  Experiments. Essay on Principles (with discussion). Section 9 (translated)}.
\newblock {\em Statistical Science\/}, {\bf 5}(4), 465--472.

\bibitem[Ogburn(2018)Ogburn]{ogburn2017challenges}
Ogburn, E.~L. (2018).
\newblock {Challenges to Estimating Contagion Effects from Observational Data}.
\newblock In {\em Complex Spreading Phenomena in Social Systems\/}, pages
  47--64. Springer.

\bibitem[Ogburn and VanderWeele(2014)Ogburn and VanderWeele]{ogburn2014DAG}
Ogburn, E.~L. and VanderWeele, T.~J. (2014).
\newblock {Causal Diagrams for Interference}.
\newblock {\em Statistical Science\/}, {\bf 29}(4), 559--578.

\bibitem[Ogburn {\em et~al.}(2017)Ogburn, Sofrygin, Diaz, and van~der
  Laan]{ogburn2017causal}
Ogburn, E.~L., Sofrygin, O., Diaz, I., and van~der Laan, M.~J. (2017).
\newblock {Causal Inference for Social Network Data}.
\newblock {\em arXiv preprint arXiv:1705.08527\/}.

\bibitem[O'Malley {\em et~al.}(2014)O'Malley, Elwert, Rosenquist, Zaslavsky,
  and Christakis]{omalley2014diff}
O'Malley, A.~J., Elwert, F., Rosenquist, J.~N., Zaslavsky, A.~M., and
  Christakis, N.~A. (2014).
\newblock {Estimating Peer Effects in Longitudinal Dyadic Data Using
  Instrumental Variables}.
\newblock {\em Biometrics\/}, {\bf 70}(3), 506--515.

\bibitem[Pearl(1995)Pearl]{pearl1995causal}
Pearl, J. (1995).
\newblock {Causal Diagrams for Empirical Research}.
\newblock {\em Biometrika\/}, {\bf 82}(4), 669--688.

\bibitem[Pearl(2000)Pearl]{pearl2000causality}
Pearl, J. (2000).
\newblock {\em Causality: Models, Reasoning and Inference\/}.
\newblock Cambridge University Press, Cambridge.

\bibitem[Pearl(2001)Pearl]{pearl2001direct}
Pearl, J. (2001).
\newblock {Direct and Indirect Effects}.
\newblock In {\em Proceedings of the Seventeenth Conference on Uncertainty in
  Artificial Intelligence\/}, pages 411--420. Morgan Kaufmann Publishers Inc.

\bibitem[Pearl and Russell(2001)Pearl and Russell]{pearl2001bayesian}
Pearl, J. and Russell, S. (2001).
\newblock {Bayesian Networks}.
\newblock In {\em Handbook of Brain Theory and Neural Networks\/}. MIT Press.

\bibitem[Robins and Greenland(1992)Robins and
  Greenland]{robins1992identifiability}
Robins, J.~M. and Greenland, S. (1992).
\newblock {Identifiability and Exchangeability for Direct and Indirect
  Effects}.
\newblock {\em Epidemiology\/}, pages 143--155.

\bibitem[Rogers(1962)Rogers]{rogers1962diffusion}
Rogers, E.~M. (1962).
\newblock {\em Diffusion of Innovations\/}.
\newblock Simon and Schuster.

\bibitem[Rubin(1974)Rubin]{rubin1974causal}
Rubin, D.~B. (1974).
\newblock {Estimating Causal Effects of Treatments in Randomized and
  Nonrandomized Studies}.
\newblock {\em Journal of Educational Psychology\/}, {\bf 66}(5), 688.

\bibitem[Sacerdote(2001)Sacerdote]{sacerdote2001peer}
Sacerdote, B. (2001).
\newblock {Peer Effects with Random Assignment: Results for Dartmouth
  Roommates}.
\newblock {\em The Quarterly Journal of Economics\/}, {\bf 116}(2), 681--704.

\bibitem[S{\"a}vje {\em et~al.}(2017)S{\"a}vje, Aronow, and
  Hudgens]{savje2017average}
S{\"a}vje, F., Aronow, P.~M., and Hudgens, M.~G. (2017).
\newblock {Average Treatment Effects in the Presence of Unknown Interference}.
\newblock {\em arXiv preprint arXiv:1711.06399\/}.

\bibitem[Shalizi and Thomas(2011)Shalizi and Thomas]{shalizi2011homophily}
Shalizi, C.~R. and Thomas, A.~C. (2011).
\newblock {Homophily and Contagion are Generically Confounded in Observational
  Social Network Studies}.
\newblock {\em Sociological Methods \& Research\/}, {\bf 40}(2), 211--239.

\bibitem[Shpitser {\em et~al.}(2012)Shpitser, VanderWeele, and
  Robins]{shpitser2012validity}
Shpitser, I., VanderWeele, T., and Robins, J.~M. (2012).
\newblock {On the Validity of Covariate Adjustment for Estimating Causal
  Effects}.
\newblock In {\em Proceedings of the 26th Conference on Uncertainty and
  Artificial Intelligence\/}, pages 527--536, Corvallis, WA. AUAI Press.

\bibitem[Sinclair(2012)Sinclair]{sinclair2012social}
Sinclair, B. (2012).
\newblock {\em The Social Citizen: Peer Networks and Political Behavior\/}.
\newblock University of Chicago Press.

\bibitem[Sofer {\em et~al.}(2016)Sofer, Richardson, Colicino, Schwartz, and
  Tchetgen~Tchetgen]{tchetgen2016negative}
Sofer, T., Richardson, D.~B., Colicino, E., Schwartz, J., and
  Tchetgen~Tchetgen, E.~J. (2016).
\newblock {On Negative Outcome Control of Unobserved Confounding as a
  Generalization of Difference-in-Differences}.
\newblock {\em Statistical Science\/}, {\bf 31}(3), 348.

\bibitem[Spirtes {\em et~al.}(2000)Spirtes, Glymour, and
  Scheines]{spirtes2000causation}
Spirtes, P., Glymour, C.~N., and Scheines, R. (2000).
\newblock {\em {Causation, Prediction, and Search}\/}.
\newblock MIT press.

\bibitem[Su and White(2008)Su and White]{su2008nonpara}
Su, L. and White, H. (2008).
\newblock {A Nonparametric Hellinger Metric Test for Conditional Independence}.
\newblock {\em Econometric Theory\/}, {\bf 24}(4), 829--864.

\bibitem[Taylor and Eckles(2017)Taylor and Eckles]{eckles2017Online}
Taylor, S.~J. and Eckles, D. (2017).
\newblock {Randomized Experiments to Detect and Estimate Social Influence in
  Networks}.
\newblock In S.~Lehmann and Y.-Y. Ahn, editors, {\em Spreading Dynamics in
  Social Systems\/}. Springer.

\bibitem[Tchetgen~Tchetgen(2013)Tchetgen~Tchetgen]{tchetgen2013control}
Tchetgen~Tchetgen, E. (2013).
\newblock {The Control Outcome Calibration Approach for Causal Inference with
  Unobserved Confounding}.
\newblock {\em American Journal of Epidemiology\/}, {\bf 179}(5), 633--640.

\bibitem[Tchetgen~Tchetgen {\em et~al.}(2017)Tchetgen~Tchetgen, Fulcher, and
  Shpitser]{tchetgen2017auto}
Tchetgen~Tchetgen, E.~J., Fulcher, I., and Shpitser, I. (2017).
\newblock {Auto-G-Computation of Causal Effects on a Network}.
\newblock {\em arXiv preprint arXiv:1709.01577\/}.

\bibitem[{United Nations High Commissioner for Refugees.}(2017){United Nations
  High Commissioner for Refugees.}]{unhcr2017}
{United Nations High Commissioner for Refugees.} (2017).
\newblock {Global Trends: Forced Displacement in 2017}.

\bibitem[van~der Laan(2014)van~der Laan]{van2014causal}
van~der Laan, M.~J. (2014).
\newblock {Causal Inference for A Population of Causally Connected Units}.
\newblock {\em Journal of Causal Inference\/}, {\bf 2}(1), 13--74.

\bibitem[VanderWeele(2009)VanderWeele]{vanderweele2009concerning}
VanderWeele, T.~J. (2009).
\newblock {Concerning the Consistency Assumption in Causal Inference}.
\newblock {\em Epidemiology\/}, {\bf 20}(6), 880--883.

\bibitem[VanderWeele(2011)VanderWeele]{vanderweele2011sen}
VanderWeele, T.~J. (2011).
\newblock {Sensitivity Analysis for Contagion Effects in Social Networks}.
\newblock {\em Sociological Methods \& Research\/}, {\bf 40}(2), 240--255.

\bibitem[VanderWeele(2015)VanderWeele]{vander2014explanation}
VanderWeele, T.~J. (2015).
\newblock {\em {Explanation in Causal Analysis: Methods for Mediation and
  Interaction}\/}.
\newblock Oxford University Press.
\newblock Forthcoming.

\bibitem[VanderWeele and An(2013)VanderWeele and An]{vanderweele2013social}
VanderWeele, T.~J. and An, W. (2013).
\newblock {Social Networks and Causal Inference}.
\newblock In {\em Handbook of Causal Analysis for Social Research\/}, pages
  353--374. Springer.

\bibitem[VanderWeele {\em et~al.}(2012)VanderWeele, Ogburn, and
  Tchetgen~Tchetgen]{vanderweele2012why}
VanderWeele, T.~J., Ogburn, E.~L., and Tchetgen~Tchetgen, E.~J. (2012).
\newblock {Why and When ``Flawed" Social Network Analyses Still Yield Valid
  Tests of No Contagion}.
\newblock {\em Statistics, Politics and Policy\/}, {\bf 3}(1).

\bibitem[Ver~Steeg and Galstyan(2010)Ver~Steeg and
  Galstyan]{versteeg2010difftest}
Ver~Steeg, G. and Galstyan, A. (2010).
\newblock {Ruling out Latent Homophily in Social Networks}.
\newblock {\em NIPS Workshop on Social Computing\/}.

\bibitem[Ver~Steeg and Galstyan(2013)Ver~Steeg and
  Galstyan]{versteeg2013statistical}
Ver~Steeg, G. and Galstyan, A. (2013).
\newblock {Statistical Tests for Contagion in Observational Social Network
  Studies}.
\newblock In {\em the 16th International Conference on Artificial Intelligence
  and Statistics\/}, pages 563--571.

\bibitem[Wilson and Kelling(1982)Wilson and Kelling]{wilson1982broken}
Wilson, J.~Q. and Kelling, G.~L. (1982).
\newblock {Broken Windows}.
\newblock {\em Atlantic Monthly\/}, {\bf 249}(3), 29--38.

\bibitem[Zhang {\em et~al.}(2012)Zhang, Peters, Janzing, and
  Sch{\"o}lkopf]{zhang2012kernel}
Zhang, K., Peters, J., Janzing, D., and Sch{\"o}lkopf, B. (2012).
\newblock {Kernel-based Conditional Independence Test and Application in Causal
  Discovery}.
\newblock In {\em 27th Conference on Uncertainty in Artificial Intelligence\/}.

\bibitem[Zhang {\em et~al.}(2011)Zhang, Joffe, and Small]{joffe2011control}
Zhang, M., Joffe, M.~M., and Small, D.~S. (2011).
\newblock {Causal Inference for Continuous-Time Processes When Covariates are
  Observed Only at Discrete Times}.
\newblock {\em Annals of Statistics\/}, {\bf 39}(1), 131 -- 173.

\end{thebibliography}

  }

  \clearpage
  \appendix

  \spacingset{1.6}
  \setcounter{table}{0}
  \setcounter{equation}{0}
  \setcounter{figure}{0}
  \renewcommand {\thetable} {A\arabic{table}}
  \renewcommand {\thefigure} {A\arabic{figure}}

  \vspace{-0.5in}

  \begin{center}
    {\LARGE \bf Supplementary Appendix}
  \end{center}

  \section{Proofs}
  \subsection{Proof of Theorem~\ref{placebo}}
  \label{subsec:placebo-proof}
  In this proof, we use $\bC$ and $\bC^P$ to denote $\bC_{i,t+1}$ and
  $\bC^P_{i,t+1}$ for notational simplicity. 
  \subsubsection{Setup}
  Given that control set $\bC$ are defined to be pre-treatment,
  theoretical results on causal DAGs \citep{pearl1995causal,
    shpitser2012validity} imply that $ Y_{i,t+1} (d) \ \indep \ \{Y_{jt}\}_{j \in \cN_i} \mid \bC$ is equivalent to no unblocked back-door
  paths from $\{Y_{jt}\}_{j \in \cN_i}$ to $Y_{i, t+1}$ with
  respect to $\bC$ in causal DAG $\cG$ (see Lemma~\ref{pearl}). Additionally, 
  $ Y_{it} (d) \ \indep \ \{Y_{jt}\}_{j \in \cN_i} \mid \bC^P$ is equivalent to no unblocked back-door
  paths from $\{Y_{jt}\}_{j \in \cN_i}$ to $Y_{it}$ with
  respect to $\bC^P$ in causal DAG $\cG$. Under the sequential
  consistency assumption (Assumption~\ref{seq}), $Y_{it} = Y_{it}(d)$
  for any $d$. Therefore, 
  $ Y_{it} \ \indep \ \{Y_{jt}\}_{j \in \cN_i} \mid \bC^P$ is equivalent to no unblocked
  back-door paths from $\{Y_{jt}\}_{j \in \cN_i}$ to $Y_{it}$
  with respect to $\bC^P$ in causal DAG $\cG$.

  The theorem requires one regularity condition -- the violation
  of the no omitted confounders assumption, if any, is {\it
    proper}. Intuitively, it means that bias (i.e., the violation of the no omitted confounders assumption) is in fact driven by
  omitted variables. Bias is not proper when the only source of bias is the misadjustment of the lag
  structure of observed covariates. Importantly, contextual confounding and homophily bias 
  are proper, and hence within the scope of this theorem. 
  \begin{definition}[Proper Bias]
    \label{proper}
    \spacingset{1.2}{Suppose control set $\bC$ does not satisfy
      Assumption~\ref{ig}. This violation (bias) is defined
      to be proper when it satisfies the following condition: If control
      set $\bC_{i,t+1}$ cannot block all back-door paths from
      $\{Y_{jt}\}_{j \in \cN_i}$ to $Y_{i,t+1}$, there is at least one
      back-door path that any subset of the following set cannot block.   
      \begin{equation*} 
        \{ \bC_{i,t+1}, \bC_{i,t+1}^{(-1)}, \bC_{i,t+1}^{(+1)}, \{Y_{j, t-1}\}_{j \in \cN_i}\},   
      \end{equation*}
      where $\bC_{i,t+1}^{(-1)}$ and $\bC_{i,t+1}^{(+1)}$ are a lag
      and a lead of the time-dependent variables in $\bC_{i,t+1}$. }
  \end{definition}





  \subsubsection{Bias $\rightarrow$ Dependence in Placebo Test}
  \label{subsubsec:app1}
  Here, we show that when set $\bC$ cannot block all back-door paths from $\{Y_{jt}\}_{j \in \cN_i}$ to $Y_{i, t+1}$, set $\bC^P$ cannot
  block all back-door paths from $\{Y_{jt}\}_{j \in \cN_i}$ to  $Y_{it}$. 

  \paragraph{Step 1 (Proper Bias):}
  Given the assumption that the set $\bC$ is proper, set $\bC^P$ cannot block all
  back-door paths from $\{Y_{jt}\}_{j \in \cN_i}$ to $Y_{i, t+1}$
  because $\bC^P$ is a subset of  $\{ \bC, \bC^{(-1)}, \bC^{(+1)}, \{Y_{j, t-1}\}_{j \in \cN_i}\}.$  

  \paragraph{Step 2 (Set up the main unblocked back-door path to investigate):} 
  Let $\pi$ be a back-door path from $\{Y_{jt}\}_{j \in \cN_i}$ to
  $Y_{i, t+1}$ that both $\bC$ and $\bC^P$ and any subset of
  $\{ \bC, \bC^{(-1)}, \bC^{(+1)}, \{Y_{j, t-1}\}_{j \in \cN_i}\}$
  cannot block. Without loss of generality, we assume that
  this unblocked back-door path starts with an arrow pointing to $Y_{kt}$ where $k \in \cN_i$ and it ends with an arrow pointing to
  $Y_{i,t+1}$. 

  \paragraph{Step 3 (Case I. the last node of
    the unblocked back-door path is time-independent):} 
  First, consider a case in which the last variable in an
  unblocked back-door path has a directed arrow pointing to $Y_{i,t+1}$
  and time-independent. Let ($Z$, $Y_{i,t+1}$) denote the last two node path 
  segment on $\pi$ where $Z$ is a time-independent variable and there
  exists a directed arrow from $Z$ to $Y_{i,t+1}$. Note that we do not
  put any individual index to $Z$ because the proof holds for any index. Since this is an
  unblocked path, $Z$ is not in $\bC^P$ and there is
  an unblocked back-door path from $Y_{kt}$ to $Z$. Since $Z$ is time-independent,
  there is a directed arrow from $Z$ to $Y_{it}$ by the structural
  stationarity (Assumption~\ref{ss}). 
  Therefore, set $\bC^P$ cannot block this back-door path from $Y_{kt}$ to $Y_{it}$. 

  \paragraph{Step 4 (Case II. the last node of
    the unblocked back-door path is time-dependent):} 
  Next, consider the case in which the last variable in an
  unblocked back-door path points to $Y_{i, t+1}$ and time-dependent. 
  Let ($B$, $X_{t+1}$, $Y_{i, t+1}$) denote the last three node path 
  segment on $\pi$ where $X_{t+1}$ is a time-dependent direct cause of
  $Y_{i, t+1}$. Note that we do not put any individual index to
  $X_{t+1}$ because the proof holds for any index. $X_{t}, X_{t+1} \not
  \in \bC^P$ because $X_{t+1} \not \in \bC$ (see Lemma~\ref{sub4} in Section~\ref{subsec:lemma}).

  \paragraph{Step 4.1 (sub-Case: the second last node is time-independent):} 
  First, assume $B$ is time-independent. Then, because a causal DAG satisfies
  the structural stationarity (Assumption~\ref{ss}), $X_t$ and $B$ have the same
  relationship as the one between $X_{t+1}$ and $B$. In
  addition, since there is an unblocked path from $Y_{kt}$ to $X_{t+1}$ to through
  $B$, there exists an unblocked path from $Y_{kt}$ to $X_t$ through
  $B$. 
  Given that there exists a directed arrow from $X_{t+1}$ to $Y_{i, t+1}$,
  there exists a directed arrow from $X_{t}$ to $Y_{it}$. Therefore, there is an unblocked back-door path from $Y_{kt}$ to $Y_{it}$. 

  \paragraph{Step 4.2 (sub-Case: the second last node is time-dependent):} 
  Next, assume $B$ is time-dependent and therefore we use
  $B_{t+1}$. First, we show that whenever $B$ is time-dependent, 
  then the directed arrow is always from $X_{t+1}$ to $B_{t+1}.$ Suppose there is a directed arrow from $B_{t+1}$ to
  $X_{t+1}.$ If $B_{t+1}$ in $\bC^P$, then this back-door is blocked
  (therefore, choose another $\pi$). So, $B_{t+1}$ is not in $\bC^P$. 
  Therefore, we can collapse $B_{t+1}$ into $X_{t+1}$, meaning that if
  $B$ is time dependent, then the directed arrow is always from $X_{t+1}$ to $B_{t+1}.$ 

  Now, suppose there is a directed arrow from $X_{t+1}$
  to $B_{t+1}.$ We know there exists an unblocked path from $Y_{kt}$ 
  to $X_{t+1}$ through $B_{t+1}$. Now, because $Y_{it} \leftarrow
  X_{t} \rightarrow X_{t+1} \rightarrow B_{t+1}$, there is an unblocked back-door
  path from $Y_{kt}$ to $Y_{it}$ because the underlying causal DAG
  satisfies the structural stationarity. 
  \qed

  \subsubsection{No Bias $\rightarrow$ Independence in Placebo Test}
  \label{subsubsec:app2}
  Next, we prove that when set $\bC$ can block all back-door paths from $\{Y_{jt}\}_{j \in \cN_i}$ to
  $Y_{i, t+1}$, set $\bC^P$ can block all back-door paths from
  $\{Y_{jt}\}_{j \in \cN_i}$ to $Y_{it}.$ We show the contraposition:
  when there is a back-door path from $\{Y_{jt}\}_{j \in \cN_i}$ to $Y_{it}$ that set $\bC^P$ cannot block, set $\bC$
  cannot block all back-door paths from $\{Y_{jt}\}_{j \in \cN_i}$
  to $Y_{i, t+1}$. Since $\bC$ does not include any $\mbox{Des}(Y_{kt})$, we
  know $\bC^P$ also does not include any $\mbox{Des}(Y_{kt}).$ Also, by
  definition, $\bC^P$ does not include any $\mbox{Des}(Y_{it}).$ Therefore, 
  without loss of generality, we can focus on 
  unblocked back-door paths that start with an arrow pointing to
  $Y_{kt}$ where $k \in \cN_i$ and end with an arrow pointing to $Y_{it}$. 

  \paragraph{Step 1 (Control Set cannot block all back-door paths to the
    Placebo outcome):} 
  First, we show that when there is a back-door path from $Y_{kt}$ to $Y_{it}$ that set $\bC^P$ cannot block, set $\bC$ 
  cannot block all back-door paths from $Y_{kt}$ to $Y_{it}$. 
  From set $\bC^P$ to set $\bC$, we need to (1) add $\mbox{Des}(Y_{it})$
  and (2) remove $\bC^{(-1)}$ and $\{Y_{j, t-1}\}_{j \in \cN_i}$. We
  show here that this process cannot block a back-door path
  that set $\bC^P$ cannot block. The step (1) cannot block the back-door
  path because adding $\mbox{Des}(Y_{it})$ cannot block a back-door path from $Y_{kt}$
  to $Y_{it}$ unblocked by set $\bC^P$ (see Lemma~\ref{des} in Section~\ref{subsec:lemma}). 
  For (2), we first check whether removing $X_t \in \bC^{(-1)}$ can block a
  back-door path that set $\bC^P$ cannot block. To begin with, we can remove $X_t$
  because $X_{t+1} \in \bC$. Removing variables $X_t$ can be
  helpful if $X_t$ is a collider or a descendant of a collider for a
  back-door path. However, if so, $X_{t+1}$ is a descendant of a collider
  and it is in set $\bC$ and therefore, removing $X_t$ cannot block any
  additional paths. Next, we need to check whether removing a variable
  $B \in \{Y_{j, t-1}\}_{j \in \cN_i}$ can block the back-door path that
  the set $\bC^P$ cannot block. Removing variable $B$ can be
  helpful if $B$ is a collider or a descendant of a collider for a
  back-door path. If so, there is an unblocked back-door path (with
  respect to $\bC^P$) that starts with an arrow pointing to $B$ and ends
  with an arrow pointing to $Y_{it}$, i.e., $B \leftarrow \ldots \rightarrow Y_{it}.$ Since $B$ has a directed
  arrow pointing to $Y_{kt}$, removing $B$ unblock a new back-door
  path from $Y_{kt}$ through $B$, which points to $Y_{it}$. Although
  this unblocked back-door path with respect to $\bC$ is different from
  the unblocked back-door path with respect to $\bC^P$, the paths are
  the same after node $B$ and therefore at least the last three nodes
  are the same. Therefore, we can use $\pi$ to be a back-door from $Y_{kt}$ to $Y_{it}$
  that both sets $\bC$ and $\bC^P$ cannot block. 

  \paragraph{Step 2 (Case I: the last node of
    the unblocked back-door path is time-independent):} 
  Consider the case in which the last two nodes are $(Z \rightarrow Y_{it})$ and $Z$
  is time-independent. Then, since $Z \rightarrow Y_{i,t+1}$ from the
  structural stationarity (Assumption~\ref{ss}), set $\bC$ cannot block this back-door. 

  \paragraph{Step 3 (Case II: the last node of
    the unblocked back-door path is time-dependent):} 
  Next, consider the case in which the last two nodes are $(X_{t}
  \rightarrow Y_{it}).$ Since $X_t \not \in \bC^P$ and $X_t \not \in
  \mbox{Des}(Y_{it})$, $X_t, X_{t+1} \not \in \bC.$ Therefore, set $\bC$
  cannot block $Y_{kt} \leftarrow \cdots X_{t} \rightarrow X_{t+1} \rightarrow Y_{i,t+1}.$
  \qed

  \subsection{Proof of Lemmas used for Theorem~\ref{placebo}}
  \label{subsec:lemma}
  Here, we prove all the lemmas used to prove Theorem~\ref{placebo}.
  \begin{lemma}[Equivalence between Back-Door Criteria and No Omitted
    Confounder Assumption \citep{shpitser2012validity}]
    \label{pearl}
    For a pretreatment control set $\bC$, the following two statements hold.
    \begin{enumerate}
    \item If a set $\bC$ satisfies the back-door criterion with respect to
      $(Y_{i, t+1}, \{Y_{jt}\}_{j \in \cN_i})$ in causal DAG $\cG$, then $Y_{i,t+1} (d) \ \indep \ \{Y_{jt}\}_{j \in \cN_i}\ \mid
      \bC$ holds in every causal model inducing causal DAG $\cG$ \citep{pearl1995causal}. 
    \item If $ Y_{i,t+1} (d) \ \indep \ \{Y_{jt}\}_{j \in \cN_i} \mid \bC$
      holds in every causal model inducing causal DAG $\cG$, then 
      a set $\bC$ satisfies the back-door criterion with respect to
      $(Y_{i,t+1}, \{Y_{jt}\}_{j \in \cN_i})$ in causal DAG $\cG$ \citep{shpitser2012validity}. 
    \end{enumerate}
  \end{lemma}

  \begin{lemma}
    \label{sub4}
    $X_{t+1} \not \in \bC \rightarrow  X_{t}, X_{t+1} \not \in \bC^P.$ \vspace{-0.075in}
  \end{lemma}
  \paragraph{Proof}
  First, we show that $X_{t}, X_{t+1}, X_{t+2} \not \in \bC$ because 
  set $\bC$ is proper. It is because if $X_{t}$ or $X_{t+2}$ are in $\bC$, 
  then the lag adjustment of the control set $\bC$ can block this path. If this path is the only
  back-door path, then $\bC$ is not proper. If there is another
  back-door path that any subset of $\{ \bC, \bC^{(-1)},
  \bC^{(+1)}, \{Y_{j, t-1}\}_{j \in \cN_i}\}$ cannot block, choose it as $\pi$. 

  Next, we show that $X_{t}, X_{t+1} \not \in \bC^P$. There are three ways
  for a variable to be in the placebo set $\bC^P$. We discuss them in order. First,
  a variable can be in the placebo set because it was already in the control
  set. We know $X_{t}, X_{t+1} \not \in \bC$, so this option is not feasible.
  Second, a variable can be in the placebo set because it is a lag of
  the original control variables. Given that $X_{t+1}, X_{t+2}$ are not
  in the control set, this option is also not feasible. Finally, a
  variable can be in the placebo set because it is a lag of the
  treatment variable. (a) It is important to notice
  that $X_{t} \notin \{Y_{j, t-1}\}_{j \in \cN_i}$ because  $X_{t+1}
  \notin \{Y_{jt}\}_{j \in \cN_i}$ (i.e., the treatment cannot be
  the last node of the unblocked back-door path). (b) Now, we verify $X_{t+1} \notin \{Y_{j,
    t-1}\}_{j \in \cN_i}$. First, this
  back-door path can be blocked by a subset of $\{ \bC, \bC^{(-1)},
  \bC^{(+1)}, \{Y_{j, t-1}\}_{j \in \cN_i}\}$. If this back-door is
  the only unblocked back-door, set $\bC$ is not proper, therefore this
  is contradictory. If there is another
  back-door path that both $\bC$ and $\bC^P$ cannot block, choose it as $\pi$. \qed

  \begin{lemma}
    \label{des}
    Adding $\mbox{Des}(Y_{it})$ cannot block a back-door path from $Y_{kt}$
    to $Y_{it}$ unblocked by set $\bC^P$. \vspace{-0.075in}
  \end{lemma}
  \paragraph{Proof}
  Suppose controlling for $\mbox{Des}(Y_{it})$ can block a
  back-door path from $Y_{kt}$ to $Y_{it}$ that the original set $\bC^P$
  cannot block. Since $\bC^P$ does not include any $\mbox{Des}(Y_{kt})$ or $\mbox{Des}(Y_{it})$, this
  unblocked back-door path contains an arrow pointing to $Y_{it}$. 

  \paragraph{Step 1 (Set up the main node $B$):} At least one of $\mbox{Des}(Y_{it})$ is a non-collider on this path
  given that controlling for $\mbox{Des}(Y_{it})$ can block this path. Let $B$
  be such a variable and focus on one arrow pointing out from the node $B$. 

  \paragraph{Step 2 (Case I. Consider one side of the main node $B$):} First, suppose this direction leads to
  $Y_{it}$. Then, since $B$ is a $\mbox{Des}(Y_{it}),$ a directed path
  from node $B$ to $Y_{it}$ cannot exist and therefore, there must be a
  collider on this direction of the path.  Since this collider is also in $\mbox{Des}(Y_{it})$
  and therefore not controlled in the original $\bC^P$, this  back-door is blocked by set $\bC^P.$ 

  \paragraph{Step 3 (Case II. Consider the other side of the main node $B$):}
  Next, consider the direction that leads to $Y_{kt}.$ Then, since $Y_{it}$ is not a cause of $Y_{kt},$ 
  a directed path from node $B$ to $Y_{kt}$ cannot exist and therefore, there must be a
  collider on this direction of the path. Since this collider is also in
  $\mbox{Des}(Y_{it})$ and therefore not controlled in the original $\bC^P$, this
  back-door is blocked by set $\bC^P.$ Hence, this is
  contradiction. This proves that controlling for $\mbox{Des}(Y_{it})$ cannot
  block a back-door path from $Y_{kt}$ to $Y_{it}$ that set $\bC^P$
  cannot block. \qed

  \subsection{Placebo Test as Joint Test}
  \label{subsec:joint}
  In this section, we clarify a relationship between the placebo test and the
  sequential consistency (Assumption~\ref{seq}). While we assume the
  sequential consistency in Theorem~\ref{placebo} to assess the no
  omitted confounders assumption, a simple proof can show that the
  proposed placebo test can also be viewed as a joint test of the
  sequential consistency assumption and the no omitted confounders
  assumption under the structural stationarity. Formally, 
  \begin{lemma}[\small Equivalence between Identification Assumptions 
    and Conditional Independence of Simultaneous Outcomes]
    \label{placebo2} \spacingset{1.2}{
      Under Assumption~\ref{ss}, 
      \begin{eqnarray*}
        \hspace{-0.3in} &&
                           \begin{cases}
                             \ \text{Sequential Consistency (Assumption~\ref{seq})} \\[2pt]
                             Y_{i, t+1}(d)   \ \indep \ \{Y_{jt}\}_{j \in \cN_i} \mid \bC_{i,t+1} 
                           \end{cases} 
        {\vspace{-0.05in} \Longleftrightarrow}  \ \ \  Y_{it} \ \indep \  \{Y_{jt}\}_{j \in \cN_i} \mid \bC_{i,t+1}^P. \label{eq:placebotest2}   
      \end{eqnarray*}}
  \end{lemma}
  This lemma shows that researchers can assess not only the assumption of
  no omitted confounders (Assumption~\ref{ig}) but also the sequential
  consistency assumption (Assumption~\ref{seq}) together. That is, 
  researchers can jointly detect simultaneity bias and omitted variable
  bias. When the conditional independence of simultaneous outcomes holds, it provides strong statistical
  evidence for both identification assumptions, i.e., the absence of
  simultaneity bias and omitted variable bias. In contrast, when we
  reject the null hypothesis of the placebo test, we cannot tell
  which assumption is violated. When the sequential consistency
  assumption is violated, the problem is more severe than omitted variable
  bias -- causal diffusion effects are not well defined. 

  The proof of this lemma is essentially the same as the one for
  Theorem~\ref{placebo} and thus is omitted. One additional idea is that when the
  sequential consistency assumption is violated, there is no set of
  variables that can make simultaneous outcomes conditionally
  independent -- the null hypothesis of the placebo test is always
  rejected.

  \subsection{Proof of Theorem~\ref{biascorrect}}
  \label{subsec:proof-correct}
  Below, we describe two lemmas useful for proving
  Theorem~\ref{biascorrect}. For completeness, their proofs follow.

  \begin{lemma}
    \label{bias-lemma0}
    \begin{equation*} 
      Y_{i,t+1} (d^L) \  \indep   \  \{Y_{jt}\}_{j \in \cN_i} \mid
      U_{i,t+1}, \bC_{i,t+1} \Longrightarrow Y_{i,t+1} (d^L)  \  \indep   \  \{Y_{jt}\}_{j \in \cN_i} \mid
      U_{i,t+1}, \bX_{i,t+1}, \bC_{i,t+1}^B
    \end{equation*}
  \end{lemma}
  \begin{lemma}
    \label{bias-lemma}
    Under Assumption~\ref{timeInv},
    \begin{eqnarray*}
      && \E[ Y_{i, t+1} (d^L) \mid D_{it}=d^H, \bX_{i,t+1} = \bx,
         \bC_{i, t+1}^B = \bc]  -   \E[ Y_{i, t+1} (d^L) \mid D_{it}=d^L,
         \bX_{i,t+1} = \bx, \bC_{i,t+1}^B  = \bc]  \\
      & = &   \E[ Y_{it} (d^L) \mid D_{it}=d^H, \bX_{it} = \bx, \bC_{i,t+1}^B =
            \bc]  -   \E[ Y_{it} (d^L) \mid D_{it}=d^L, \bX_{it} = \bx, \bC_{i,t+1}^B = \bc].  
    \end{eqnarray*}
  \end{lemma}
  \paragraph{Proof of the theorem}
  Based on Lemma~\ref{bias-lemma} and Assumptions~\ref{seq} and~\ref{timeInv}, 
  \begin{eqnarray*}
    && \E[ Y_{i, t+1} (d^L) \mid D_{it}=d^H, \bX_{i,t+1} = \bx, \bC_{i,t+1}^B = \bc]  \\
    & = & \E[ Y_{i, t+1} (d^L) \mid D_{it}=d^L, \bX_{i,t+1} = \bx, \bC_{i,t+1}^B  = \bc]  \\
    && \quad + \E[ Y_{it} (d^L) \mid D_{it}=d^H, \bX_{it} = \bx, \bC_{i,t+1}^B =
       \bc]  -   \E[ Y_{it} (d^L) \mid D_{it}=d^L, \bX_{it} = \bx,
       \bC_{i,t+1}^B = \bc] \\ 
    & = & \E[ Y_{i, t+1} \mid D_{it}=d^L, \bX_{i,t+1} = \bx, \bC^B_{i,t+1}  = \bc]  \\
    && \quad + \E[ Y_{it} \mid D_{it}=d^H, \bX_{it} = \bx, \bC^B_{i,t+1} =
       \bc]  -   \E[ Y_{it} \mid D_{it}=d^L, \bX_{it} = \bx, \bC^B_{i,t+1} = \bc].  
  \end{eqnarray*}

  Therefore, {\small
    \begin{eqnarray*}
      && \E[Y_{i, t+1} (d^H)  - Y_{i, t+1} (d^L) \mid D_{it}=d^H] \\ 
      &= & \int \{\E[Y_{i, t+1} (d^H) \mid D_{it}=d^H, \bX_{i,t+1}, \bC^B_{i,t+1}] \\
      && \qquad - \E[ Y_{i, t+1} (d^L) \mid D_{it}=d^H, \bX_{i,t+1}, \bC^B_{i,t+1}]\} d F_{\bX_{i,t+1}, \bC^B_{i,t+1} \mid D_{it}=d^H}(\bx, \bc) \\
      &= & \int \E[Y_{i, t+1} \mid D_{it}=d^H, \bX_{i,t+1}, \bC^B_{i,t+1}] d F_{\bX_{i,t+1}, \bC^B_{i,t+1} \mid D_{it}=d^H}(\bx, \bc)\\
      && - \bigl\{ \E[ Y_{i, t+1} \mid D_{it}=d^L, \bX_{i,t+1} = \bx, \bC^B_{i,t+1}  = \bc]  + \E[ Y_{it} \mid D_{it}=d^H, \bX_{it} = \bx, \bC^B_{i,t+1} =
         \bc]  \\
      && \quad -   \E[ Y_{it} \mid D_{it}=d^L, \bX_{it} = \bx, \bC^B_{i,t+1} =
         \bc] \bigr\} d F_{\bX_{i,t+1}, \bC^B_{i,t+1} \mid D_{it}=d^H}(\bx, \bc)\\
      &= & \int \bigl\{ \E [Y_{i, t+1} \mid D_{it}=d^H,
           \bX_{i,t+1}, \bC^B_{i,t+1}]  - \E[ Y_{i, t+1}
           \mid  D_{it} = d^L, \bX_{i,t+1}, \bC^B_{i,t+1}] \bigr\} d F_{\bX_{i,t+1}, \bC^B_{i,t+1} \mid D_{it}=d^H}(\bx, \bc)\\
      && - \int \bigl\{ \E[Y_{it} \mid D_{it}=d^H, \bX_{it}, \bC^B_{i,t+1}]  -  \E[Y_{it} \mid D_{it}=d^L, \bX_{it},
         \bC^B_{i,t+1}] \bigr\} d F_{\bX_{i,t+1}, \bC^B_{i,t+1} \mid D_{it}=d^H}(\bx, \bc).
    \end{eqnarray*}}
  This completes the proof of Theorem~\ref{biascorrect} in cases where $U_{i,t+1}$ is
  time-dependent and affected by the outcome at time $t$. In
  Section~\ref{subsubsec:extend}, we 
  extend results to two other cases (1) when $U_{i,t+1}$ is time-dependent but is not affected by the outcome at time $t$ and (2) when
  unobserved confounder is time-independent $Z_i$. \qed

  \subsubsection{Proof of Lemma~\ref{bias-lemma0}}
  If we write out control set $\bC$, the lemma can be rewritten as 
  \begin{eqnarray*} 
    && Y_{i,t+1} (d^L) \  \indep   \  \{Y_{jt}\}_{j \in \cN_i} \mid
       U_{i,t+1}, \bX_{i,t+1}, \bV_{i,t+1},\bZ_i \\ 
    & \Longrightarrow & Y_{i,t+1} (d^L)  \  \indep   \  \{Y_{jt}\}_{j \in \cN_i}
                        \mid U_{i,t+1}, \bX_{i,t+1}, \bV_{i,t+1}, \bV_{it}, \bZ_i, \{Y_{j,t-1}\}_{j \in \cN_i}.
  \end{eqnarray*}
  First, note that all variables in set $\{U_{i,t+1},
  \bX_{i,t+1}, \bV_{i,t+1}, \bV_{it}, \bZ_i, \{Y_{j,t-1}\}_{j \in
    \cN_i}\}$ are neither affected by the potential outcome, $Y_{i,t+1}(d^L)$,
  nor affected by the treatment $\{Y_{jt}\}_{j \in \cN_i}$. The difference
  between the conditioning sets in the right- and left-hand sides is
  $\bV_{it}$ and $ \{Y_{j,t-1}\}_{j \in \cN_i}.$ Including these
  variables can open back-door paths only when these
  variables are colliders for these new back-door paths. However, because
  a descendant of $\bV_{it}$, $\bV_{i,t+1}$, is in the conditioning
  set, it is contradictory if conditioning on $\bV_{it}$ can open a
  new back-door path. Additionally, because $\{Y_{j,t-1}\}_{j \in \cN_i}$ is a
  parent of the treatment $\{Y_{jt}\}_{j \in \cN_i}$, it is contradictory
  if conditioning on $\{Y_{j,t-1}\}_{j \in \cN_i}$ can open a new
  back-door path. Therefore,  including $\bV_{it}$ and $
  \{Y_{j,t-1}\}_{j \in \cN_i}$ don't open any back-door path, which
  completes the proof. \qed
  \subsubsection{Proof of Lemma~\ref{bias-lemma}}
  Under Assumption~\ref{timeInv}, {\small
    \begin{eqnarray*}
      && \int_{\mathcal{C}}  \{\E[Y_{i, t+1} (d^L) | U_{i,t+1} = u_1,
         \bX_{i,t+1}=\bx, \bC^B_{i,t+1}=\bc] - \E[Y_{i, t+1} (d^L) | U_{i,t+1} =
         u_0, \bX_{i,t+1}=\bx, \bC^B_{i,t+1}=\bc] \} \\
      && \qquad \times \{ dF_{U_{i,t+1} \mid D_{it} = d^H,
         \bX_{i,t+1}=\bx, \bC^B_{i,t+1}=\bc} (u_1) - d F_{U_{i,t+1} \mid
         D_{it} = d^L, \bX_{i,t+1}=\bx, \bC^B_{i,t+1}=\bc} (u_1) \}\\ 
      & = & \int_{\mathcal{C}}  \{\E[Y_{it} (d^L) | U_{it} = u_1,
            \bX_{it}=\bx, \bC^B_{i,t+1}=\bc] - \E[Y_{it} (d^L) | U_{it} =
            u_0, \bX_{it}=\bx, \bC^B_{i,t+1}=\bc] \} \\
      && \qquad \times \{ dF_{U_{it} \mid D_{it} = d^H,
         \bX_{it}=\bx, \bC^B_{i,t+1}=\bc} (u_1) - d F_{U_{it} \mid
         D_{it} = d^L, \bX_{it}=\bx, \bC^B_{i,t+1}=\bc} (u_1) \}.
    \end{eqnarray*}}
  Now we analyze each side of the equation. {\small
    \begin{eqnarray*}
      && \int_{\mathcal{C}}  \{\E[Y_{i, t+1} (d^L) | U_{i,t+1} = u_1,
         \bX_{i,t+1}=\bx, \bC^B_{i,t+1}=\bc] - \E[Y_{i, t+1} (d^L) | U_{i,t+1} =
         u_0, \bX_{i,t+1}=\bx, \bC^B_{i,t+1}=\bc] \} \\
      && \qquad \times \{ dF_{U_{i,t+1} \mid D_{it} = d^H,
         \bX_{i,t+1}=\bx, \bC^B_{i,t+1}=\bc} (u_1) - d F_{U_{i,t+1} \mid
         D_{it} = d^L, \bX_{i,t+1}=\bx, \bC^B_{i,t+1}=\bc} (u_1) \}\\ 
      & = & \int_{\mathcal{C}}  \E[Y_{i, t+1} (d^L) | U_{i,t+1} = u_1,
            \bX_{i,t+1}=\bx, \bC^B_{i,t+1}=\bc]\\
      && \qquad \times \{ dF_{U_{i,t+1} \mid D_{it} = d^H,
         \bX_{i,t+1}=\bx, \bC^B_{i,t+1}=\bc} (u_1) - d F_{U_{i,t+1} \mid
         D_{it} = d^L, \bX_{i,t+1}=\bx, \bC^B_{i,t+1}=\bc} (u_1) \}\\ 
      & = & \int_{\mathcal{C}}  \E[Y_{i, t+1} (d^L) | D_{it} = d^H, U_{i,t+1} = u_1,
            \bX_{i,t+1}=\bx, \bC^B_{i,t+1}=\bc] dF_{U_{i,t+1} \mid D_{it} = d^H,
            \bX_{i,t+1}=\bx, \bC^B_{i,t+1}=\bc} (u_1) \\
      && \qquad - \int_{\mathcal{C}}  \E[Y_{i, t+1} (d^L) | D_{it} = d^L,
         U_{i,t+1} = u_1,
         \bX_{i,t+1}=\bx, \bC^B_{i,t+1}=\bc]  d F_{U_{i,t+1} \mid
         D_{it} = d^L, \bX_{i,t+1}=\bx, \bC^B_{i,t+1}=\bc} (u_1) \\ 
      & = & \E[Y_{i, t+1} (d^L) | D_{it} = d^H, 
            \bX_{i,t+1}=\bx, \bC^B_{i,t+1}=\bc] - \E[Y_{i, t+1} (d^L) | D_{it} = d^L,
            \bX_{i,t+1}=\bx, \bC^B_{i,t+1}=\bc],
    \end{eqnarray*}}
  where the first equality follows from the fact that $\E[Y_{i, t+1} (d^L) | U_{i,t+1} =
  u_0, \bX_{i,t+1}=\bx, \bC^B_{i,t+1}=\bc]$ does not include $u_1$, the
  second equality comes from Lemma~\ref{bias-lemma0}, and the final
  from the rule of conditional expectations. 
  Similarly, {\small
    \begin{eqnarray*}
      && \int_{\mathcal{C}}  \{\E[Y_{it} (d^L) | U_{it} = u_1,
         \bX_{it}=\bx, \bC^B_{i,t+1}=\bc] - \E[Y_{it} (d^L) | U_{it} =
         u_0, \bX_{it}=\bx, \bC^B_{i,t+1}=\bc] \} \\
      && \qquad \times \{ dF_{U_{it} \mid D_{it} = d^H,
         \bX_{it}=\bx, \bC^B_{i,t+1}=\bc} (u_1) - d F_{U_{it} \mid
         D_{it} = d^L, \bX_{it}=\bx, \bC^B_{i,t+1}=\bc} (u_1) \}.\\
      & = &   \E[ Y_{it} (d^L) \mid D_{it}=d^H, \bX_{it} = \bx, \bC^B_{i,t+1} =
            \bc]  -   \E[ Y_{it} (d^L) \mid D_{it}=d^L, \bX_{it} = \bx, \bC^B_{i,t+1} = \bc].  
    \end{eqnarray*}}
  Taken together, {\small 
    \begin{eqnarray*}
      && \E[ Y_{i, t+1} (d^L) \mid D_{it}=d^H, \bX_{i,t+1} = \bx,
         \bC^B_{i,t+1} = \bc]  -   \E[ Y_{i, t+1} (d^L) \mid D_{it}=d^L,
         \bX_{i,t+1} = \bx, \bC^B_{i,t+1}  = \bc]  \\
      & = &   \E[ Y_{it} (d^L) \mid D_{it}=d^H, \bX_{it} = \bx, \bC^B_{i,t+1} =
            \bc]  -   \E[ Y_{it} (d^L) \mid D_{it}=d^L, \bX_{it} = \bx, \bC^B_{i,t+1} = \bc].  
    \end{eqnarray*}}
  \qed

  \subsubsection{Other cases}
  \label{subsubsec:extend}
  In Theorem~\ref{biascorrect}, we consider cases in which $U_{i,t+1}$ is
  time-dependent and affected by the outcome at time $t$. Now we study two other cases (1)
  when $U_{i,t+1}$ is time-dependent but is not affected by the outcome at time $t$ and (2) when
  unobserved confounder is time-independent $Z_i$. For both cases, Assumption~\ref{timeInv}
  needs to be modified accordingly, although their substantive meanings
  stay the same. The definition of the bias-corrected estimator is also 
  the same. For case (1), define $\widetilde{U}_{i} \equiv
  (U_{i,t+1}, U_{it})$ and for case (2), define $\widetilde{U}_{i}
  \equiv Z_{i}.$ Then, Assumption~\ref{timeInv} is modified as follows.
  \begin{enumerate}
  \item Time-invariant effect of unobserved confounder $\widetilde{U}$: For all $u_1, u_0, \bx$ and $\bc,$
    \begin{eqnarray*}
      \hspace{-0.6in} && \E[Y_{i, t+1} (d^L) \mid  \widetilde{U}_{i}  = u_1, \bX_{i,t+1}=\bx, \bC^B_{i,t+1}=\bc] - 
                         \E[Y_{i, t+1} (d^L) \mid \widetilde{U}_{i} = u_0,\bX_{i,t+1}=\bx, \bC^B_{i,t+1}=\bc] \\
      \hspace{-0.6in} & = & \E[Y_{it} (d^L) \mid  \widetilde{U}_{i} = u_1, \bX_{it}=\bx, \bC^B_{i,t+1}=\bc] - \E[Y_{it} (d^L) \mid \widetilde{U}_{i}  = u_0, \bX_{it}=\bx, \bC^B_{i,t+1}=\bc].
    \end{eqnarray*}
  \item Time-invariant imbalance of unobserved confounder $\widetilde{U}$: For
    all $u, \bx$ and $\bc,$
    \begin{eqnarray*}
      \hspace{-0.6in} && \Pr (\widetilde{U}_{i} \leq u \mid D_{it} = d^H, \bX_{i,t+1}=\bx,
                         \bC^B_{i,t+1}=\bc) - \Pr (\widetilde{U}_{i} \leq u \mid D_{it} = d^L, \bX_{i,t+1}=\bx, 
                         \bC^B_{i,t+1}=\bc)  \\ 
      \hspace{-0.6in} & = &   \Pr (\widetilde{U}_{i} \leq u \mid D_{it} = d^H, \bX_{it}=\bx,
                            \bC^B_{i,t+1}=\bc) -\Pr (\widetilde{U}_{i} \leq u \mid D_{it} = d^L, \bX_{it}=\bx,
                            \bC^B_{i,t+1}=\bc).
    \end{eqnarray*}
  \end{enumerate}

  \subsection{Extensions}
  \label{subsec:bc-ex}
  \subsubsection{Sensitivity Analysis}
  As Lemma~\ref{bias-lemma} shows, Assumption~\ref{timeInv} is
  equivalent to the following equality. 
  \begin{eqnarray*}
    && \E[ Y_{i, t+1} (d^L) \mid D_{it}=d^H, \bX_{i,t+1} = \bx,
       \bC_{i, t+1}^B = \bc]  -   \E[ Y_{i, t+1} (d^L) \mid D_{it}=d^L,
       \bX_{i,t+1} = \bx, \bC_{i,t+1}^B  = \bc]  \\
    & = &   \E[ Y_{it} (d^L) \mid D_{it}=d^H, \bX_{it} = \bx, \bC_{i,t+1}^B =
          \bc]  -   \E[ Y_{it} (d^L) \mid D_{it}=d^L, \bX_{it} = \bx, \bC_{i,t+1}^B = \bc], 
  \end{eqnarray*}
  which substantively means the time-invariant bias. However, this
  assumption might hold only approximately in applied settings. To
  assess the robustness of the bias-corrected estimates, we consider
  a sensitivity analysis. In particular, we introduce sensitivity
  parameter $\lambda$ as follows. 
  \begin{eqnarray*}
    \frac{\mbox{B}_{t+1}(\bx, \bc)}{\mbox{B}_{t}(\bx, \bc)} = \lambda
  \end{eqnarray*}
  where 
  \begin{eqnarray*}
    \hspace{-0.6in} \mbox{B}_{t+1}(\bx, \bc) & = & \E[ Y_{i, t+1} (d^L) \mid D_{it}=d^H, \bX_{i,t+1} = \bx,
                                                   \bC_{i, t+1}^B = \bc]  -   \E[ Y_{i, t+1} (d^L) \mid D_{it}=d^L,
                                                   \bX_{i,t+1} = \bx, \bC_{i,t+1}^B  = \bc],  \\
    \hspace{-0.6in} \mbox{B}_{t}(\bx, \bc) & = &   \E[ Y_{it} (d^L) \mid D_{it}=d^H, \bX_{it} = \bx, \bC_{i,t+1}^B =
                                                 \bc]  -   \E[ Y_{it} (d^L) \mid D_{it}=d^L, \bX_{it} = \bx, \bC_{i,t+1}^B = \bc]. 
  \end{eqnarray*}
  The time-invariance assumption (Assumption~\ref{timeInv}) corresponds
  to $\lambda = 1.$ Using this sensitivity parameter, we can re-define
  the bias-corrected estimator as follows.
  \begin{eqnarray*}
    \hat{\tau}_{\text{Main}}  - \lambda \times \hat{\delta}_{\text{Placebo}} \label{eq:bc-sen}
  \end{eqnarray*}
  Therefore, a sensitivity analysis is to compute the bias-corrected estimator for a range of plausible values of $\lambda$ and investigate
  whether substantive conclusions vary according to the choice of the sensitivity parameter.   
  
  \subsubsection{Assumptions for Identification of ACDE}
  As we show in Section~\ref{subsubsec:BCtheory2},
  Assumption~\ref{timeInv} is sufficient for the identification of the
  ACDE for the treated. Here, we consider an extension of this
  assumption sufficient for the identification of the ACDE. In
  particular, we additionally assume the following equality, which is
  an extension of Assumption~\ref{timeInv}.1 to the case of potential outcomes $Y_i(d^H).$    
  \paragraph{Assumption~\ref{timeInv}.3} (Time-invariant effect of
  unobserved confounder $U$ on potential outcomes $Y_{i, t+1} (d^H)$)
  \begin{eqnarray*}
    \hspace{-0.6in} && \E[Y_{i, t+1} (d^H) | U_{i,t+1} = u_1, \bX_{i,t+1}=\bx, \bC_{i,t+1}^B=\bc] - \E[Y_{i, t+1} (d^H) | U_{i,t+1} = u_0, \bX_{i,t+1}=\bx, \bC_{i,t+1}^B=\bc] \\
    \hspace{-0.6in} & = & \E[Y_{it} (d^H) | U_{it} = u_1, \bX_{it}=\bx, \bC_{i,t+1}^B=\bc] - \E[Y_{it} (d^H) | U_{it} = u_0, \bX_{it}=\bx, \bC_{i,t+1}^B=\bc],
  \end{eqnarray*}
  for all $u_1, u_0, \bx$ and $\bc.$ 

  \noindent Combining this assumption and
  Assumption~\ref{timeInv}.2, we obtain 
  {\small 
    \begin{eqnarray*}
      && \E[ Y_{i, t+1} (d^H) \mid D_{it}=d^H, \bX_{i,t+1} = \bx,
         \bC^B_{i,t+1} = \bc]  -   \E[ Y_{i, t+1} (d^H) \mid D_{it}=d^L,
         \bX_{i,t+1} = \bx, \bC^B_{i,t+1}  = \bc]  \\
      & = &   \E[ Y_{it} (d^H) \mid D_{it}=d^H, \bX_{it} = \bx, \bC^B_{i,t+1} =
            \bc]  -   \E[ Y_{it} (d^H) \mid D_{it}=d^L, \bX_{it} = \bx, \bC^B_{i,t+1} = \bc],
    \end{eqnarray*}}
  where the proof follows from Lemma~\ref{bias-lemma}. Using this
  result, we can additionally show the identification of the ACDE for
  units who received $d^L.$
  \begin{eqnarray*}
    \E[\hat{\tau}^{d^L}_{\text{BC}}]  & = & \tau^{d^L}_{t+1} (d^H, d^L),
  \end{eqnarray*}
  under Assumption~\ref{timeInv}.2, and
  Assumption~\ref{timeInv}.3, where 
  \begin{eqnarray*}
    \hspace{-0.4in} \hat{\tau}^{d^L}_{\text{BC}}  &= & \int \bigl\{ \widehat{\E} [Y_{i, t+1} \mid D_{it}=d^H,
                                                       \bX_{i,t+1}, \bC^B_{i,t+1}]  - \widehat{\E}[ Y_{i, t+1}
                                                       \mid  D_{it} = d^L, \bX_{i,t+1}, \bC^B_{i,t+1}] \bigr\} d F_{\bX_{i,t+1}, \bC^B_{i,t+1} \mid D_{it}=d^L}(\bx, \bc)\\
    \hspace{-0.4in}  && - \int \bigl\{ \widehat{\E}[Y_{it} \mid D_{it}=d^H, \bX_{it}, \bC^B_{i,t+1}]  -  \widehat{\E}[Y_{it} \mid D_{it}=d^L, \bX_{it},
                        \bC^B_{i,t+1}] \bigr\} d F_{\bX_{i,t+1}, \bC^B_{i,t+1} \mid D_{it}=d^L}(\bx, \bc).
  \end{eqnarray*}
  The proof is analogous to Theorem~\ref{biascorrect}.  Finally, by combining this result and Theorem~\ref{biascorrect} with
  weights $\Pr(D_{it} = d^H)$ and $\Pr(D_{it} = d^L)$, we can get
  \begin{eqnarray*}
    \E[\hat{\tau}^\ast_{\text{BC}}]  & = & \tau_{t+1} (d^H, d^L).
  \end{eqnarray*}
  under Assumption~\ref{timeInv}.1, Assumption~\ref{timeInv}.2, and
  Assumption~\ref{timeInv}.3, where
  \begin{eqnarray*}
    \hat{\tau}^\ast_{\text{BC}}  \equiv  \Pr(D_{it} = d^H) \hat{\tau}_{\text{BC}}  + \Pr(D_{it} = d^L) \hat{\tau}^{d^L}_{\text{BC}}.\label{eq:bc_final}
  \end{eqnarray*}
  \section{Causal Directed Acyclic Graphs: Review}
  \label{sec:dag_review}
\begin{figure}[!h]
    \begin{center}
      \begin{tabular}{c}
        \begin{minipage}[b]{0.45\hsize}
          \begin{center}
            \includegraphics[width=3.3cm]{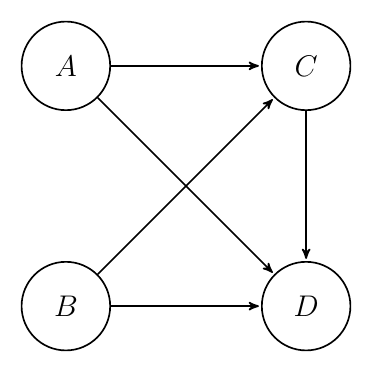} \vspace{-0.2in} 
          \end{center}
          (a) A causal directed acyclic graph
        \end{minipage}
        \begin{minipage}[b]{0.45\hsize}
          \centering     
          \begin{eqnarray*}
            A & = & f_{A} (\epsilon_{A})\\
            B & = & f_{B} (\epsilon_{B})\\
            C & = & f_{C} (A, B, \epsilon_{C})\\
            D & = & f_{D} (A, B, C, \epsilon_{D})
          \end{eqnarray*}
          (b) A structural equation model
        \end{minipage}
      \end{tabular}
    \end{center}
    \vspace{-0.2in}\caption{An Example of Causal DAGs and SEMs}\label{ex}
  \end{figure}

  In the paper, we use a causal directed acyclic graph and
  nonparametric structural equations to represent causal
  relationships. Here, we review basic definitions and results. See
  \cite{pearl2000causality} for a comprehensive review. Following
  \cite{pearl1995causal}, we define a causal directed acyclic
  graph (causal DAG) to be a set of nodes and directed edges among nodes such
  that the graph has no cycles and each node corresponds to a
  univariate random variable. Each random variable is given by its nonparametric structural
  equation. When there is a directed edge from one variable to another
  variable, the latter variable is a function of the former variable. 
  For example, in a causal DAG in
  Figure~\ref{ex} (a), four random variables $(A, B, C, D)$ are given by 
  nonparametric structural equations in Figure~\ref{ex} (b);  $A  =  f_{A}
  (\epsilon_{A}), B  =  f_{B} (\epsilon_{B}), C  =  f_{C} (A, B,
  \epsilon_{C})$, and $D  =  f_{D} (A, B, C, \epsilon_{D}),$ where $f_A, f_B, f_C$ and $f_D$ are unknown nonparametric structural
  equations and $(\epsilon_{A}, \epsilon_B, \epsilon_C, \epsilon_{D})$ are
  mutually independent errors. The node that a directed edge starts from
  is called the {\it parent} of the node that the edge goes into. The
  node that the edge goes into is the {\it child} of the node it comes
  from. If two nodes are connected by a directed path, the first node is
  the {\it ancestor} of every node on the path, and every node on the
  path is the {\it descendant} of the first node
  \citep{pearl2000causality}. For example, node A is a parent of node C, 
  and nodes C and D are descendants of node B. The requirement that the errors be mutually 
  independent essentially means that there is no variable absent from
  the graph which, if included on the graph, would be a parent of two or more variables. 

  The nonparametric structural equations are general -- random
  variables may depend on any function of their parents
  and variable-specific errors. They encode counterfactual relationships between the 
  variables on the graph by recursively representing
  one-step-ahead counterfactuals. Under a hypothetical intervention
  setting $A$ to $a$, the distribution of the variables $B, C$, and $D$ are then
  recursively given by the nonparametric structural equations with  $A  =
  f_{A}(\epsilon_{A})$ replaced by $A=a$. Specifically, $B =  f_B
  (\epsilon_B),  \ C = C(a)  =  f_{C} (A=a, B, \epsilon_{C})$, and $D  =
  D(a) = f_{D} (A=a, B, C=C(a), \epsilon_{D})$ where $C(a), D(a)$ are the counterfactual
  values of $C$ and $D$ when $A$ is set to $a$. 

  \section{Simulation Study}
  \label{sec:sim}
  In this section, we consider the performance of the proposed placebo
  test and bias-corrected estimator in a simulation study calibrated to
  the real hate crime data. In Section~\ref{subsec:simpl}, we show that (1)  a placebo estimator is
  consistent for zero under the no omitted confounders assumption as
  Theorem~\ref{placebo} implies and (2) the statistical power of the
  proposed placebo test is comparable to an ``oracle'' test --- test whether an estimated ACDE is statistically
  distinguishable from the true ACDE, which is available only in
  simulations. In Section~\ref{subsec:simbc}, we demonstrate that the
  bias-corrected estimator reduces bias and root mean squared
  error (RMSE) even under a slight violation of the time-invariance
  assumption (Assumption~\ref{timeInv}). 

  \paragraph{Setup.}
  To approximate realistic data generating processes, we use the same
  hate crime data as in the main application but focus on another
  important outcome, the number of attacks against refugee housing,
  which is also an important aspect of hate crimes studied in the
  literature. As for observed covariates, we include five major
  contextual variables; the number of refugees, the number of crimes per
  100,000 inhabitants, per capita income, the unemployment rate, and the
  share of school leavers without lower secondary education
  graduation. We fit a linear regression with these five covariates, as
  in equation~\eqref{eq:main}, to estimate the basic parameters of the
  data generating process.  

  We simulate a distance
  matrix $\bW$ based on the stochastic block model
  \citep{holland1983stochastic} for each of the sample size $n \in
  \{100, 500, 1000, 2000\}$. Each group consists of ten units and there
  exist $K = n/10$ groups. $K$ groups are divided into $L = K/5$
  blocks. If units $i$ and $j$ are within the same group, $\Pr(W_{ij} =
  1) = 0.8.$ If units $i$ and $j$ are within the same block but not in
  the same group, $\Pr(W_{ij} = 1) = 0.2.$ If units $i$ and $j$ are in
  different blocks, $\Pr(W_{ij} = 1) = 0.$ This setup is designed to
  ensure that the network dependency does not keep growing as the
  sample size grows. See \cite{savje2017average} and
  \cite{ogburn2017causal} for general discussions on network
  asymptotics. 

  We then simulate an unobserved contextual variable $U_{it}.$ In particular, we consider
  two scenarios; (1) time-invariant confounding where 
  assumptions for both the placebo test and the bias-corrected estimator
  hold, and (2) structural stationarity where assumptions hold for the
  placebo test but the time-invariance assumption required for the
  bias-correction is violated. For the first scenario, we set unobserved
  contextual variable $U$ to be time-invariant where $U_{i} =
  \widetilde{U}_{k[i]}$ where $\tilde{U}_k \sim \cN(0, 0.5)$ and $k[i]$ is a
  group indicator for unit $i$.  For the second scenario, we draw
  unobserved contextual variable $U$ as follows. $U_{it} =
  \widetilde{U}_{k[i], t}$ where $U_{k, t} = 0.9U_{k, t-1} + \cN(0, 0.1)$ where $U_{k0} \sim \cN(0, 0.5).$ 

  Given this setup, we sample potential outcomes using the following data generating process.  
  \begin{equation}
    Y_{i,t+1}(D_{it})  =  \alpha + \tau D_{it} + \bX_{i,t+1}^\top\beta + \gamma U_{i,t+1} + \epsilon_{i,t+1},
  \end{equation}
  for sample size in each time period $n \in \{100, 500, 1000, 2000\}$
  and the total number of time periods $T = 20$. $D_{it} \equiv
  \bW_{i}^\top \bY_t$ indicates the treatment variable,
  five-dimensional vector $\bX_{i,t+1}$ represents five observed covariates
  from the real hate crime data, $U_{i,t+1}$ is the unobserved
  contextual confounder affecting multiple units, and the error term
  $\epsilon_{i,t+1}$ follows the normal distribution, $\epsilon_{i,t+1} \sim \cN(0, 0.1)$. Coefficients
  \{$\alpha = 0.59, \tau = 0.74, \beta = (0.75, -0.11, -0.28, -3.38,
  3.90)$\} are based on estimated parameters from the real hate crime
  data. The effect of unobserved contextual confounder $U$ is set to
  $\gamma = 0.1.$ Based on this data generating process, we conduct 5000
  independent Monte Carlo simulations. 

  \subsection{Placebo Test}
  \label{subsec:simpl}
\begin{figure}[!t]
    \begin{center}
      \includegraphics[width = 0.7\textwidth]{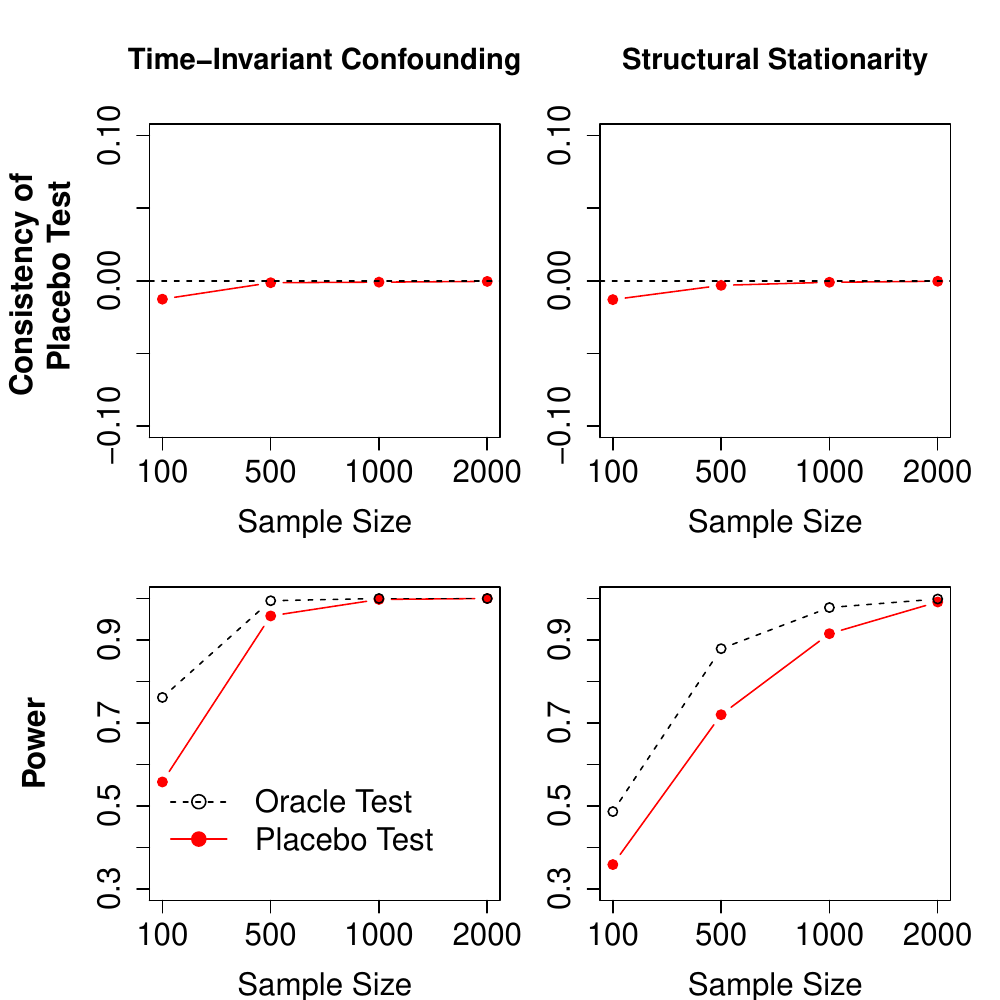}
    \end{center}
    \vspace{-0.2in} 
    \spacingset{1}{\caption{Simulation Results on Placebo Test. {\it
          Note:} The first row considers the consistency of the
        placebo test under the no omitted confounders assumption. 
        The second row compares the
        statistical power of the proposed placebo test (solid
        red line) and the oracle test (dotted black line). The first and
        second columns correspond to the time-invariant confounding and
        the structural stationarity, respectively. Results are based on 5000 Monte Carlo draws using
        four sample sizes.}\label{fig:sim-pl}}
  \end{figure}

  First, we consider the consistency of the proposed placebo test under the
  no omitted confounders assumption. Theorem~\ref{placebo} implies that
  when the no omitted confounders assumption holds, the treatment
  variable and the lagged dependent variable are conditionally
  independent. In particular, we fit a placebo regression: 
  \begin{equation}
    Y_{it}  =  \alpha_0 + \delta D_{it} + \tau_0 D_{i,t-1} + \bX_{it}^\top\beta_0 + \gamma_0 U_{it} + \epsilon_{it}.
  \end{equation}
  We expect that a test statistic $\widehat{\delta}$ is consistent for zero under the no omitted
  confounders assumption. The first row in Figure~\ref{fig:sim-pl}
  presents the results. As Theorem~\ref{placebo} shows, under the no omitted
  confounders assumption, the placebo estimator $\widehat{\delta}$
  converges to zero as the sample size grows. Because Theorem~\ref{placebo}
  only requires the structural stationarity, the placebo test is
  consistent under both scenarios.  

  We also investigate the statistical power of the proposed placebo test
  when the no omitted confounders assumption is violated. We fit a placebo regression: 
  \begin{equation}
    Y_{it}  =  \tilde{\alpha}_0 + \tilde{\delta} D_{it} + \tilde{\tau}_0 D_{i,t-1} + \bX_{it}^\top\tilde{\beta}_0 + \tilde{\epsilon}_{it}.
  \end{equation}
  The key difference is that this regression now ignores contextual
  confounder $U_{it}$. Here, $\widehat{\tilde{\delta}}$ serves as a test
  statistic for the placebo test. We compare this to an oracle test
  where we fit the following main linear regression,
  \begin{equation}
    Y_{i,t+1}  =  \alpha_m + \tau_m D_{it} + \bX_{i,t+1}^\top\beta_m  + \xi_{i,t+1},
  \end{equation} 
  and test $H_0: \tau_m = \tau.$ This test is an ``oracle'' test 
  because it is available only in the simulation where we know the true
  ACDE $\tau.$  The second row in Figure~\ref{fig:sim-pl}
  presents the results. Even when the sample size is small, the proposed placebo
  test achieves more than 70\% of the oracle test's
  power. As the sample size grows, the proposed placebo test attains the
  statistical power as high as that of the oracle test. Given that
  the oracle test is available only in simulations where the true ACDE
  is known, these results suggest that the placebo test can serve as a
  powerful practical tool to detect biases in applied settings.  

\clearpage
  \subsection{Bias-Corrected Estimator}
  \label{subsec:simbc}
  In Section~\ref{subsec:calibrate}, we show that the proposed
  bias-corrected estimator can identify the ACDE for the treated under
  Assumption~\ref{timeInv}. Here, we investigate how much
  the bias-corrected estimator can reduce bias and RMSE even in settings where
  this required time-invariance assumption is slightly violated. 

  In particular, we compare an uncorrected estimator, which ignores
  unobserved contextual confounder $U$, and the proposed bias-corrected
  estimator under two scenarios; (1) time-invariant confounding and (2)
  structural stationarity. The time-invariance assumption required for
  the bias correction (Assumption~\ref{timeInv}) holds in the first but not in the second scenario. 

\begin{figure}[!h]
    \begin{center}
      \includegraphics[width = 0.8\textwidth]{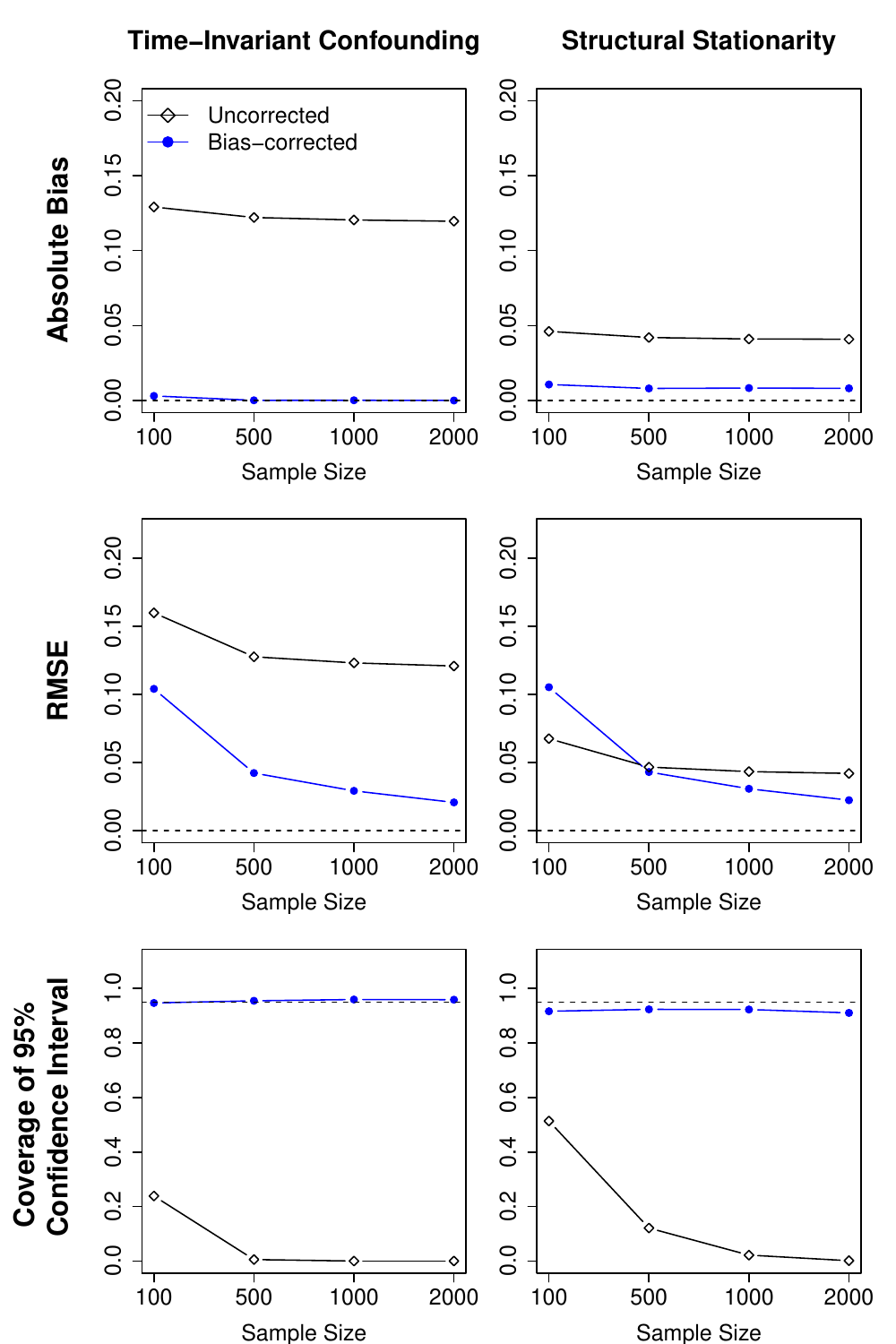}
    \end{center}
    \vspace{-0.2in} 
    \spacingset{1}{\caption{Simulation Results on Bias-Corrected
        Estimator. {\it Note:} 
        The first row compares the absolute bias of the uncorrected
        estimator (empty black square) and the bias-corrected estimator
        (solid blue circle). The second row examines the root mean squared
        error (RMSE) and the third row shows the coverage of the 95\%
        confidence interval. The first and second columns correspond to the time-invariant confounding and
        the structural stationarity, respectively. Results are based on 5000 Monte Carlo draws using
        four sample sizes.}\label{fig:sim-bc}}
  \end{figure}

  Figure~\ref{fig:sim-bc} presents the simulation results. In the
  time-invariant confounding case (the first column), whereas the bias in the
  conventional uncorrected estimator is about $0.12$, the bias in the proposed
  bias-corrected estimator is essentially $0$. The bias is corrected as
  Theorem~\ref{biascorrect} implies. The RMSE also significantly
  improves upon the uncorrected conventional estimator. The 95\%
  confidence interval is close to its nominal coverage rate in contrast
  to that of the uncorrected estimator. 

  More
  importantly, even in the structural stationarity case (the second
  column in Figure~\ref{fig:sim-bc}) where the required assumption for the bias correction is
  slightly violated, the bias-corrected estimator shows reasonable
  performance. While the bias in the
  conventional uncorrected estimator is about $0.04$, the bias in the proposed
  bias-corrected estimator is less than $0.01$. Although the bias does
  not vanish, it reduces by about 80\%. This benefit is also clear in
  the results of RMSE. Because the bias-corrected estimator tends to have a
  larger standard error, the RMSE of the bias-corrected estimator is
  bigger than the one of the uncorrected estimator when the sample size is
  small. However, as the sample size grows, the bias-corrected estimator
  outperforms the uncorrected estimator. Finally, as the required
  time-invariance assumption is violated, the coverage of the 95\%
  confidence interval for the bias-corrected estimator is slightly smaller than its nominal coverage
  rate, but it attains more than 90\% in contrast to the performance of
  the uncorrected estimator. These results suggest that the proposed bias-corrected
  estimator can reduce bias and RMSE in applied settings where the necessary
  assumption might hold only approximately. 

  \clearpage
  \section{Empirical Analysis in Section~\ref{sec:app}}
  \label{sec:app_em}
  \subsection{Control Sets and Placebo Sets}
  \label{subsec:hate-control}
  We investigate five different control sets to illustrate how to use the
  proposed placebo test and bias-corrected estimator. Table~\ref{tab:control}
  describes types of variables we use for those five control sets
  and their corresponding placebo sets. The column of ``Main model''
  indicates variables used for control sets and the column of ``Placebo
  model'' indicates corresponding variables in placebo sets. 

  \vspace{0.2in}
  \renewcommand{\arraystretch}{1.5}
  \begin{table}[!h]
    \begin{center}
      \scriptsize
      \scalebox{0.9}{
        \begin{tabular}{l|l|l}
          Type &  {\bf Main Model} &  {\bf Placebo Model} \\
          \hline
          {\bf Outcome} &   Physical Attack$_{t+1}$ & Physical Attack$_{t}$\\
          \hline
          {\bf Treatment} &   Physical Attack$_t$ in Neighbors 
                                   &  Physical Attack$_{t}$ in Neighbors\\
          \hline
          \multicolumn{1}{l|}{{\bf A Control Set/A Placebo Set}} & & \\
          \cdashline{1-3} 
          \ {\bf Basic Variables} & Physical Attack$_t$ & Physical Attack$_{t-1}$ \\
               &   Physical Attack$_{t-1}$ in Neighbors & Physical
                                                          Attack$_{t-1,
                                                          t-2}$ in Neighbors \\
               &   the number of neighbors & the number of neighbors \\
               &   variance of $\bW_i$ & variance of $\bW_i$ \\
          \cdashline{1-3} 
          \ {\bf Two-month Lags}   & Physical Attack$_{t-1}$ & Physical Attack$_{t-2}$ \\
          \cdashline{1-3} 
          \ {\bf Contextual Variables} (annual) &  & \\    
          \ \ \ Refugee variables & Total number of refugees & Total number of refugees \\    
               & Total number of foreign born & Total number of foreign born\\
          \ \ \ Population variables & Population size & Population size \\
               & Share of male inhabitants & Share of male inhabitants \\
          \ \ \ Crime variables & Number of general crimes per 100,000 inhabitants & Number of general crimes per 100,000 inhabitants\\
               & Percent of general crimes solved & Percept of general crimes solved \\
          \ \ \ Economic variables & Number of newly registered business& Number of newly
                                                                          registered business\\
               & Number of newly deregistered business& Number of newly deregistered business\\
               & Number of insolvency & Number of insolvency \\
               & per capita income & per capita income \\
               & Number of employees with social security & Number of employees with social security \\
               & Unemployment rate & Unemployment rate \\
          \ \ \ Education variables & Share of school leavers & 
                                                                Share of school leavers \\
               & \ \ without lower secondary education
                 graduation  & \ \ without lower secondary education graduation\\
          \ \ \ Political variables & Turnout rate in 2013 & Turnout rate in 2013 \\
               & Vote share of extreme right and & Vote share of extreme right and \\
               & \ \ populist right-wing parties in 2013 & \ \ populist right-wing parties in 2013 \\
          \hline
        \end{tabular}}
      \caption{Five Control Sets and Placebo Sets: Spatial Diffusion of Hate Crimes.} \label{tab:control}
    \end{center}
  \end{table}
  \renewcommand{\arraystretch}{1.0}
  The first control set (C1) includes variables from ``Basic
  Variables.'' The second control set (C2) adds variables from
  ``Two-month Lags'' to the first control set. The third control set adds
  state fixed effects to the second control 
  set. The fourth control set adds all the variables from ``Contextual
  Variables,'' which include variables on refugees, demographics,
  general crimes, economic indicators, education, and politics. Note
  that these contextual variables are measured only annually. 
  The final fifth set adds the time trend variable as third-order polynomials to the fourth set. 

  \clearpage
  \subsection{Conditional ACDEs by Education}
  \label{subsec:hate-condition}
\begin{figure}[!h]
    \begin{center}
      \includegraphics[width=\textwidth]{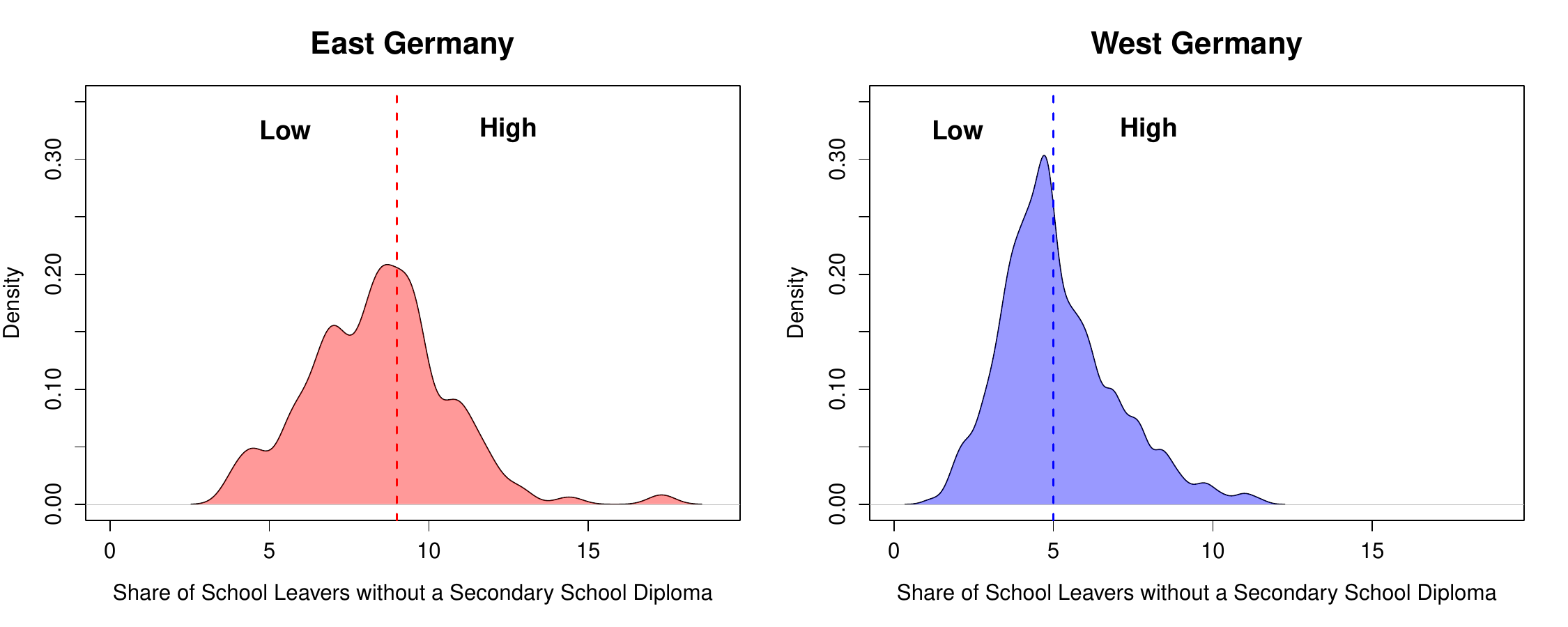}\\
    \end{center} \vspace{-0.2in}
    \spacingset{1}{\caption{Distribution of Proportions of School Dropouts. {\small Note: For East Germany, we use
          9\% as a cutoff for high and low proportions of school dropouts, which is approximately
          the median value in East Germany. For West Germany, we use
          5\% as a cutoff for high and low proportions of school dropouts, which is approximately
          the median value in West Germany.}}\label{fig:dist}} \vspace{-0.1in}
  \end{figure}

  We present the distribution of proportions of school dropouts
  without a secondary school diploma, separately for East Germany and
  West Germany. Because these distributions
  are substantially different between them (Figure~\ref{fig:dist}), we estimate the conditional
  ACDE by proportions of school dropouts, separately for the East and the West. \vspace{-0.1in}
  \\
  Next, we present the conditional ACDE for counties in East Germany with
  low proportions of school dropouts. In contrast to
  Figure~\ref{fig:marCE}, estimates are small.
  \vspace{-0.05in}
  \begin{figure}[!h]
    \begin{center}
      \includegraphics[height=2in]{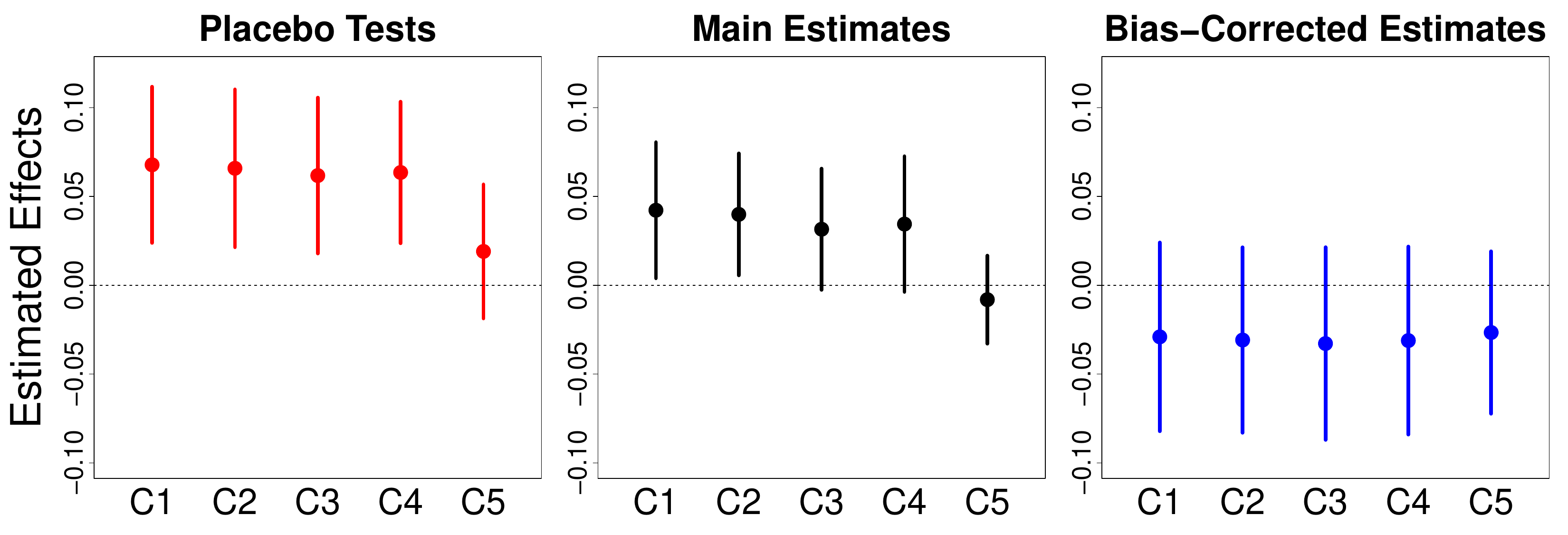}\\
      \hspace{0.05in} \ \ \ \ \ (a) \hspace{1.575in} (b) \hspace{1.575in}
      (c) \vspace{-0.15in}
    \end{center}
    \spacingset{1}{\caption{Results of the conditional ACDE (Low Proportion of School
        Dropouts, East). {\small Note: Figure (a) shows that the
          last fifth set produces the smallest placebo estimate. Focusing
          on this fifth control set, 
          a point estimate of the ACDE in Figure (b) is close to zero and
          its 95\% confidence interval covers zero. Figure (c)
          shows that bias-corrected estimates are similar regardless of the selection of
          control variables and all of their 95\% confidence intervals cover zero.}}\label{fig:hate-eastlow}}
  \end{figure}

  \clearpage
  \noindent Now, we present the conditional ACDEs for counties in West Germany with
  high and low proportions of school dropouts. Given that proportions of
  school dropouts are lower in West Germany, estimates of the
  conditional ACDEs are small, in contrast to
  Figure~\ref{fig:marCE}.
  \vspace{0.3in}
  \begin{figure}[!h]
    \begin{center}
      \includegraphics[height=2in]{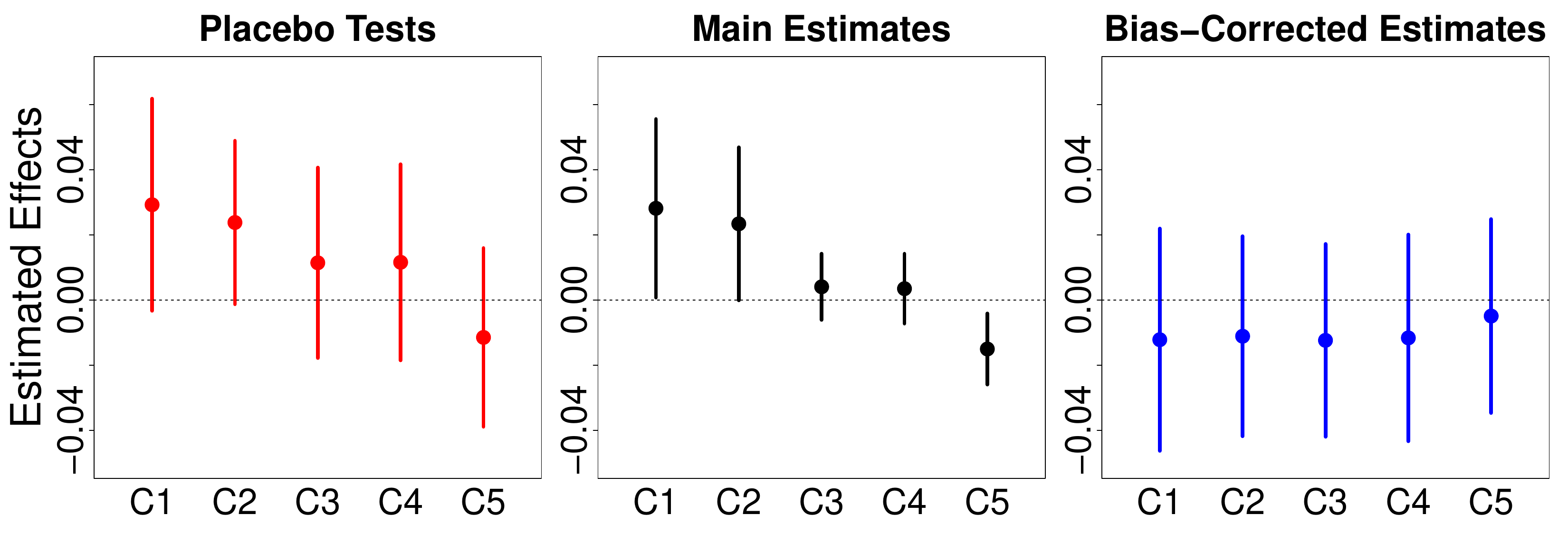}\\
      \hspace{0.05in} \ \ \ \ \ (a) \hspace{1.575in} (b) \hspace{1.575in}
      (c) \vspace{-0.15in}
    \end{center}
    \spacingset{1}{\caption{Results of the conditional ACDE (High Proportion of School
        Dropouts, West). {\small Note: Figure (a) shows that the
          third, fourth and fifth sets produce small placebo estimates. Focusing
          on these sets, point estimates of the ACDE in Figure (b) are
          close to zero and sometimes negative. Figure (c)
          shows that bias-corrected estimates are similar regardless of the selection of
          control variables and all of their 95\% confidence intervals cover zero.}}\label{fig:hate-westhigh}}
  \end{figure}
  \vspace{0.3in}
  \begin{figure}[!h]
    \begin{center}
      \includegraphics[height=2in]{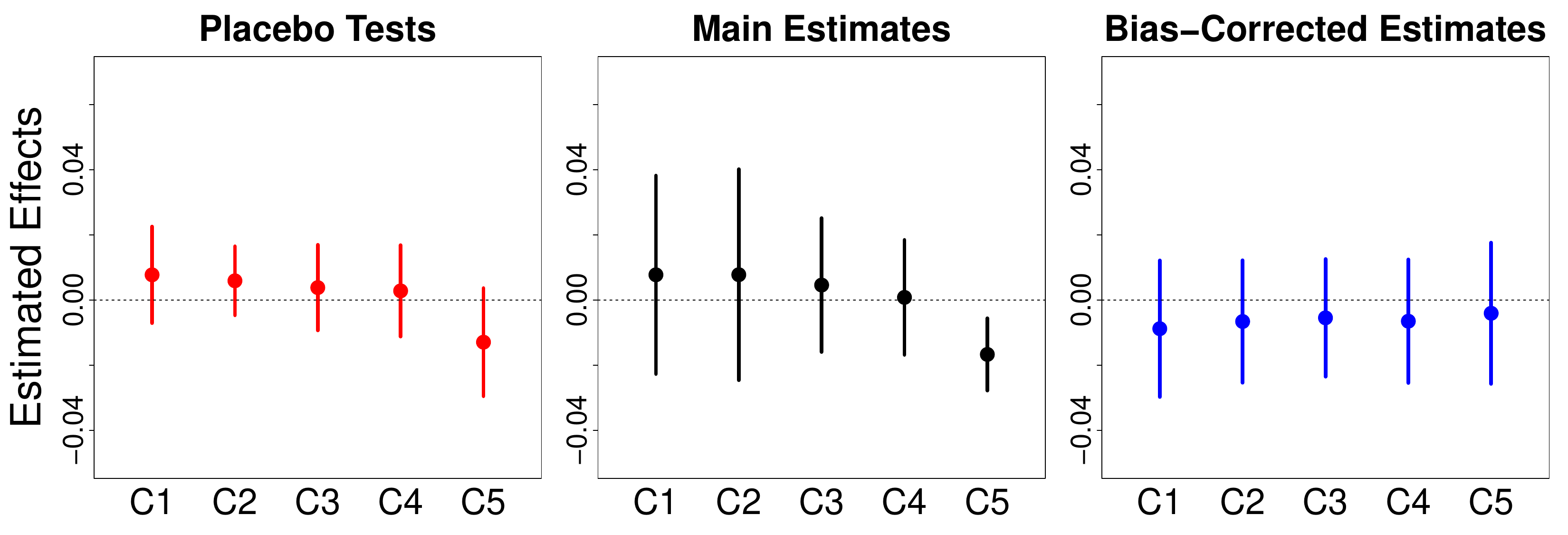}\\
      \hspace{0.05in} \ \ \ \ \ (a) \hspace{1.575in} (b) \hspace{1.575in}
      (c) \vspace{-0.15in}
    \end{center}
    \spacingset{1}{\caption{Results of the conditional ACDE (Low Proportion of School
        Dropouts, West). {\small Note: Figure (a) shows that all
          the sets produce small placebo estimates. This is partly because
          there are few hate crimes in this area and hence, there is no
          variation in outcomes and treatments. In addition, point estimates of the ACDE in Figure (b) are
          close to zero and sometimes negative. Figure (c)
          shows that bias-corrected estimates are similar regardless of the selection of
          control variables and all of their 95\% confidence intervals cover zero.}}\label{fig:hate-westlow}}
  \end{figure}

\end{document}